\documentclass[preprint,3p,12pt]{elsarticle}

\usepackage{mathrsfs}
\usepackage{amsmath}
\usepackage{stmaryrd}
\usepackage{bbding}
\usepackage{dcolumn}
\usepackage{graphicx}
\usepackage{amsfonts}
\usepackage{amssymb}
\usepackage{psfrag}
\usepackage{wrapfig}
\usepackage{subfigure}
\usepackage{makeidx}
\usepackage{bm}
\usepackage{epsf}
\usepackage{epsfig}
\usepackage{setspace}
\usepackage{graphicx}
\usepackage{epstopdf}
\usepackage{psfrag}
\usepackage{subfigure}
\usepackage{color}
\usepackage{comment}
\usepackage{algorithm}
\usepackage{algorithmic}
\newcommand \td {\mathrm{~d}}
\journal{}

\begin{document}



\title{Revisit viscous shock tube at low Reynolds number}

\author[HKUST1]{Yue Zhang}
\ead{yzhangnl@connect.ust.hk}	
\author[HKUST1,HKUST2,HKUST3]{Kun Xu\corref{cor1}}
\ead{makxu@ust.hk}

\address[HKUST1]{Department of Mathematics, Hong Kong University of Science and Technology, Clear Water Bay, Kowloon, Hong Kong}
\address[HKUST2]{Department of Mechanical and Aerospace Engineering, Hong Kong University of Science and Technology, Clear Water Bay, Kowloon, Hong Kong}
\address[HKUST3]{Shenzhen Research Institute, Hong Kong University of Science and Technology, Shenzhen, China}
\cortext[cor1]{Corresponding author}

\begin{abstract}
  The viscous shock tube is a canonical test case for assessing Navier–Stokes (NS) solvers in the continuum-flow regime, widely used to validate numerical accuracy and probe flow physics. It features a rich set of interacting structures—shock and rarefaction waves, contact discontinuities, boundary layers, and their couplings—spanning multiple spatial and temporal scales. However, NS-based modeling, which presumes near-equilibrium behavior, may fail to capture important non-equilibrium effects even in nominally continuum conditions. This study investigates the viscous shock tube at low Reynolds numbers and demonstrates the presence of non-equilibrium phenomena within the conventional continuum regime. To obtain physically consistent solutions across scales, we employ the unified gas-kinetic scheme (UGKS) and compare its results with NS solutions computed using the gas-kinetic scheme (GKS). Discrepancies between UGKS and GKS solutions reveal pronounced non-equilibrium effects in regions where shock waves interact with boundary layers. For continuum flows at high Mach and low Reynolds numbers, such multiscale non-equilibrium transport becomes important, underscoring the need for multiscale methods in analysis and prediction.
\end{abstract}
\begin{keyword}
  Viscous shock tube problem, non-equilibrium effect, GKS, UGKS, low-Reynolds-number regime.
\end{keyword}
\maketitle
%{\bf MSC Codes }  {\it(Optional)} Please enter your MSC Codes here

\section{Introduction}

The viscous shock tube is a standard test case for assessing the accuracy of numerical methods in continuum-flow regimes. This configuration features a reflected shock interacting with a developing boundary layer and is relevant to applications such as high-speed aerodynamic facilities that investigate chemical relaxation in high-temperature gas mixtures or generate high-enthalpy reservoir conditions. In this setting, the reflected shock encounters an incident boundary layer whose properties have been altered by end-wall conditions, producing a complex, unsteady coupling between the shock and the boundary layer. When the boundary-layer stagnation pressure falls below that downstream of the normal reflected shock, separation occurs and a characteristic lambda-shaped shock system forms. The seminal analysis of this interaction was provided by \cite{mark1957interaction}, who used a simplified model, followed by experiments that corroborated his predictions. Subsequent detailed experimental studies in shock tubes in (\cite{davies1969influence,matsuo1974interaction}) refined and extended Mark’s original framework. However, at low Reynolds numbers, pronounced non-equilibrium effects arise that are not captured by the traditional Navier–Stokes equations.

Investigating viscous shock tube problems at low Reynolds numbers (Re) is of paramount importance, driven by the growing demand for high-fidelity modeling in micro-scale gas dynamics and high-altitude flight regimes. In these environments, which are characterized by minute physical dimensions or extremely low gas densities, the flow physics deviates significantly from the classical continuum assumption. Prominent engineering applications include micro-thrusters in micro-electro-mechanical systems (MEMS) \cite{xu2024review}, where pressure wave propagation is inherently viscous-dominated, and the early-stage atmospheric re-entry of hypersonic vehicles \cite{li2019gas}, where rarefaction effects are critical. Unlike high-Re flows, these low-Re conditions exhibit strong coupling between shock structures and thick boundary layers, with non-equilibrium effects and kinetic-scale interactions dictating the overall flow field. Consequently, the low-Re shock tube serves as an essential canonical model for evaluating the capability of multiscale numerical schemes to capture these non-classical features. Furthermore, its relatively simple physical setup and distinct flow characteristics make it an ideal, easily standardized computational benchmark. Recent research \cite{tian2026small} demonstrates that even subtle rarefaction effects can induce significant errors in surface stress and heat flux predictions when using the Navier–Stokes (NS) equations under turbulent conditions, thereby exposing the intrinsic limitations of the NS framework in non-equilibrium flows. Motivated by these challenges, this work investigates non-equilibrium phenomena within the viscous shock tube by comparing continuum NS solvers against a multiscale approach.

Multiscale gas flows arise frequently in aerospace engineering (\cite{bird1994molecular,xu2021unified}) and in micro-electro-mechanical systems (MEMS) (\cite{senturia1997simulating,wang2022investigation}). Across all Knudsen-number regimes, accurate prediction requires properly accounting for both free-streaming and interparticle collisions. Because non-equilibrium effects demand additional degrees of freedom, they are naturally described within a kinetic formulation.
The Boltzmann equation is the fundamental governing equation of rarefied gas dynamics. In principle, it captures multiscale physics over the full range of Knudsen numbers by resolving phenomena at the mean-free-path and collision-time scales. Mainstream numerical approaches for non-equilibrium gas flows can be broadly classified as stochastic methods—such as the direct simulation Monte Carlo (DSMC) method (\cite{bird1963approach,bird1998recent})—and deterministic methods—such as the discrete velocity method (DVM) (\cite{chu1965kinetic,yang1995rarefied}). In recent years, a class of gas-kinetic methods has been developed, including the DVM-based unified gas-kinetic scheme (UGKS) (\cite{xu2010unified,juan-chen_huang_unified_2012}), the particle-based unified gas-kinetic wave–particle (UGKWP) method (\cite{liu2020ugkwp,zhu2019ugkwp}), the discrete unified gas-kinetic scheme (DUGKS) (\cite{guo2013dugks}), and the discrete UGKWP method (\cite{yang2023dugkwp}). By coupling particle collisions and free transport within the evolution process, these methods relax the stringent mesh and time-step requirements for obtaining accurate solutions across regimes, especially in collision-dominated flows. Recently, \cite{guo2024kinetic,guo2025unified} presented a unified perspective on kinetic schemes, such as UGKS, UGKWP, and DUGKS, and clarified how their governing equations differ from the classical Navier–Stokes formulation. In this paper, we examine these differences through the viscous shock-tube problem by comparing a continuum solver (GKS) with a multiscale solver (UGKS).

The paper is organized as follows. Section \ref{sec:methodology} presents the finite-volume formulation for conservation laws and the BGK model. The BGK-based gas-kinetic scheme (GKS) and unified gas-kinetic scheme (UGKS) will be presented. Section \ref{sec:verificationAndValidation} presents the verification and validation of the GKS and UGKS in the cavity flow and 1D shock tube problem. Section \ref{sec:2Dresults} describes the viscous shock-tube configuration, including the geometry, boundary conditions, and initial conditions. The GKS and UGKS results will be presented to highlight the non-equilibrium effects in the viscous shock-tube problem. Finally, Section \ref{sec:conclusion} summarizes the main conclusions.

\section{Methodology}
\label{sec:methodology}

Traditional computational fluid dynamics (CFD) methods numerically solve the Navier-Stokes equations to determine the flow field. However, these equations lose their validity in certain regimes, such as in rarefied gas flows. To bridge this gap, direct modeling-based numerical methods are employed.

The finite volume method is used in this paper. The domain is divided into cells, for each cell $\Omega_i$, the governing equation of cell-averaged conservative values $\mathbf{W}_i=(\rho_i,\rho_i \mathbf{V}_i,\rho_i E_i)^T$, i.e. density, momentum and energy, is given by

\begin{equation}
 \label{macroscopicUpdata}
 W_i^{n+1}=W_i^n -\frac{1}{\Omega_i}\int_{t^n}^{t^{n+1}}\sum_{j\in N(i)} S_{ij} \mathbf{F}_{ij}  \td t,
\end{equation}
where $W_i^n$ and $W_i^{n+1}$ are cell-averaged conservative value at time $t^n$ and $t^{n+1}$, $|\Omega_i|$ denotes the volume of cell $i$, $N(i)$ is the set of all interface-adjcent neighboring cells of cell $i$ and $j$ is one of the neighboring cells of $i$.  The interface between cells $i$ and $j$ is labeled $ij$ and has area $S_{ij}$. The $\mathbf{F}_{ij}$ is the macroscopic flux through interface $ij$. It is important to note that Eq.(\ref{macroscopicUpdata}) represents the conservation law of macroscopic conservative quantities at the discrete level, which is a fundamental physical law valid at all temporal and spatial scales. Thus, the key point is to accurately describe the macroscopic flux $\mathbf{F}_{ij}$ through interface $ij$. According to the gas kinetic theory, the macroscopic flux can be determined by taking moments of the microscopic flux at the cell interface $ij$ through

\begin{equation*}
  \mathbf{F}_{ij}=\int \mathbf{v}\cdot \mathbf{n}_{ij} {f}_{ij}\Psi \td \mathbf{v} \td \mathbf{\xi},
  \end{equation*}
where $\mathbf{n}_{ij}$ is the normal vector of interface $ij$, ${f}_{ij}$ is the distribution function at the cell interface $ij$ and $\Psi=(1,\mathbf{v},1/2(\mathbf{v}^2+\mathbf{\xi}^2))^T$ is the collision invariants. In $\mathbf{\xi}^2=\xi_1^2+\xi_2^2+\cdots+\xi_K^2$ and $\td\mathbf{\xi}=\td\xi_1\td\xi_2\cdots\td\xi_K$,  $K$ is the number of internal degrees of freedom. The $\mathbf{v}$ is the microscopic particle translation velocity, and $\mathbf{\xi}$ is the internal variable for the rotational and vibrational degrees of freedom. The conservative values $\mathbf{W}$ are obtained by

\begin{equation*}
  \mathbf{W}=\int {f}\Psi \td \mathbf{v} \td \mathbf{\xi},
\end{equation*}
Here, the Bhatnagar-Gross-Krook(BGK) relaxation model is used to describe the process of distribution function evolution at the cell interface $ij$

\begin{equation}
  \label{BGK}
  \frac{\partial f}{\partial t} + \mathbf{v}\cdot\nabla f =\frac{f^+-f}{\tau},
\end{equation}
where $f=f(\mathbf{x},t,\mathbf{v},\mathbf{\xi})$ is the distribution function for gas molecules at physical space $\mathbf{x}$ and time $t$. Here $\mathbf{\xi}$, $\tau$ is the particle collision time, and $f^+$ is the modified equilibrium distribution function. The modified equilibrium distribution function is given by the Shakhov model

\begin{equation*}
 f^+=g\left[ 1+(1-\text{Pr})2\lambda\mathbf{c}\cdot\mathbf{q}\left(2\lambda c^2-5\right)/(5p)\right]=g+g^+,
\end{equation*}
where $g$ is the Maxwellian distribution, $\text{Pr}$ is the Prandtl number, $\mathbf{c}=\mathbf{v}-\mathbf{V}$ is the peculiar velocity, $\mathbf{V}$ is the macroscopic velocity, $\mathbf{q}$ is the heat flux, $\lambda=m/(2k_B T)$, $m$ is molecule mass, $k_B$ is Boltzmann constant, $T$ is the temperature. The Maxwellian distribution is

\begin{equation*}
 g=\rho {\left(\frac{\lambda}{\pi}\right)}^\frac{K+D}{2}e^{-\lambda ((\mathbf{v}-\mathbf{V})^2+\mathbf{\xi}^2)},
\end{equation*}
where $D$ is the spatial dimension. The heat flux is obtained through distribution function
\begin{equation*}
  \mathbf{q}=\frac{1}{2}\int (\mathbf{v}-\mathbf{V})(|\mathbf{v}-\mathbf{V}|^2+\mathbf{\xi}^2)f\td\mathbf{v} \td \mathbf{\xi}.
\end{equation*}
The collision term satisfies the compatibility condition

\begin{equation*}
 \int (f^+-f)\Psi\td \mathbf{v}\td\mathbf{\xi}=0.
\end{equation*}

 To construct the numerical flux at cell interface $\mathbf{x}_0=(0,0,0)^T$, the integral solution along the characteristic line $\mathbf{x}^\prime =\mathbf{x}_0-\mathbf{v}(t-t^\prime)$ of the BGK equation (\ref{BGK}) gives
\begin{equation*}
 f(\mathbf{x},t,\mathbf{\xi})=\frac{1}{\tau}\int_{t_0}^t f^+(\mathbf{x}^\prime,t^\prime)e^{-(t-t^\prime)/\tau}\td t^\prime + e^{-t/\tau}f_{0}(\mathbf{x}-\mathbf{v} t),
\end{equation*}
where $f_{0}(\mathbf{x})$ is the initial gas distribution function at the beginning of each step $t_n$, and $f^+(\mathbf{x},t)$ is the effective equilibrium state distributed in space and time. The integral solution provides a multiscale model of the evolution from an initial non-equilibrium distribution $f$ to an equilibrium distribution $f^+$ via collisions. To achieve second-order accuracy, the initial distribution function $f_{0}(\mathbf{x})$ is approximated as

\begin{equation*}
 f_{0}(\mathbf{x}) = \begin{cases}
 f^l_G+\mathbf{x}\cdot \nabla f^l_G,&v_n>0,\\
 f^r_G+\mathbf{x}\cdot \nabla  f^r_G,&v_n<0,\\
 \end{cases}
\end{equation*}
where $f^l_G$ and $f^r_G$ are the reconstructed initial distribution functions at the left and right sides of the interface. The equilibrium state is approximated as

\begin{equation*}
 f^+(\mathbf{x},t) \approx g_{0}+ g^+_{0} + \mathbf{x}\cdot\nabla g_{0}+\frac{\partial g_{0}}{\partial t}t,
\end{equation*}
where $g_{0}$ is the Maxwellian distribution function obtained from the conservative flow variables of colliding particles
from both sides of the interface.
For the gas kinetic scheme, the initial distribution function $f_G$ is approximated by the Chapman-Enskog expansion,
\begin{equation*}
f_G = g - \tau( \mathbf{v}\cdot\nabla g + \frac{\partial g}{\partial t}).
\end{equation*}
In this limit, the GKS recovers the macroscopical Navier-Stokes solutions. To overcome the limitations of GKS in rarefied regions, the unified gas-kinetic scheme (UGKS) is implemented to enhance predictive accuracy. For the unified gas-kinetic scheme, the distribution function also needs to be updated with the following equation

\begin{equation*}
 f_{i,k}^{n+1} =\left(1+\frac{\Delta t}{2\tau}\right)^{-1}\left[f_{i,k}^{n} + \frac{1}{\Omega_i}\sum_{j\in N(i)} S_{ij}\int_{t^n}^{t^{n+1}}f_{ij,k}\td t+\frac{\Delta t}{2} \left(\frac{f_{i,k}^{+,n+1}}{\tau}+\frac{f_{i,k}^{+,n}-f_{i,k}^{n}}{\tau}\right)\right],
\end{equation*}
where the index $k$ denotes the discrete points in the particle velocity space. And the integral over the velocity space is approximated by the summation in the discrete velocity space. More details of the GKS and the UGKS are demonstrated in earlier work \cite{xuGKS2001,xu2010unified,xu2015direct}.

\section{Schemes validation and verification}
\label{sec:verificationAndValidation}

In this section, we evaluate the GKS and UGKS methods and validate the simulation results across a range of Mach and Reynolds numbers.

\subsection{Cavity flow}

In this subsection, the cavity flow problem, a classic benchmark for validating numerical methods, is employed to verify the accuracy of the GKS and UGKS. The computational setup consists of a two-dimensional flow within a square domain containing a centrally located square obstacle. To comprehensively assess the methods, simulations are conducted at three Reynolds numbers: $20, 50$, and $100$. The working fluid is argon gas with a Prandtl number of
$\text{Pr}=0.73$. The isothermal wall temperature is fixed at $T_w=273$K, and the upper wall moves with a velocity of $U_w=50.0$m/s. 
For both the GKS and UGKS, the physical domain is discretized using a $60\times 60$ spatial mesh, with the first layer of near-wall cells refined to a height of $1\times 10^{-3}$. Additionally, for the UGKS, the velocity space is discretized into
$60\times 60$ elements. 

\begin{figure}[!htpb]
  \centering
  \subfigure[]{\includegraphics[width=0.3\textwidth]{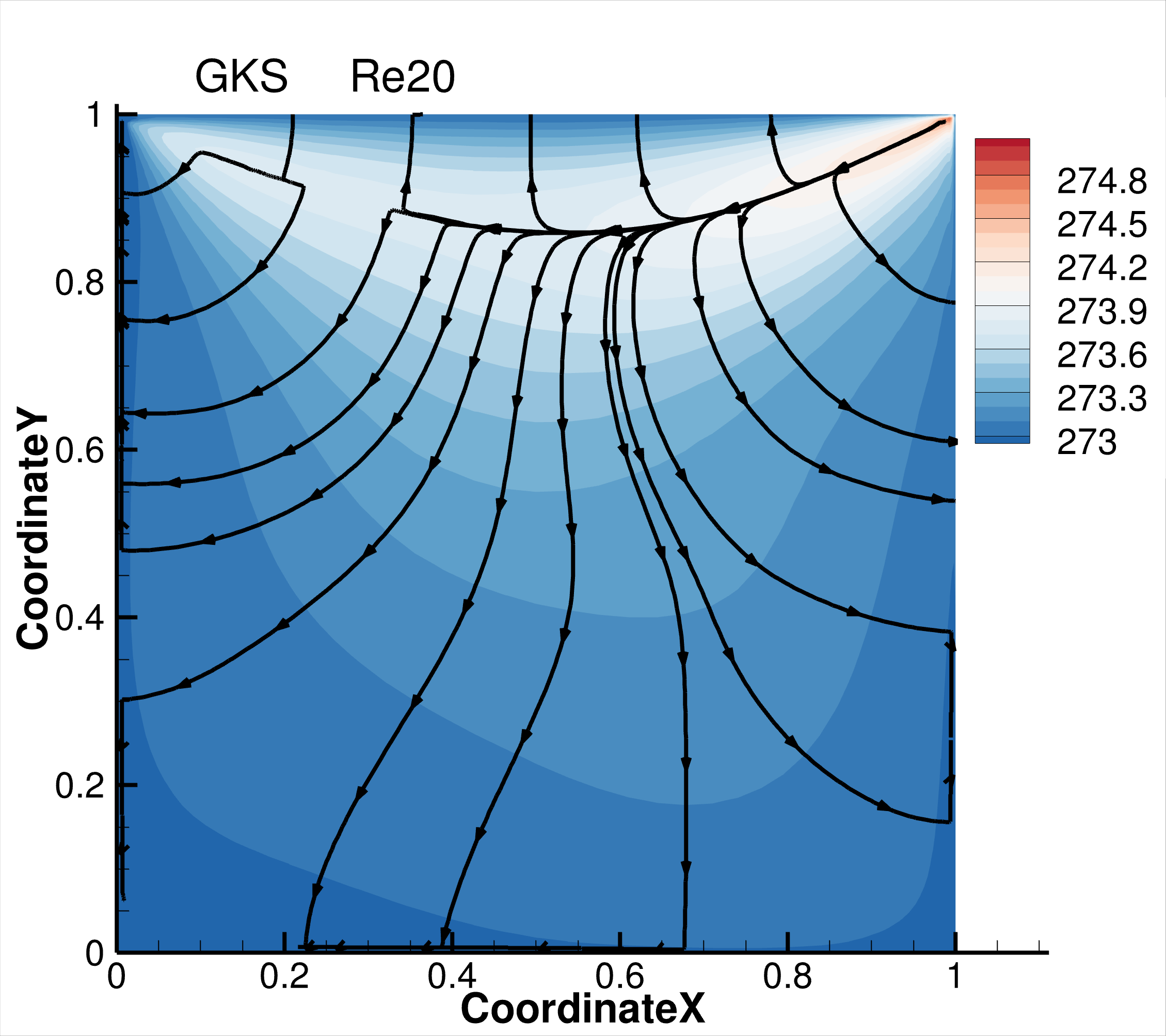}}
  \subfigure[]{\includegraphics[width=0.3\textwidth]{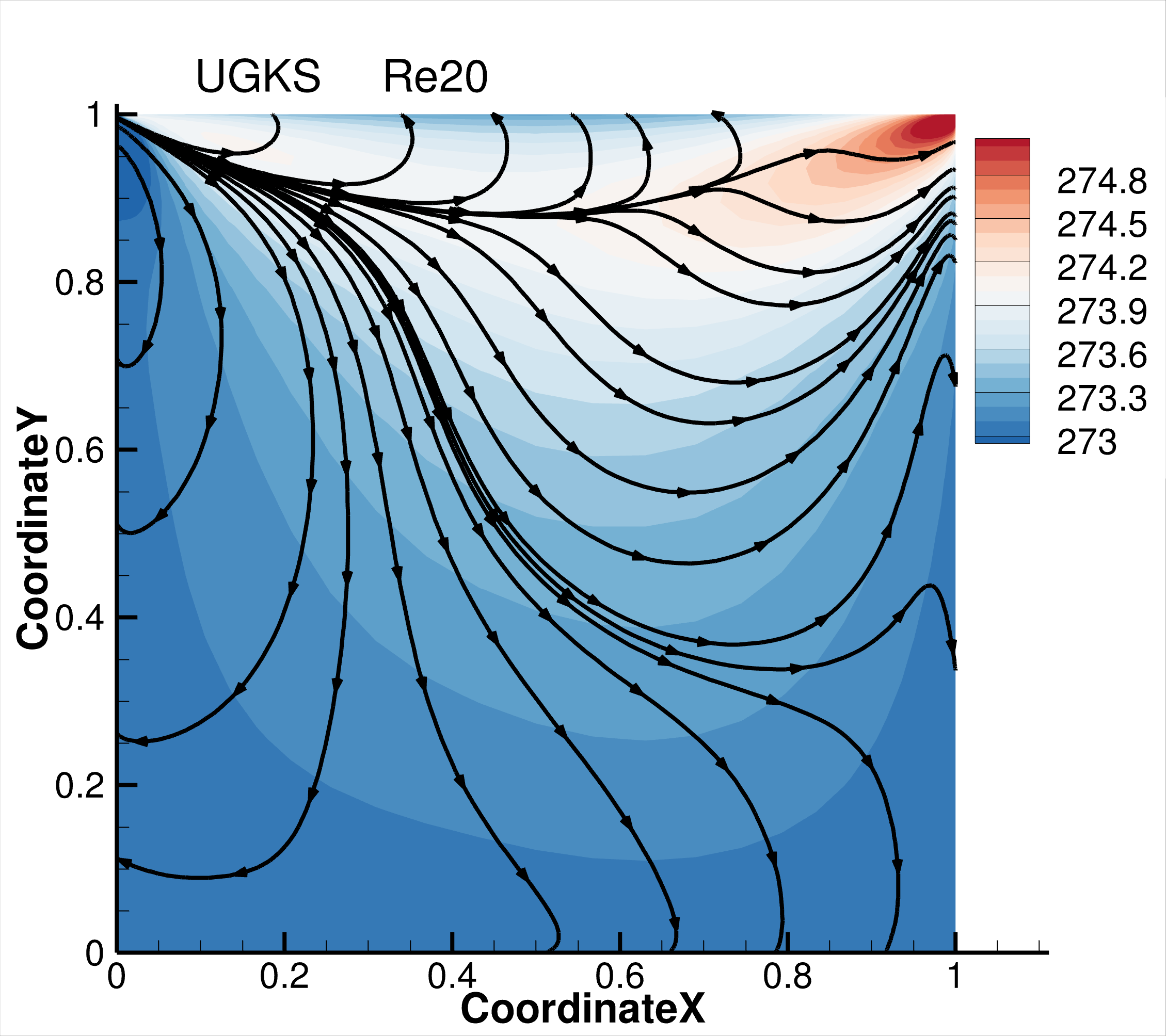}}
  \subfigure[]{\includegraphics[width=0.3\textwidth]{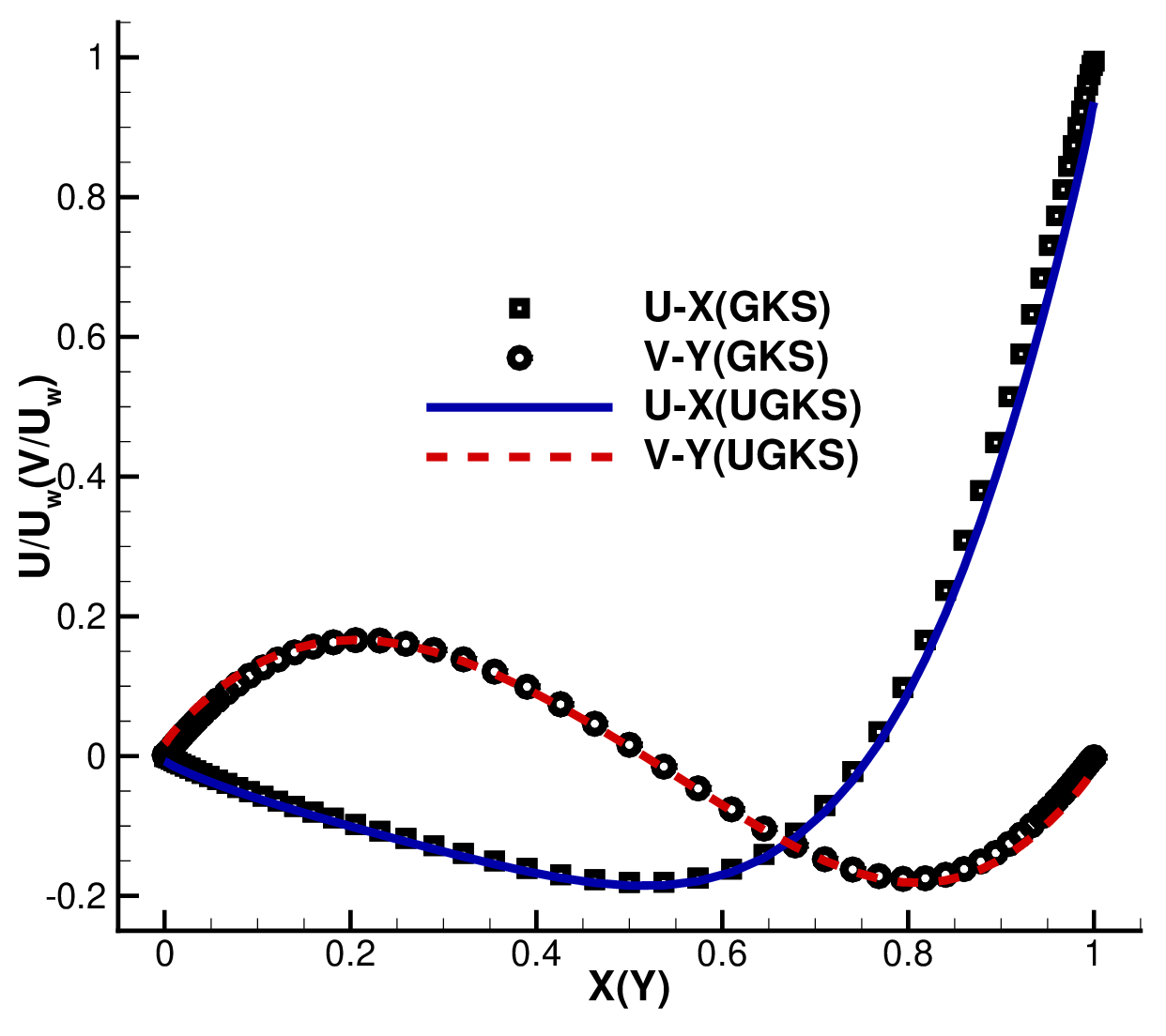}}
  \caption{Cavity simulation using GKS and UGKS at $Re=20$. (a) The temperature contour and heat flux using GKS. (b) The temperature contour and heat flux using UGKS. (c) $U$-velocity along the central vertical line and $V$-velocity along the central horizontal line, symbols:GKS, lines:UGKS.}
  \label{fig:cavitySimulationRe20}
\end{figure}

\begin{figure}[!htpb]
  \centering
  \subfigure[]{\includegraphics[width=0.3\textwidth]{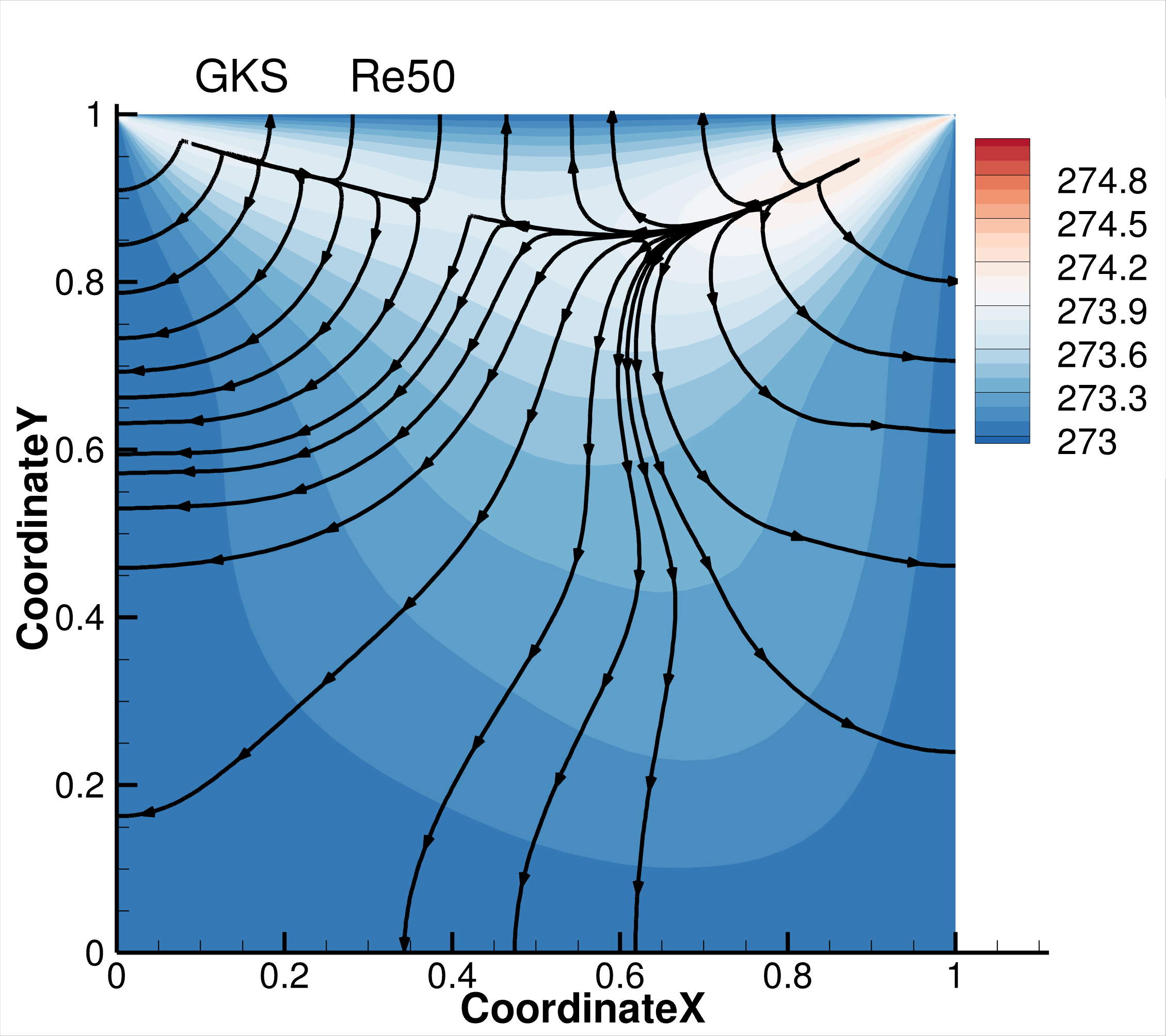}}
  \subfigure[]{\includegraphics[width=0.3\textwidth]{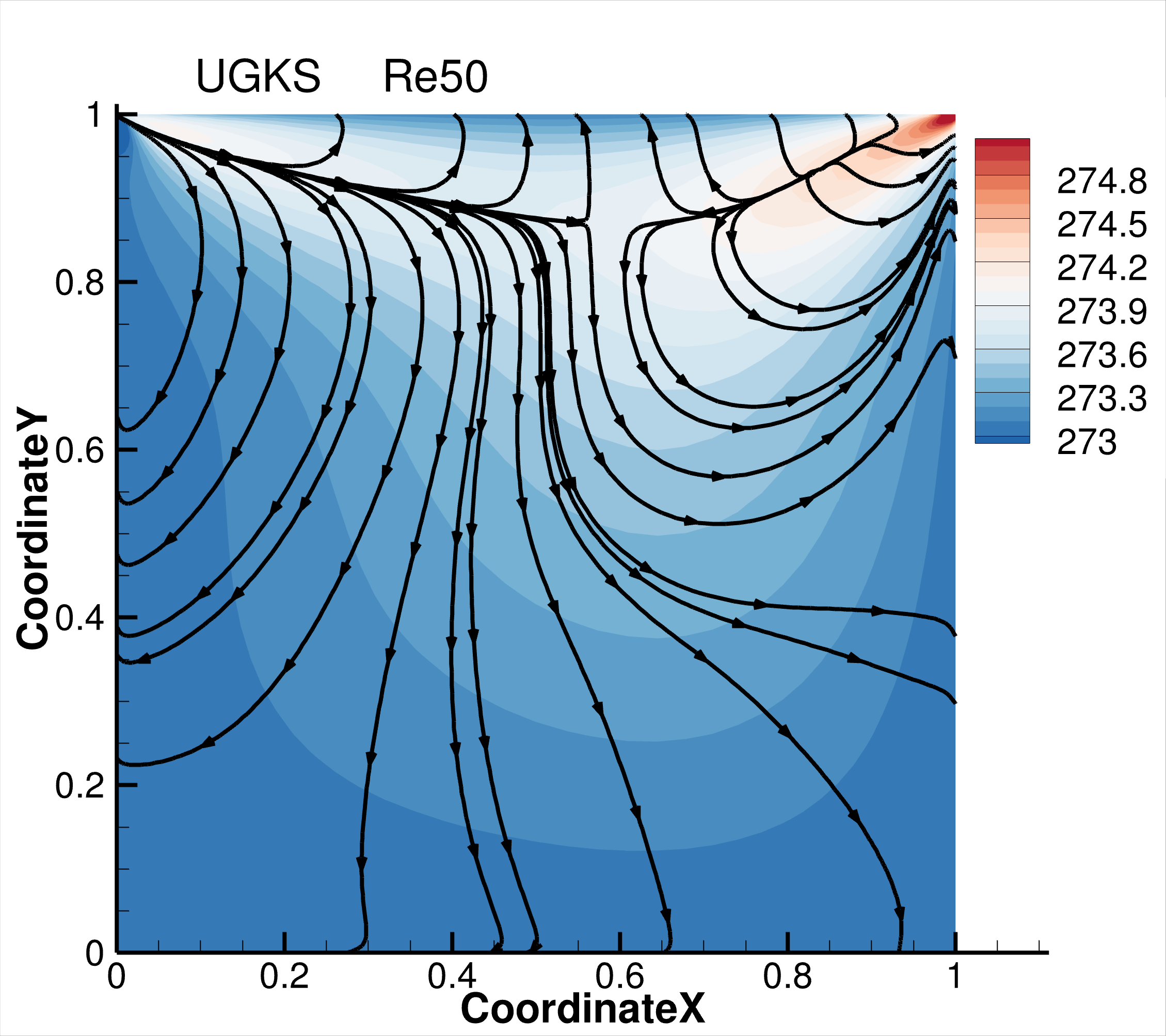}}
  \subfigure[]{\includegraphics[width=0.3\textwidth]{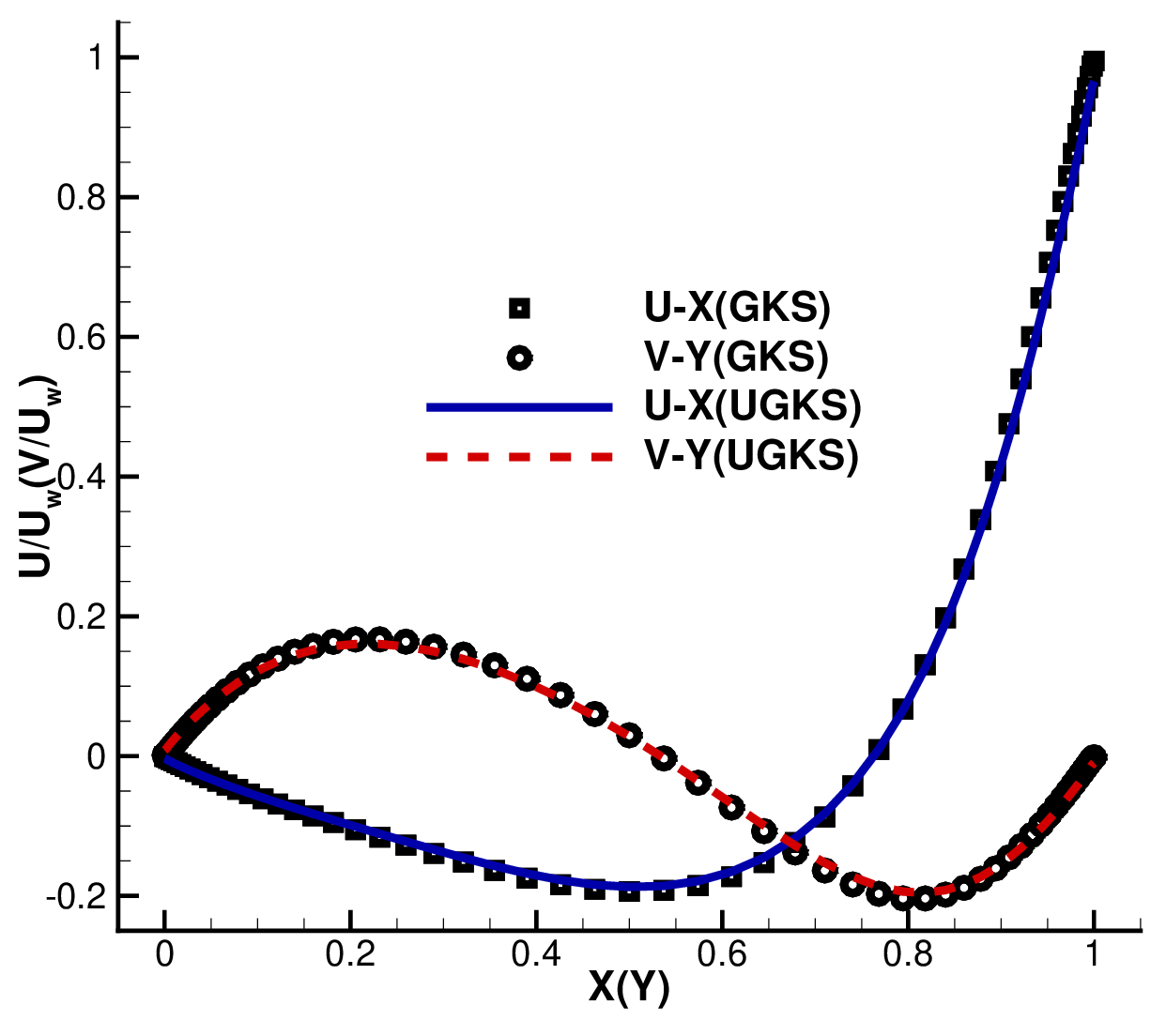}}
  \caption{Cavity simulation using GKS and UGKS at $Re=50$. (a) The temperature contour and heat flux using GKS. (b) The temperature contour and heat flux using UGKS. (c) $U$-velocity along the central vertical line and $V$-velocity along the central horizontal line, symbols:GKS, lines:UGKS.}
  \label{fig:cavitySimulationRe50}
\end{figure}

\begin{figure}[!htpb]
  \centering
  \subfigure[]{\includegraphics[width=0.3\textwidth]{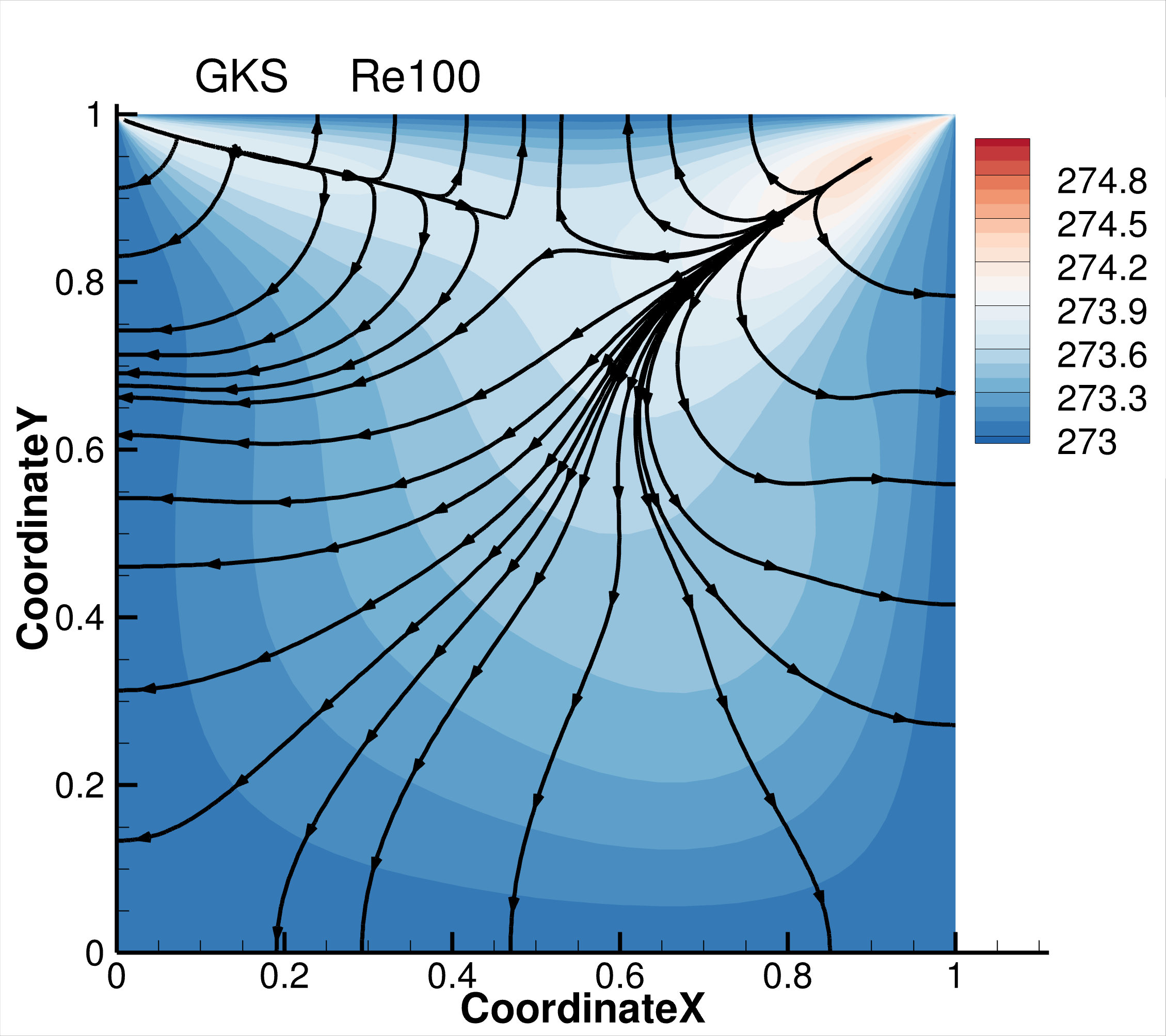}}
  \subfigure[]{\includegraphics[width=0.3\textwidth]{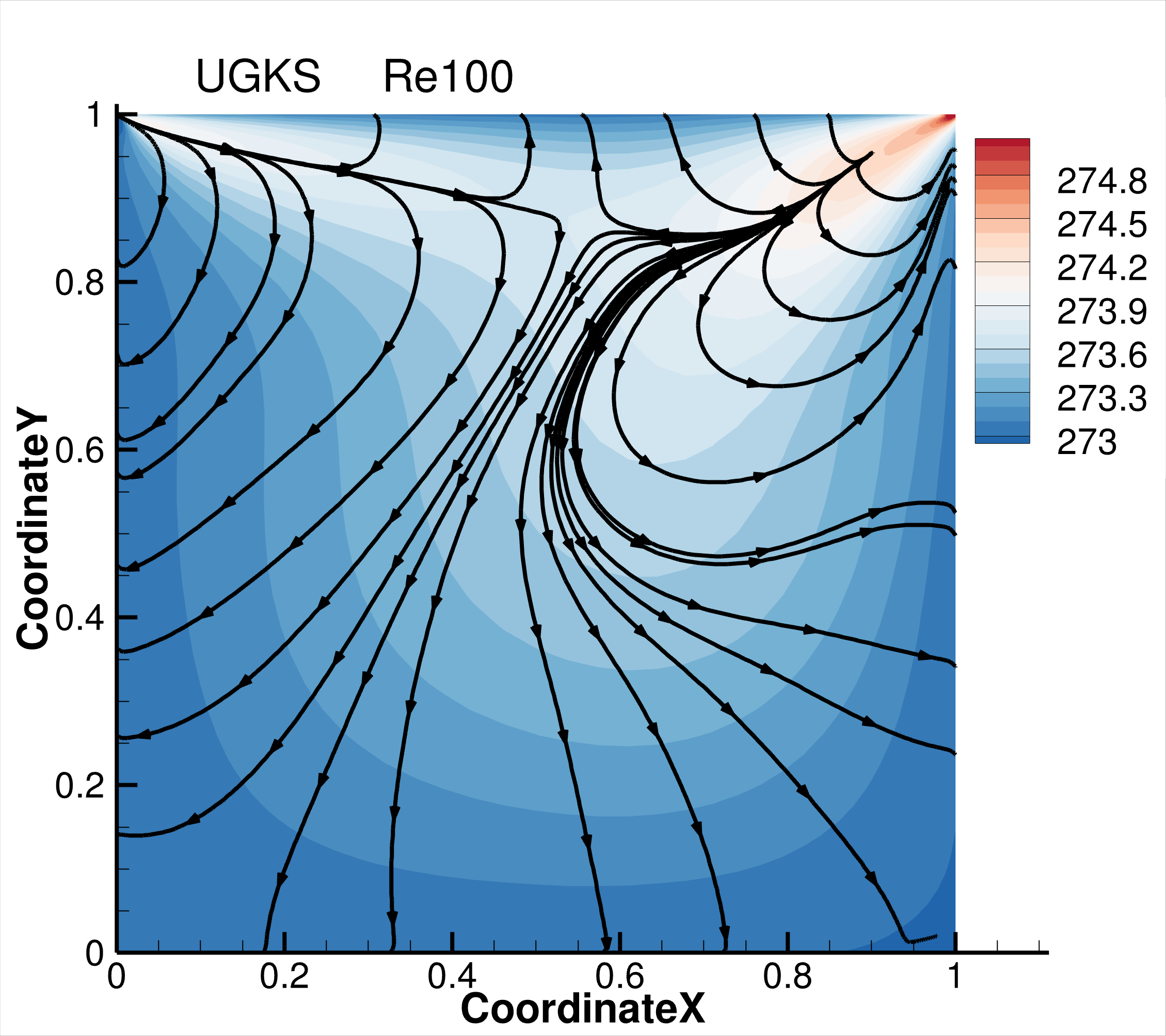}}
  \subfigure[]{\includegraphics[width=0.3\textwidth]{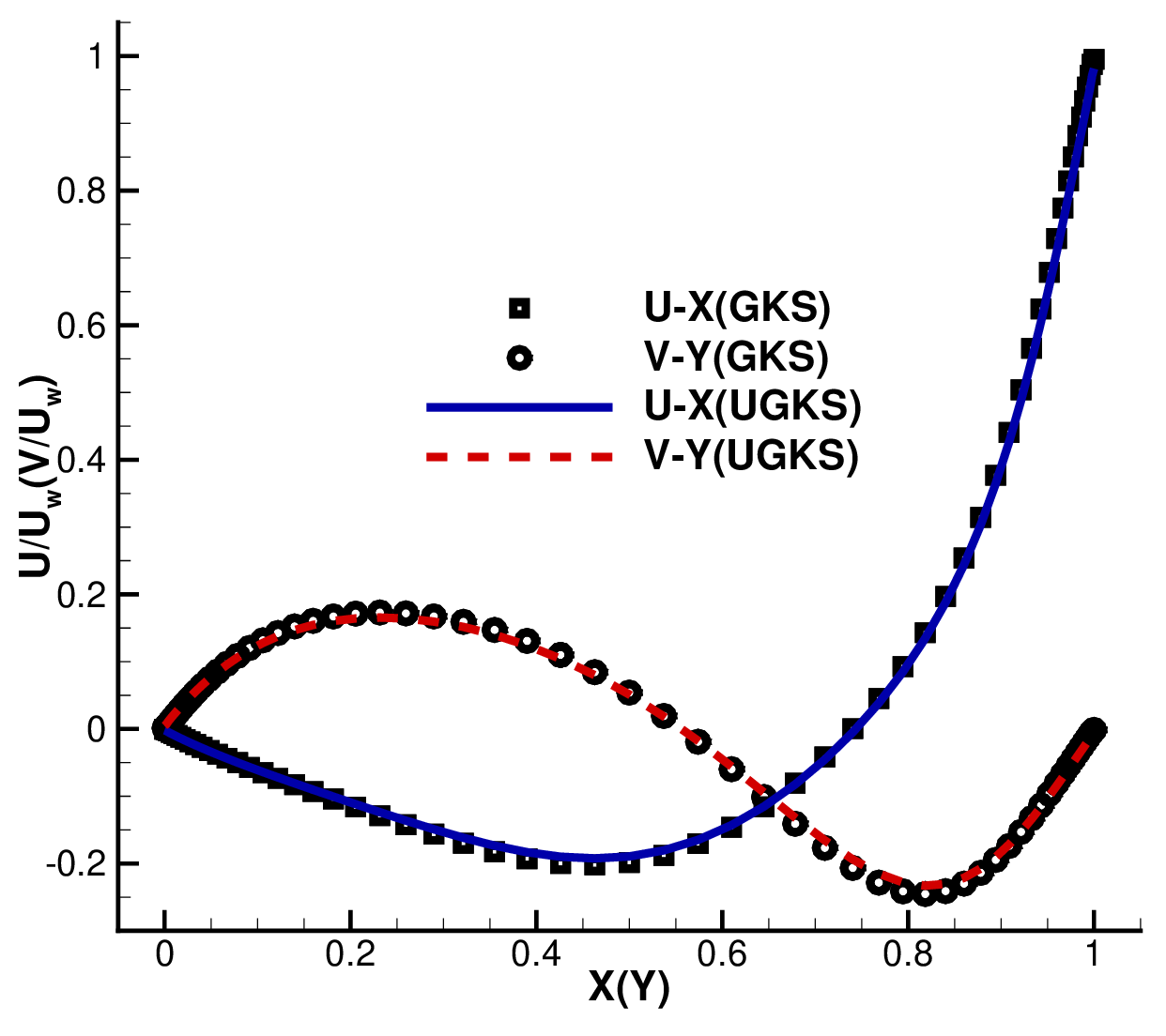}}
  \caption{Cavity simulation using GKS and UGKS at $Re=100$. (a) The temperature contour and heat flux using GKS. (b) The temperature contour and heat flux using UGKS. (c) $U$-velocity along the central vertical line and $V$-velocity along the central horizontal line, symbols:GKS, lines:UGKS.}
  \label{fig:cavitySimulationRe100}
\end{figure}

The results of the GKS and UGKS are shown in Figure \ref{fig:cavitySimulationRe20}, \ref{fig:cavitySimulationRe50}, and \ref{fig:cavitySimulationRe100}. For all three Reynolds numbers, both GKS and UGKS exhibit similar velocity distributions. However, differences arise in temperature and heat flux distributions between the two methods. Notably, at a low Reynolds number (e.g., $Re=20$), the heat flux predicted by UGKS deviates from Fourier’s law, exhibiting a counter-gradient phenomenon where heat flows from low-temperature to high-temperature regions. As the Reynolds number increases, the UGKS results gradually get back to the NS solutions.

\subsection{One-dimensional viscous shock tube}

In order to validate the grid convergence solution of UGKS, the following 1D shock tube problem will be studied.
In this study, the computational domain is defined as $[0,1]^2$ with all boundaries being adiabatic walls. The initial conditions are specified as
\begin{equation*}
(\rho,p,u,v)=\begin{cases}
  (120.0, 120.0/\gamma, 0.0, 0.0), & \text{if } x<0.5, \\
  (1.2, 1.2/\gamma, 0.0, 0.0), & \text{if } x>0.5, \\
\end{cases}
\end{equation*}
where $\rho$ is the density, $p$ is the pressure, and $(u,v)$ are the velocities in the $x$ and $y$ directions, respectively. The non-dimensional gas equation of state is $\gamma p = \rho T$ with a specific heat ratio of $\gamma=1.4$. The Reynolds number is defined based on the domain length $L=1$, reference density $\rho_0=1.0$, reference velocity $U_0=1.0$, and a constant viscosity, i.e., $\mu=1.0/{Re}$. The mean free path is given by $l = \mu/(\rho a) = 1/(\rho a{Re})$, where $a = \sqrt{T}$ denotes the speed of sound. The Prandtl number is set to ${Pr} = 0.73$. The simulation time is $t=1.0$, with the time step determined by a CFL number $0.5$.

The grid configuration for both the physical and velocity spaces is as follows. For two-dimensional simulations, the physical domain is discretized into $700\times700$ cells, and for UGKS, the velocity space $[-6, 6]^2$ is discretized into $80 \times 80$ uniform grid points. For the one-dimensional reference case, which captures the evolution of key flow features such as the shock wave, contact discontinuity, and rarefaction wave, and their interactions, the computational domain is divided into 700 cells along the $x$-axis, with the velocity space discretized into 80 elements. Only the left and right boundaries are considered in the one-dimensional case.

\begin{figure}[!htpb]
  \centering
  \subfigure[]{\includegraphics[width=0.3\textwidth]{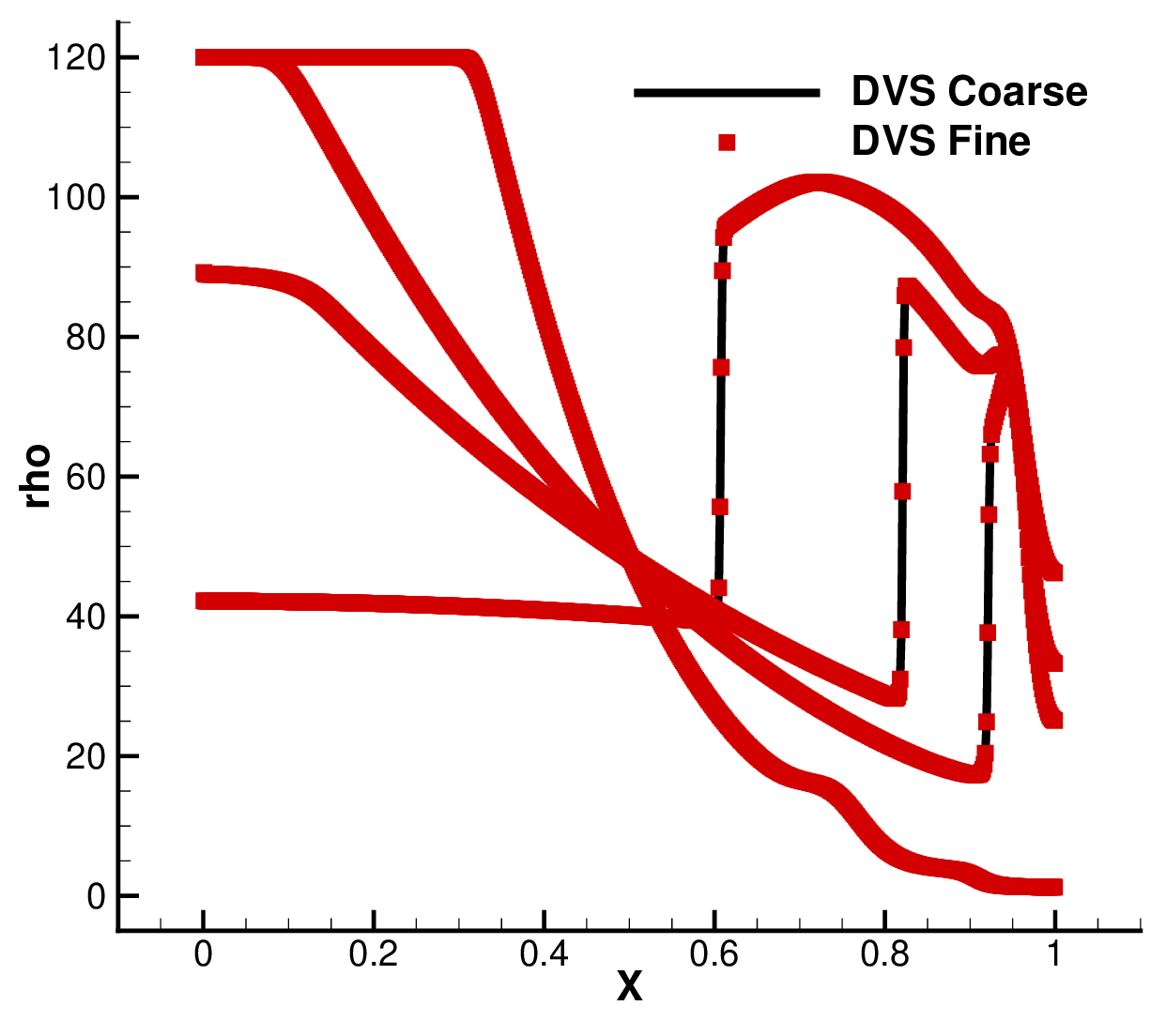}}
  \subfigure[]{\includegraphics[width=0.3\textwidth]{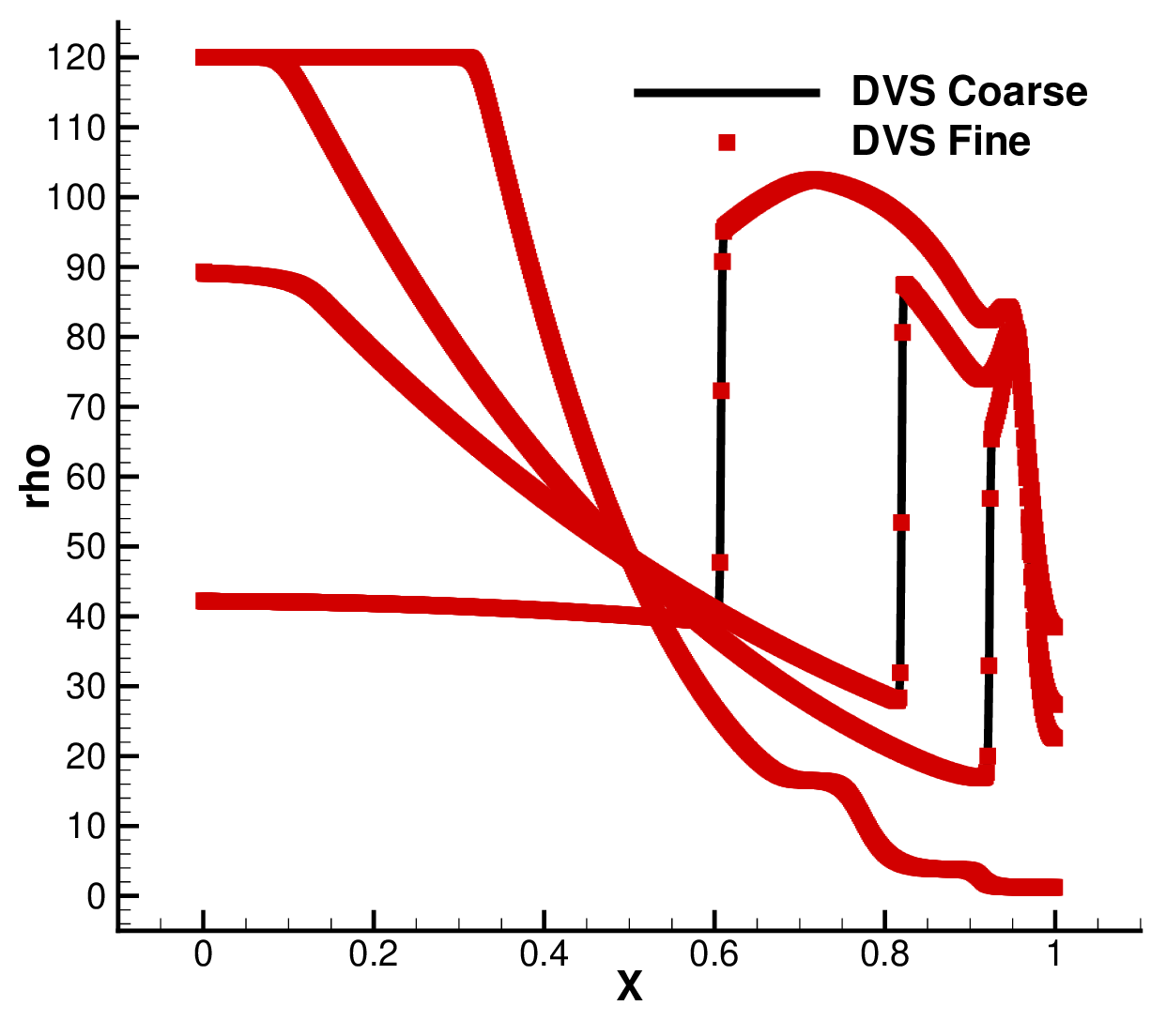}}
  \subfigure[]{\includegraphics[width=0.3\textwidth]{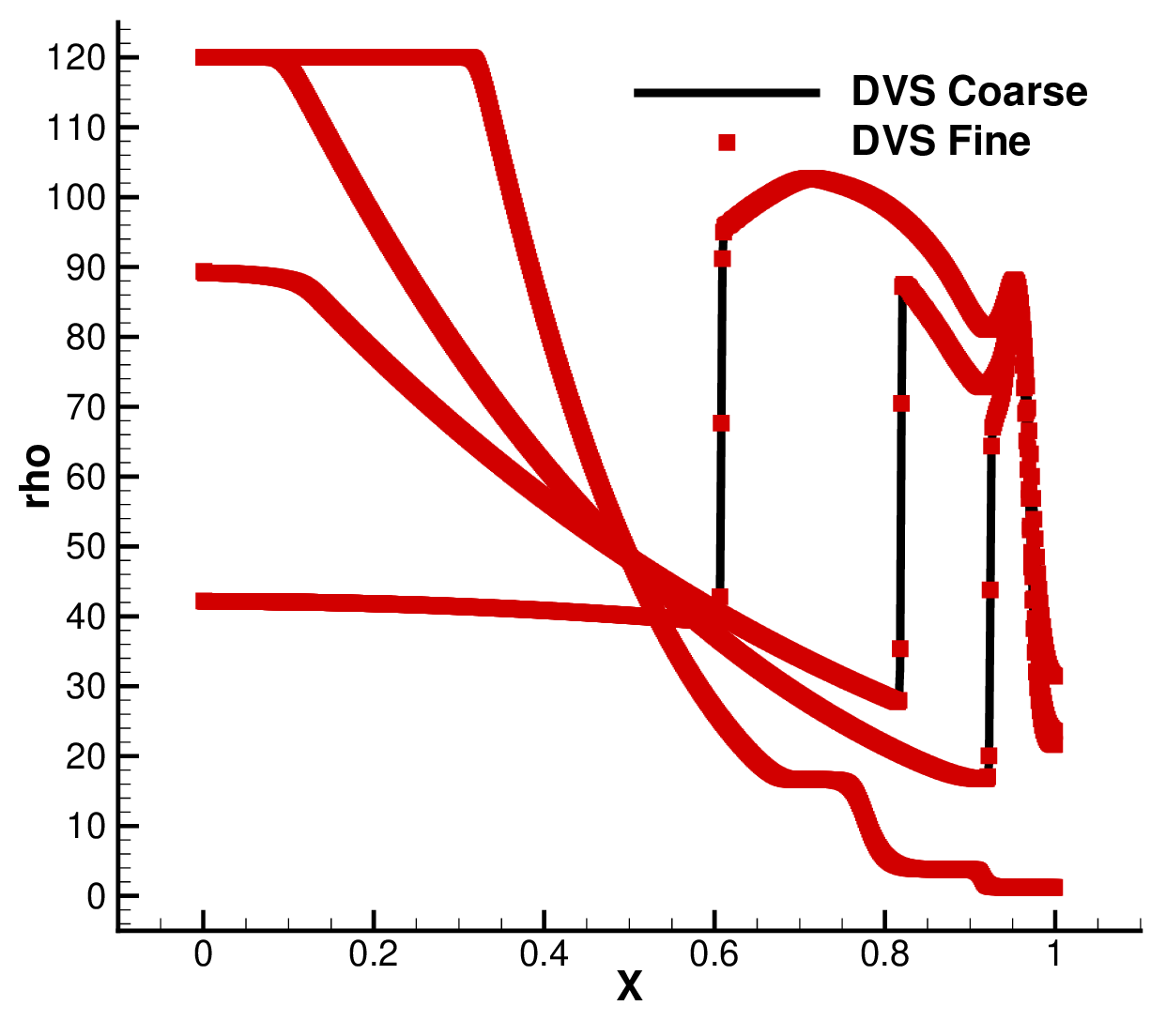}}
  \caption{Velocity-space mesh independence test: Comparison of density profiles using 80 elements in the range $[-6, 6]$ and 200 elements in the range $[-10, 10]$ for (a) $Re=50$, (b) $Re=100$, and (c) $Re=200$.}
  \label{fig:DensityCompare}
\end{figure}

To validate the adequacy of the mesh resolution and ensure mesh independence, systematic studies have been conducted for both the physical and velocity-space discretizations. Figure \ref{fig:DensityCompare} presents the density profiles at different times computed by UGKS with two levels of discrete velocity space resolutions: 80 elements in the range $[-6,6]$, and 200 elements in the range $[-10,10]$. The comparison confirms that the results are insensitive to further refinement of the velocity mesh, with good agreement between the fine and coarse discretizations.

\begin{figure}[!htpb]
  \centering
  \includegraphics[width=0.3\textwidth]{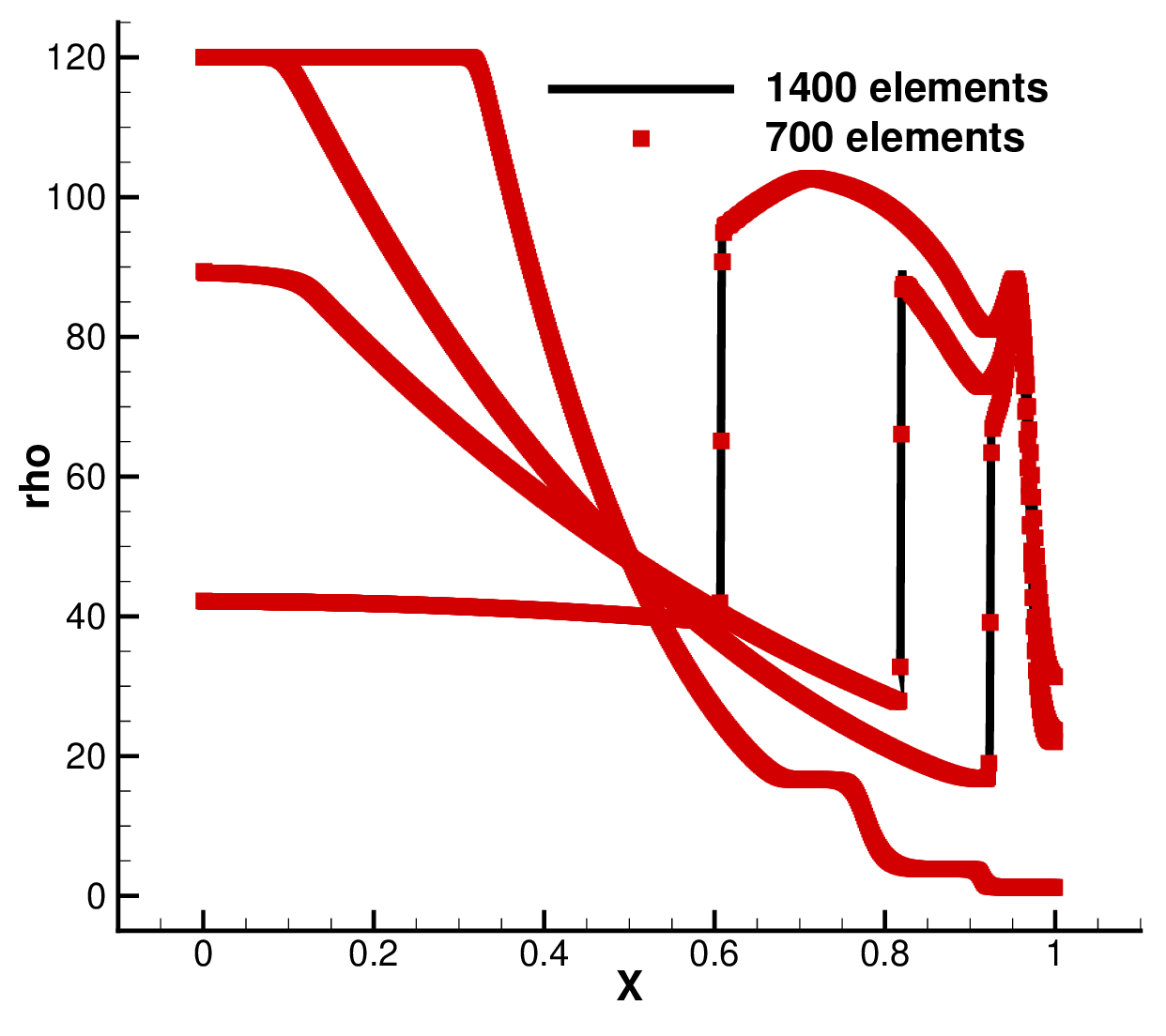}
  \caption{Physical-domain mesh independence test: Comparison of density profiles using 700 and 1400 cells in the physical domain for $Re=200$.}
  \label{fig:PhysicalDomainCompare}
\end{figure}

Figure \ref{fig:PhysicalDomainCompare} demonstrates the mesh independence solution from UGKS with 700 cells and 1400 cells in the one-dimensional physical domain. The test is performed at $\text{Re}=200$. The results show that the converged physical solution can be obtained by UGKS using 700 mesh points.

Therefore, unless otherwise stated, all one-dimensional simulations adopt 700 cells in the physical domain and 80 elements in the velocity space, while two-dimensional simulations use $700 \times 700$ cells and $80 \times 80$ grid points in the velocity space.

\begin{figure}[!htpb]
  \centering
  \subfigure[]{\includegraphics[width=0.3\textwidth]{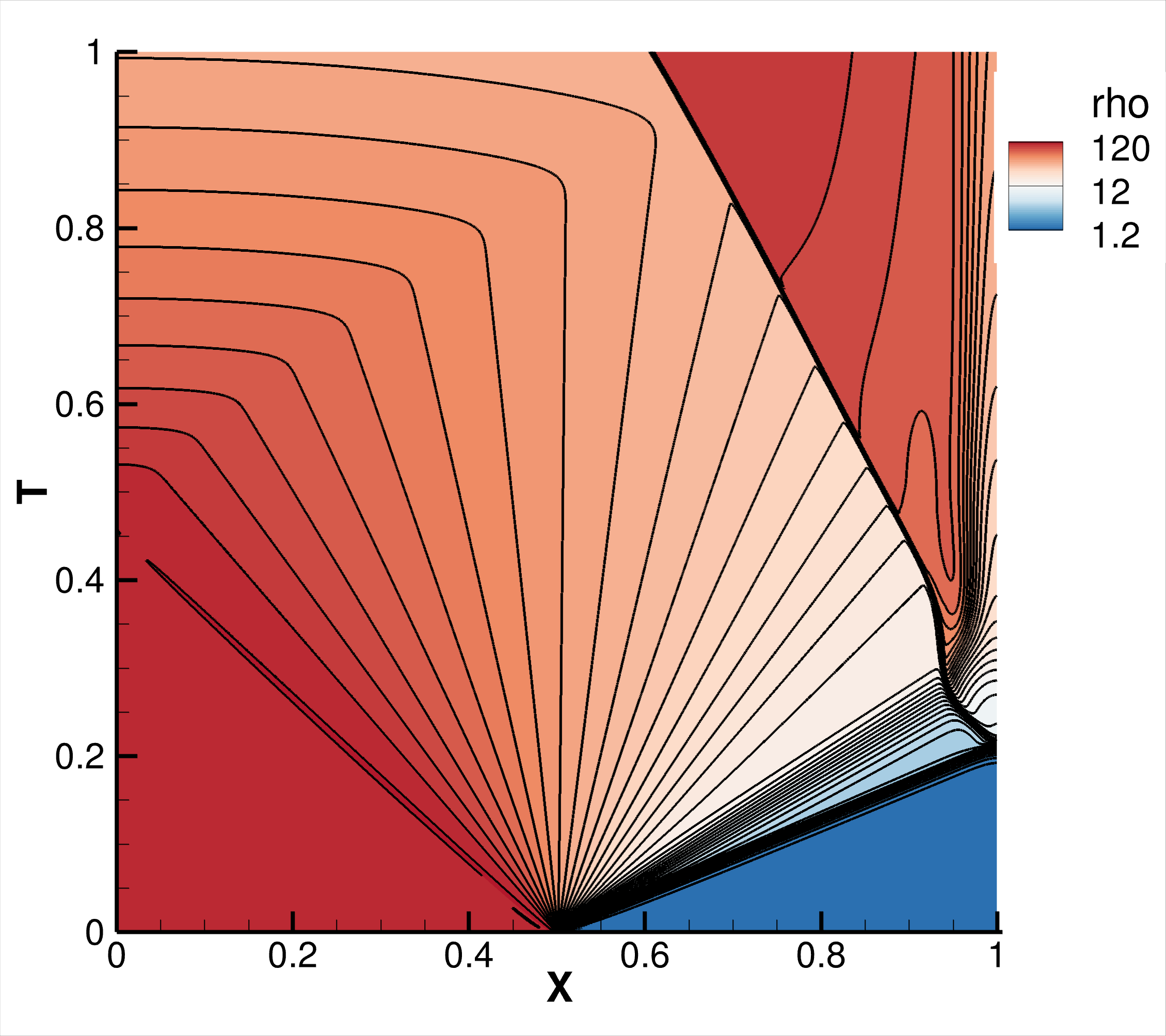}}
  \subfigure[]{\includegraphics[width=0.3\textwidth]{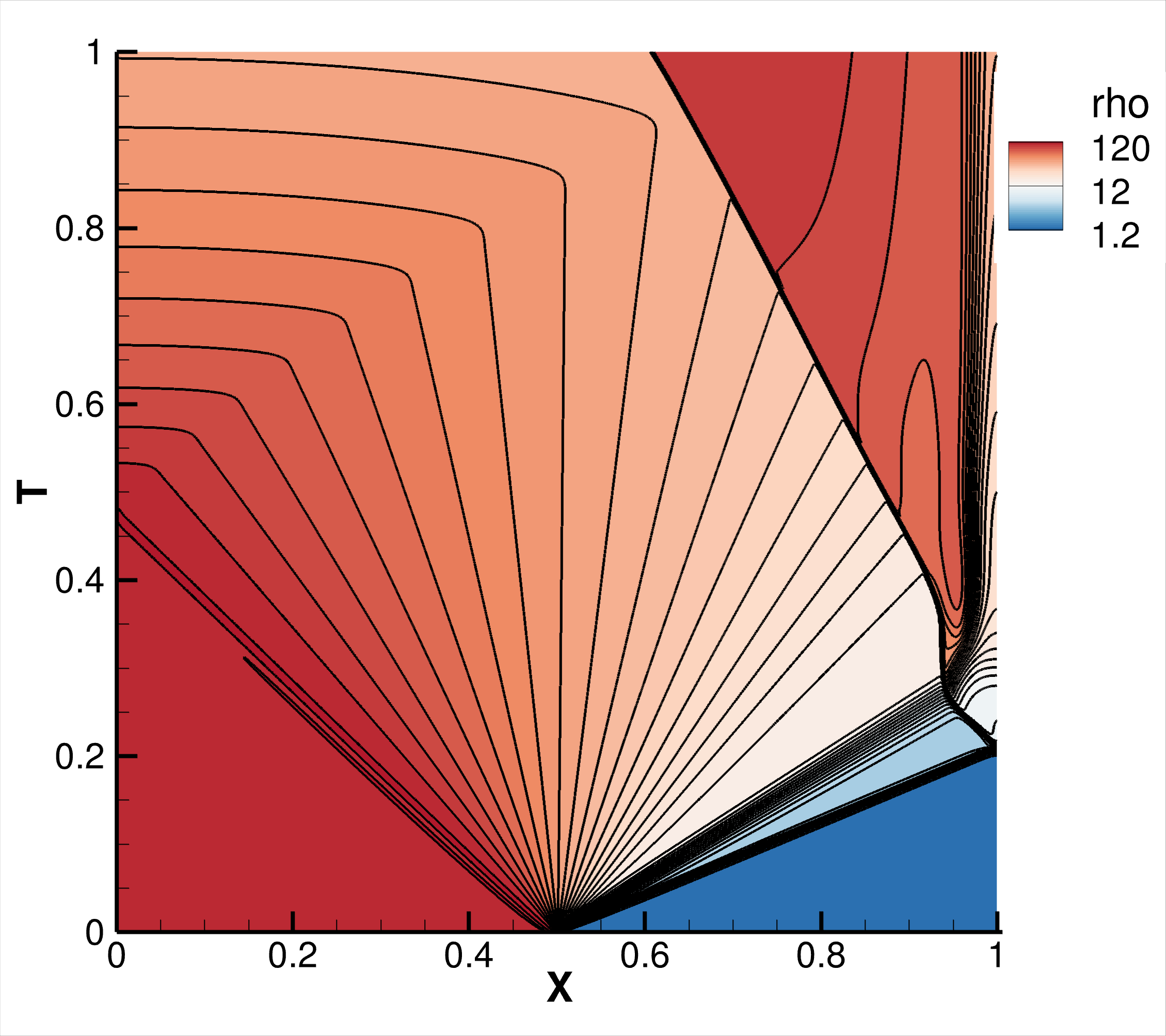}}
  \subfigure[]{\includegraphics[width=0.3\textwidth]{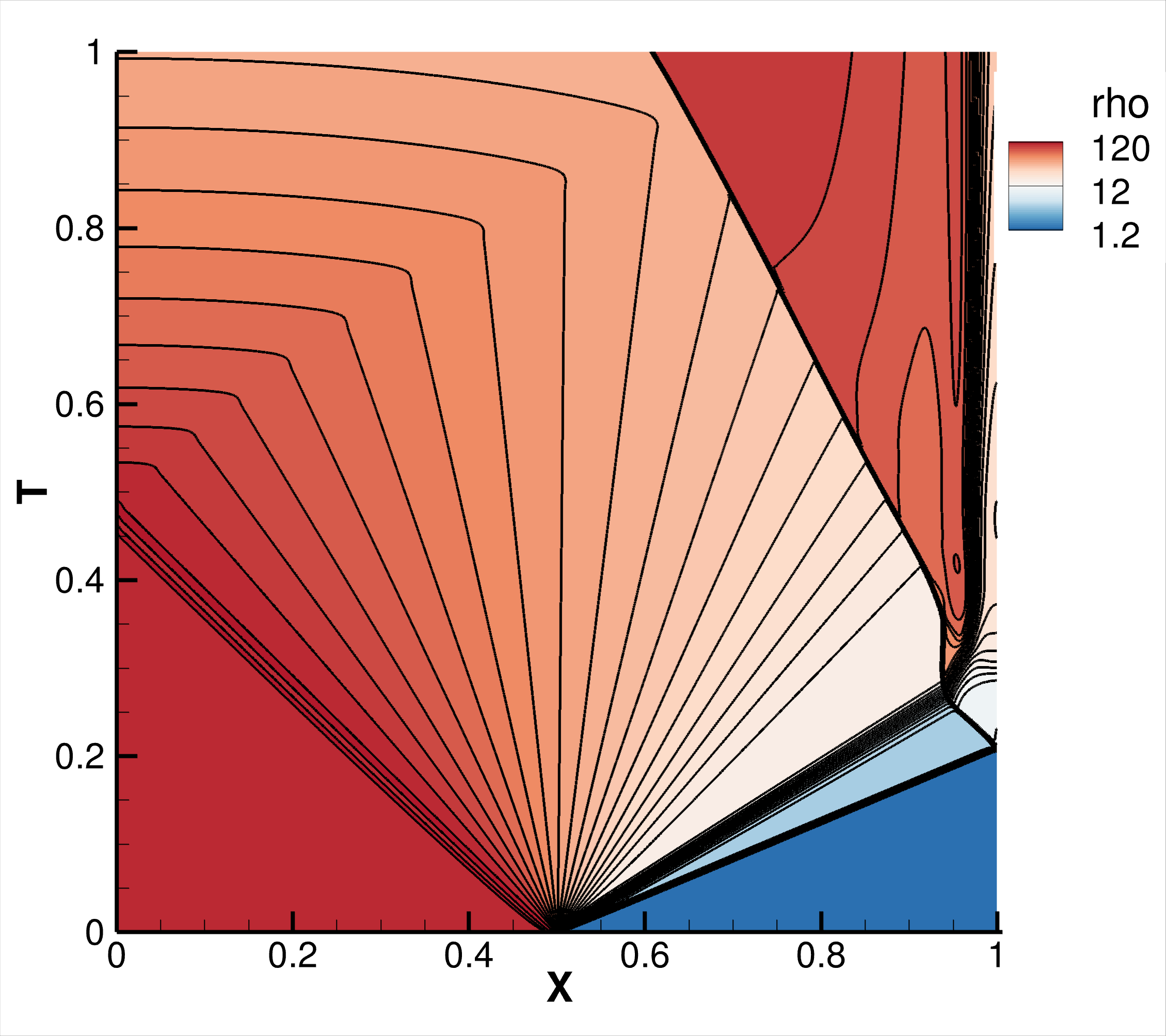}}
  \subfigure[]{\includegraphics[width=0.3\textwidth]{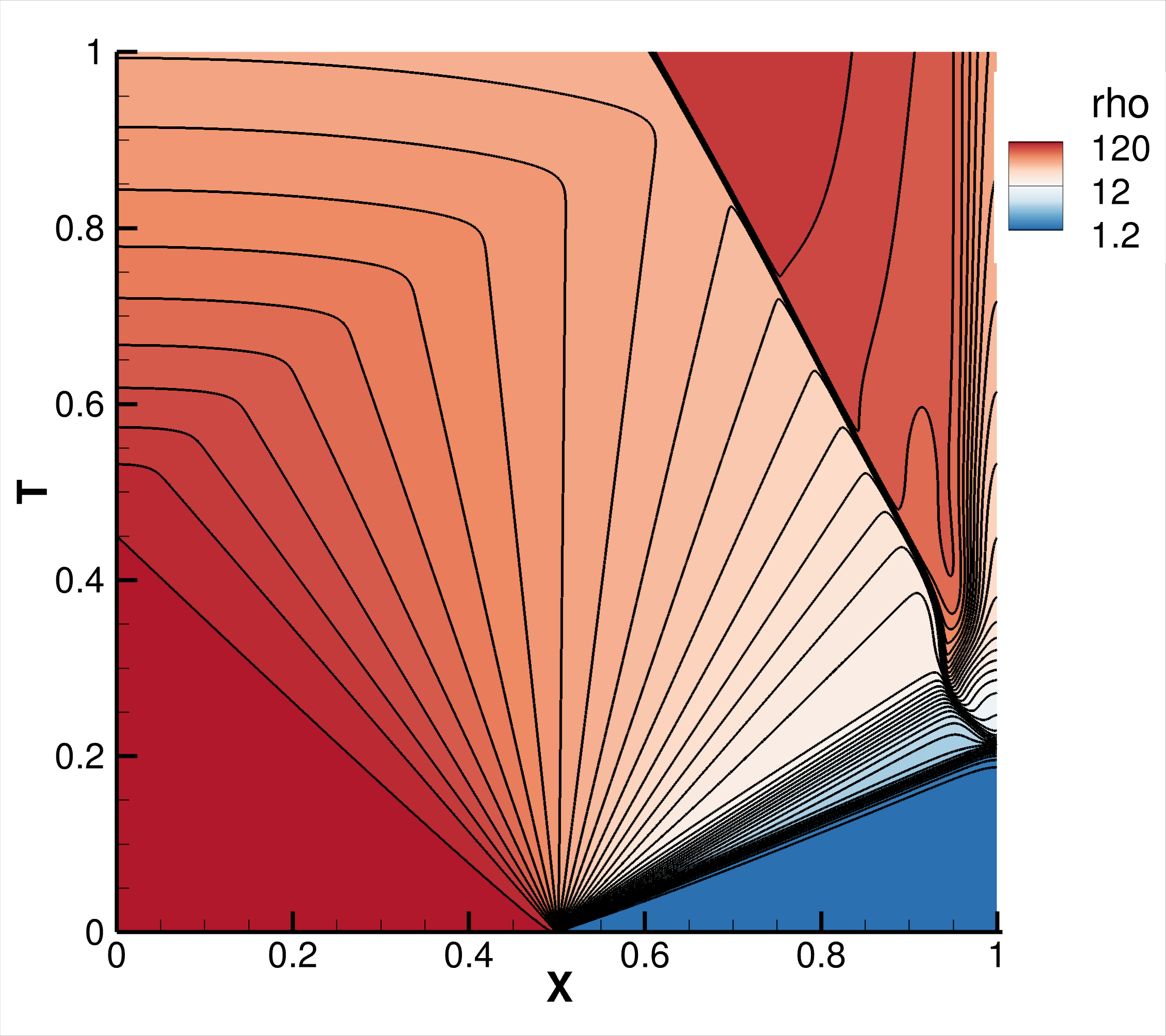}}
  \subfigure[]{\includegraphics[width=0.3\textwidth]{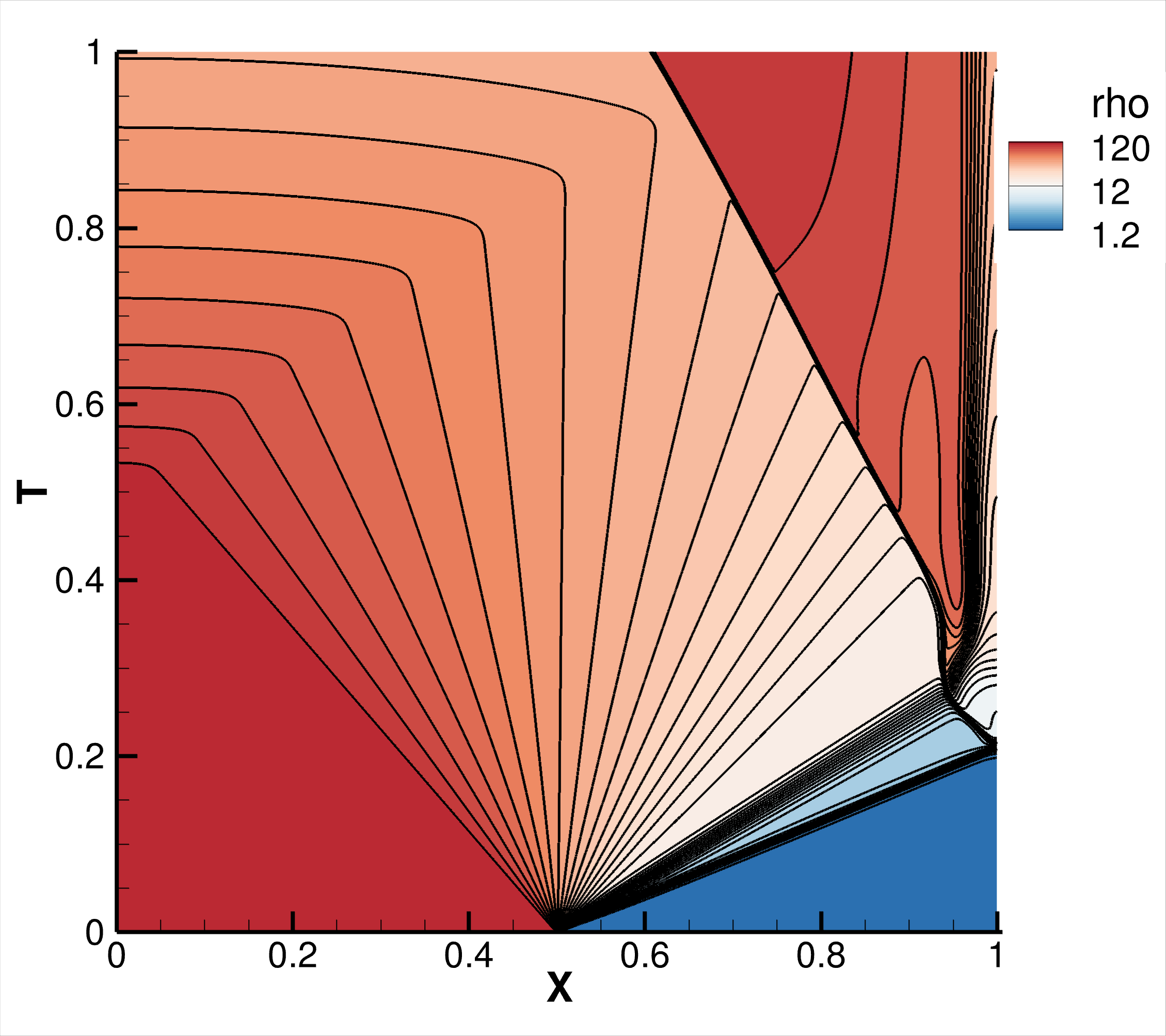}}
  \subfigure[]{\label{fig:DensityUGKSRe200}\includegraphics[width=0.3\textwidth]{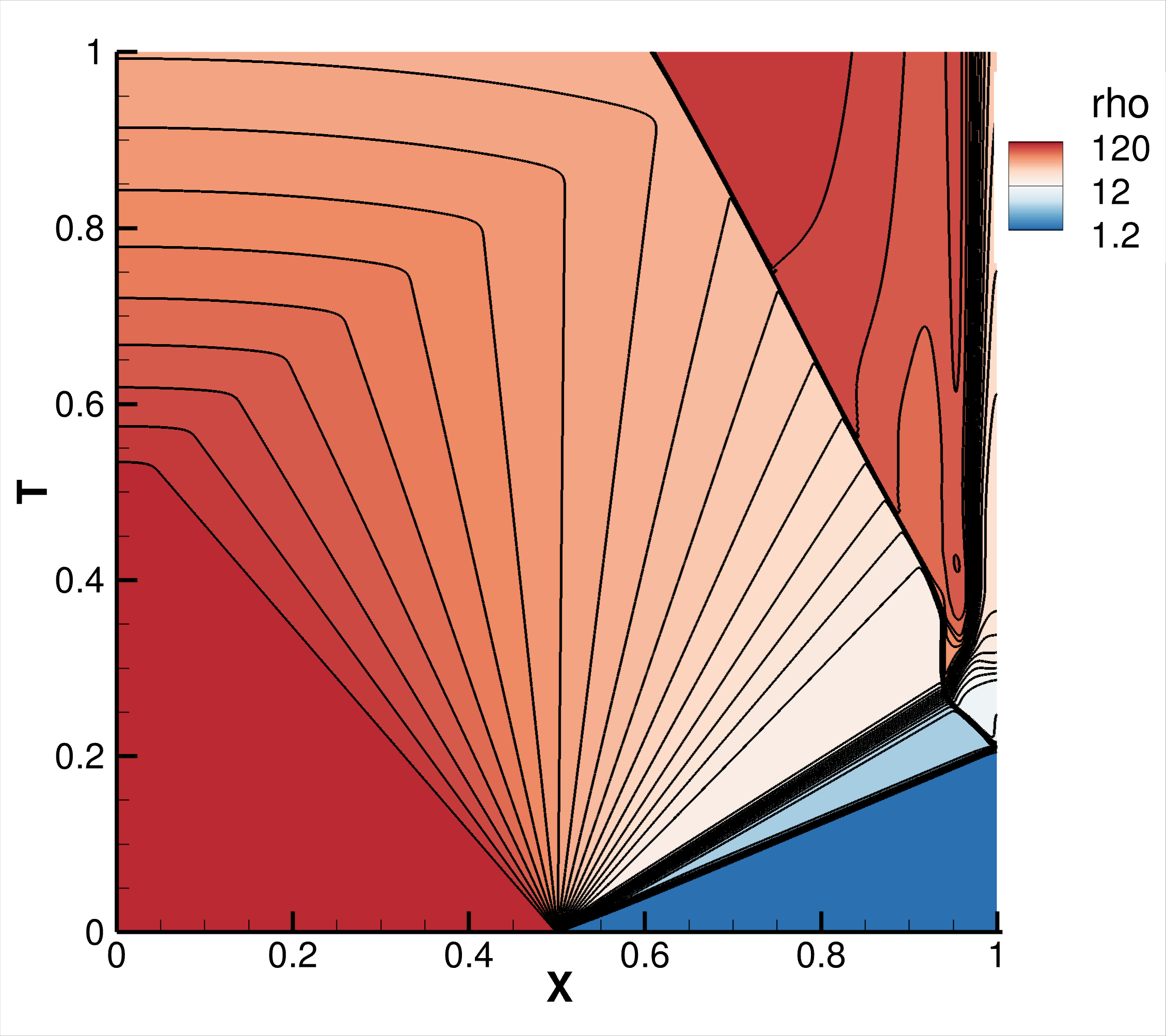}}
  \caption{$x-t$ diagram of the density by GKS and UGKS with different Reynolds numbers. The first row is the results of GKS, and the second row is the results of UGKS. From left to right, the Reynolds numbers are 50, 100, and 200.}
  \label{fig:DensityProfile}
\end{figure}

\label{sec:1Dresults}
The one-dimensional case is considered to illustrate the difference in the main flow structure between GKS and UGKS in different Reynolds numbers. The evolution of density in the $x-t$ diagram is shown in Figure \ref{fig:DensityProfile}.

A shock wave, followed by a contact discontinuity, propagates toward the low-pressure region (the right side), while a rarefaction wave propagates toward the high-pressure region (the left side). The shock Mach number is 2.37. The incident shock wave is relatively weak and reflects off the right wall at time
$t\approx0.2$. Following this reflection, it interacts with the contact discontinuity. At later times, the contact discontinuity remains stationary near the right wall. Conversely, the reflected shock wave continues to propagate, beginning to interact with the incident rarefaction wave at time
$t=0.4$, and subsequently with the reflected rarefaction wave. During these interactions, the reflected shock wave maintains an approximately constant velocity.

A primary distinction between the two methods is that the GKS generally exhibits lower numerical dissipation. For instance, at the head of the rarefaction wave, the GKS may produce an overshoot (i.e., density values exceeding expected physical limits). This suggests that the GKS lacks sufficient dissipation to accurately capture physical diffusion within extreme flow structures, such as shocks and rarefaction waves, potentially leading to non-physical numerical artifacts. In contrast, the UGKS, as a multi-scale method, more comprehensively accounts for rarefaction effects and various diffusion processes. This effectively suppresses overshoots and provides a more accurate resolution of the shock and rarefaction structures. A more detailed comparison of these methods will be presented in the subsequent sections, where numerical solutions at various times are analyzed to highlight specific differences in main flow structures, shock resolution, and dissipation characteristics.

\begin{figure}[!htpb]
  \centering
  \subfigure[$t=0.175$]{\label{fig:DensityT0.175}\includegraphics[width=0.45\textwidth]{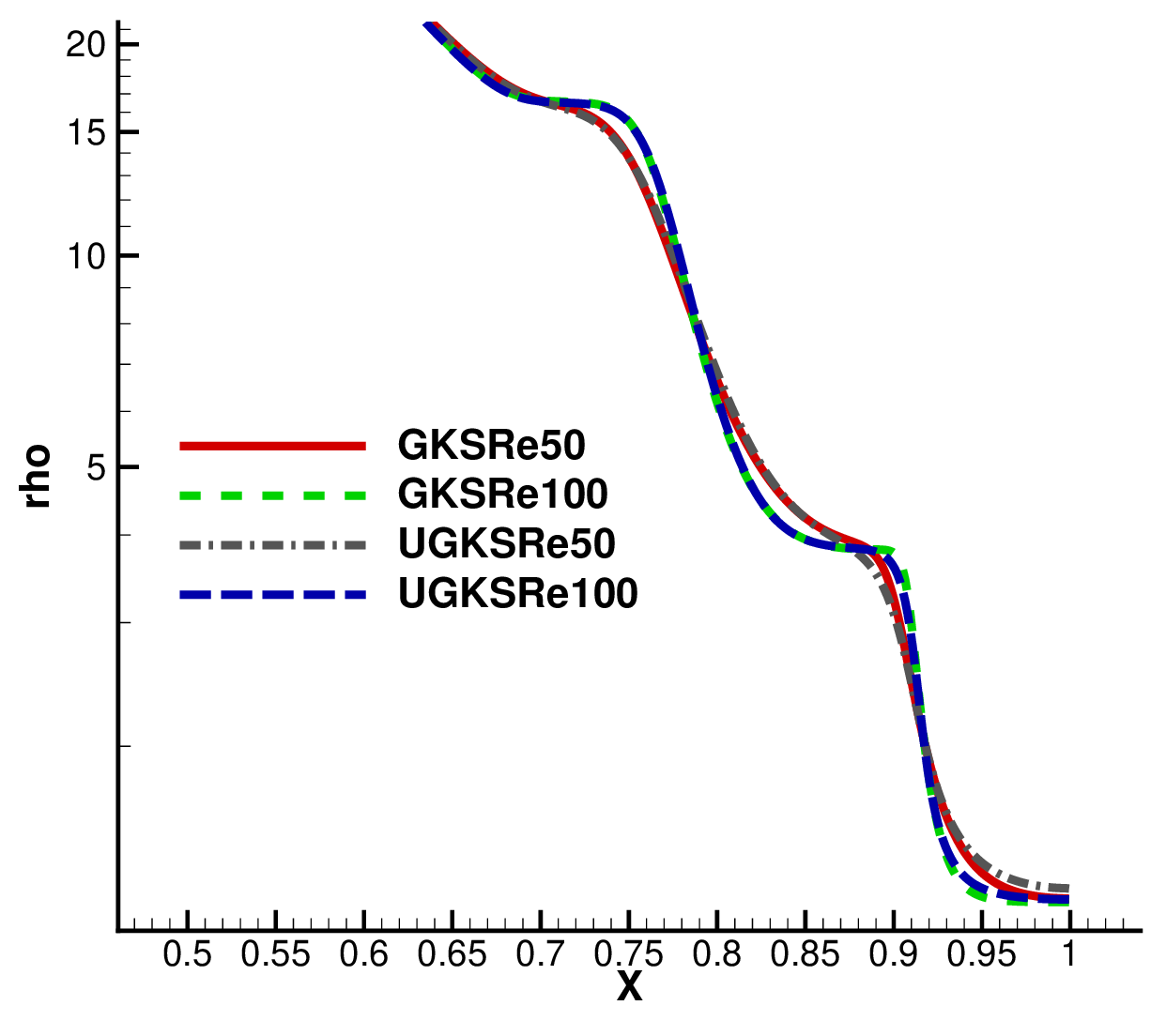}}
  \subfigure[$t=0.4$]{\label{fig:DensityT0.4}\includegraphics[width=0.45\textwidth]{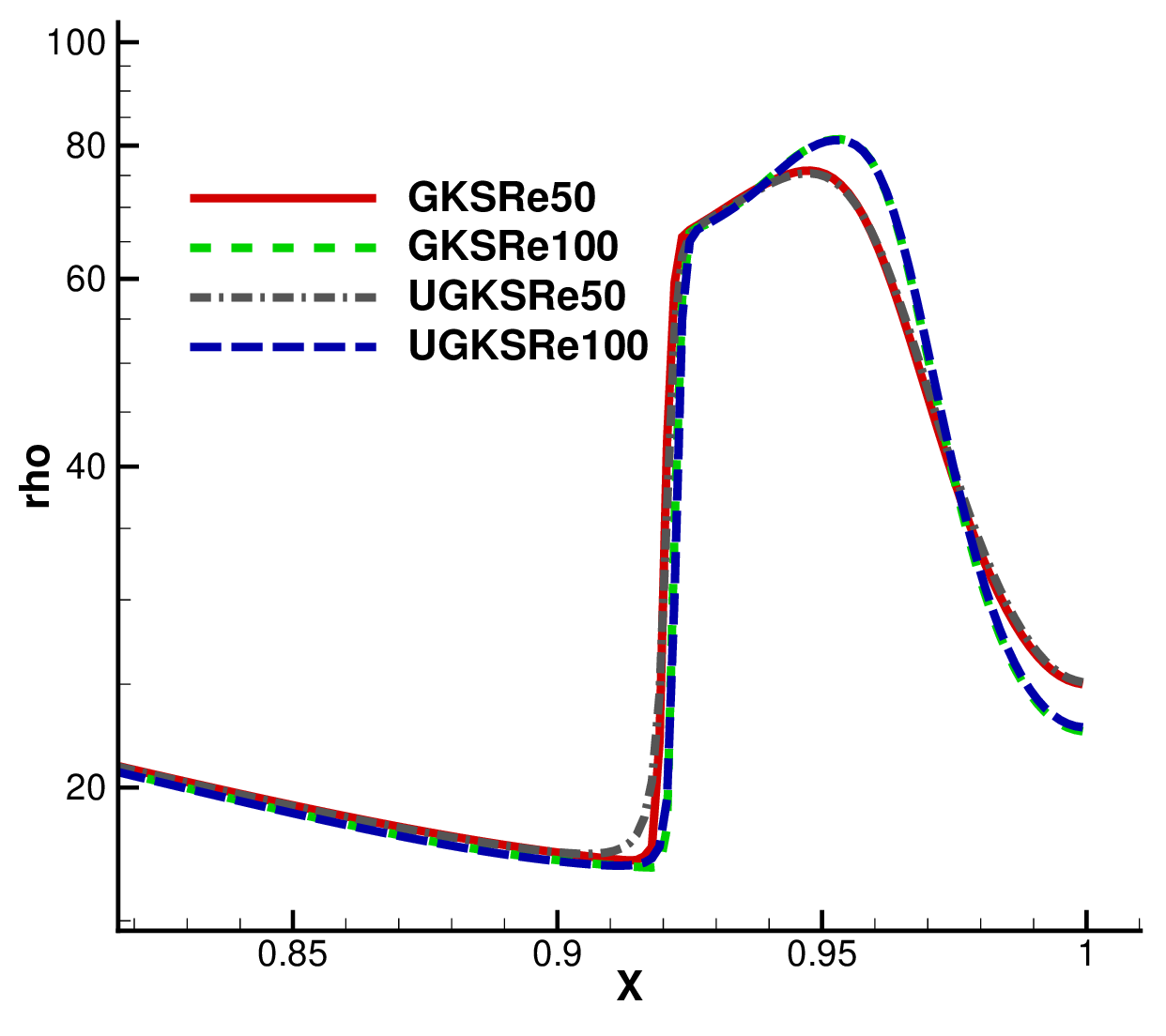}}
  \subfigure[$t=0.6$]{\label{fig:DensityT0.6}\includegraphics[width=0.45\textwidth]{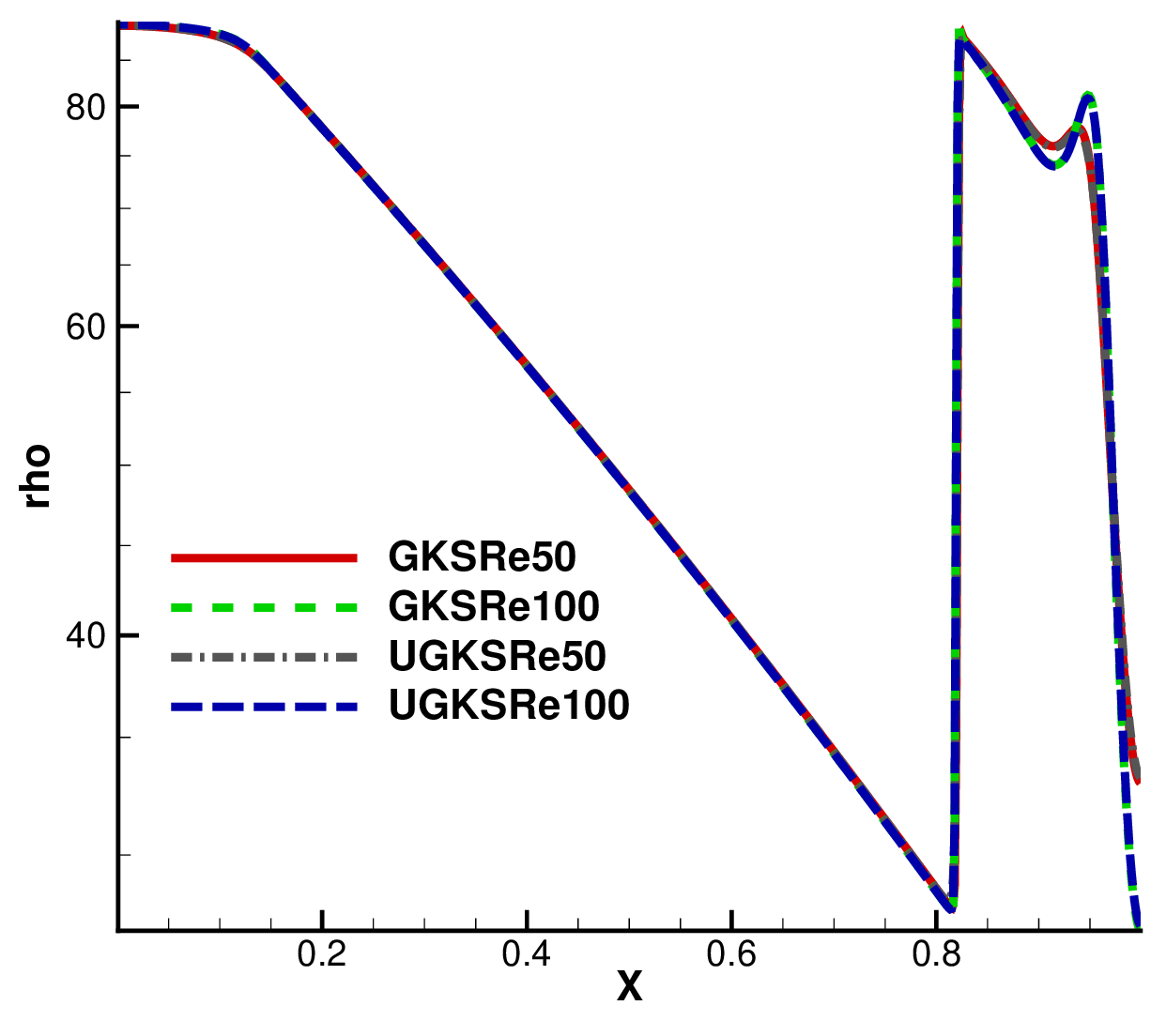}\includegraphics[width=0.45\textwidth]{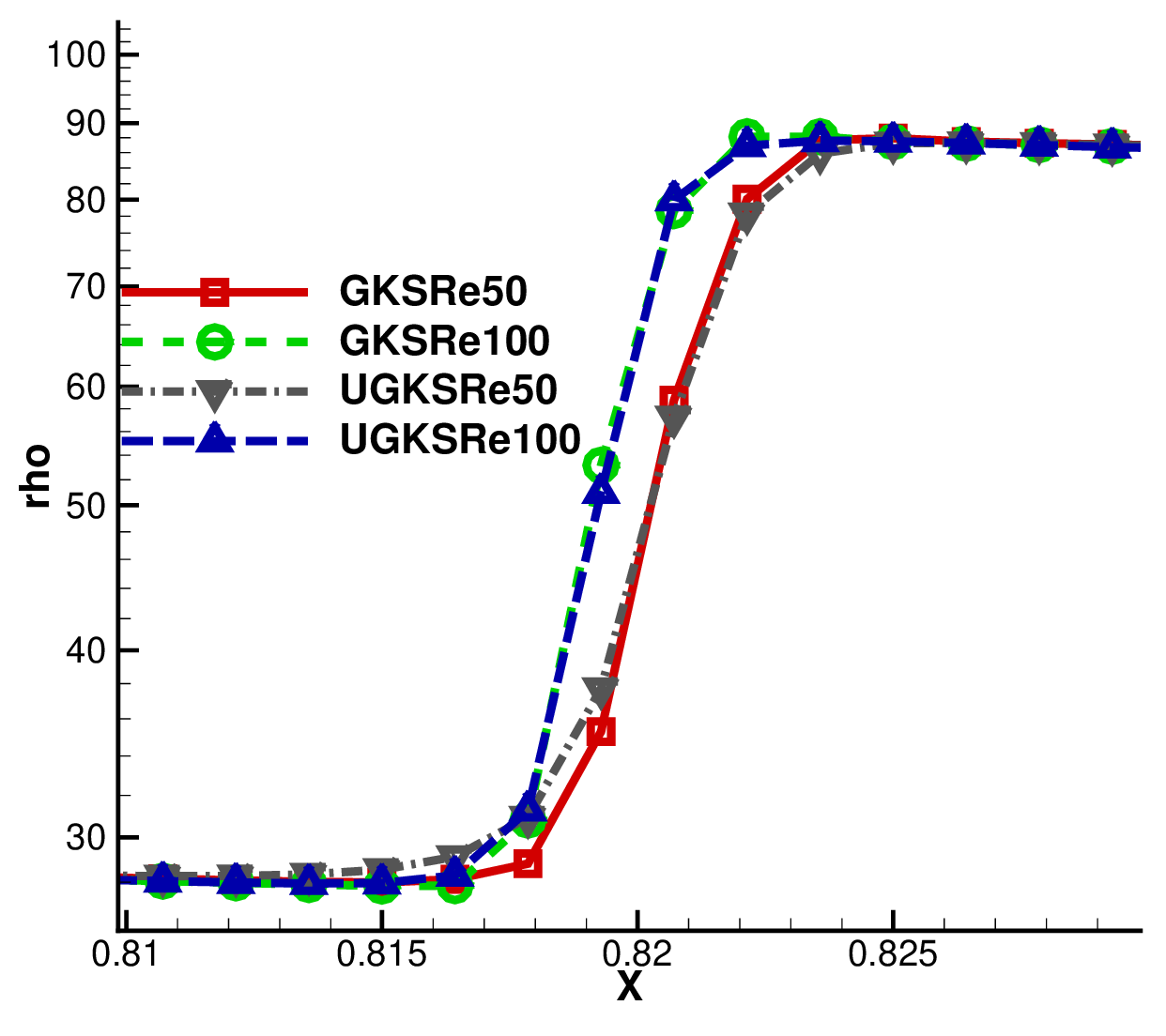}}
  \subfigure[$t=1.0$]{\label{fig:DensityT1.0}\includegraphics[width=0.45\textwidth]{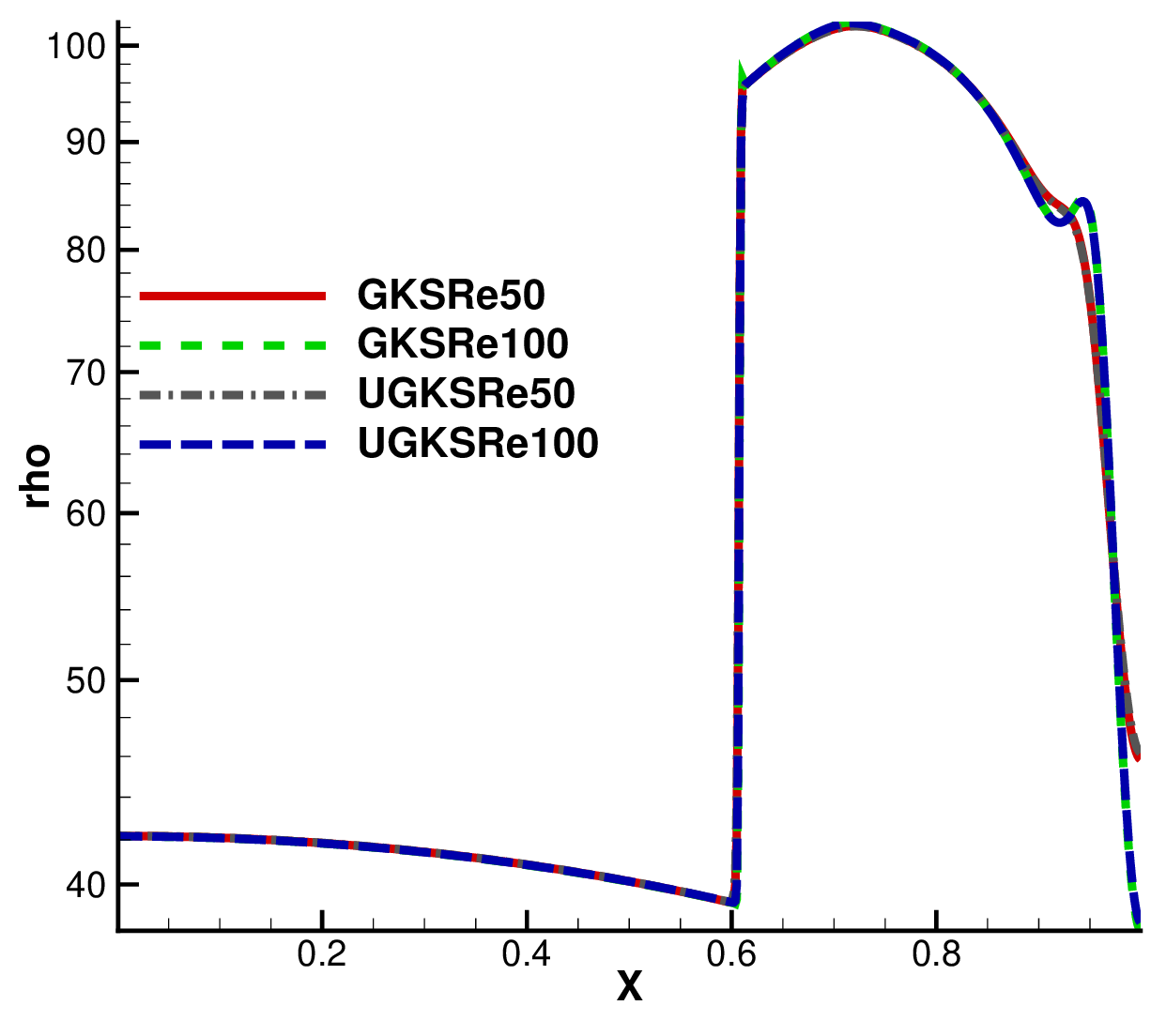}\includegraphics[width=0.45\textwidth]{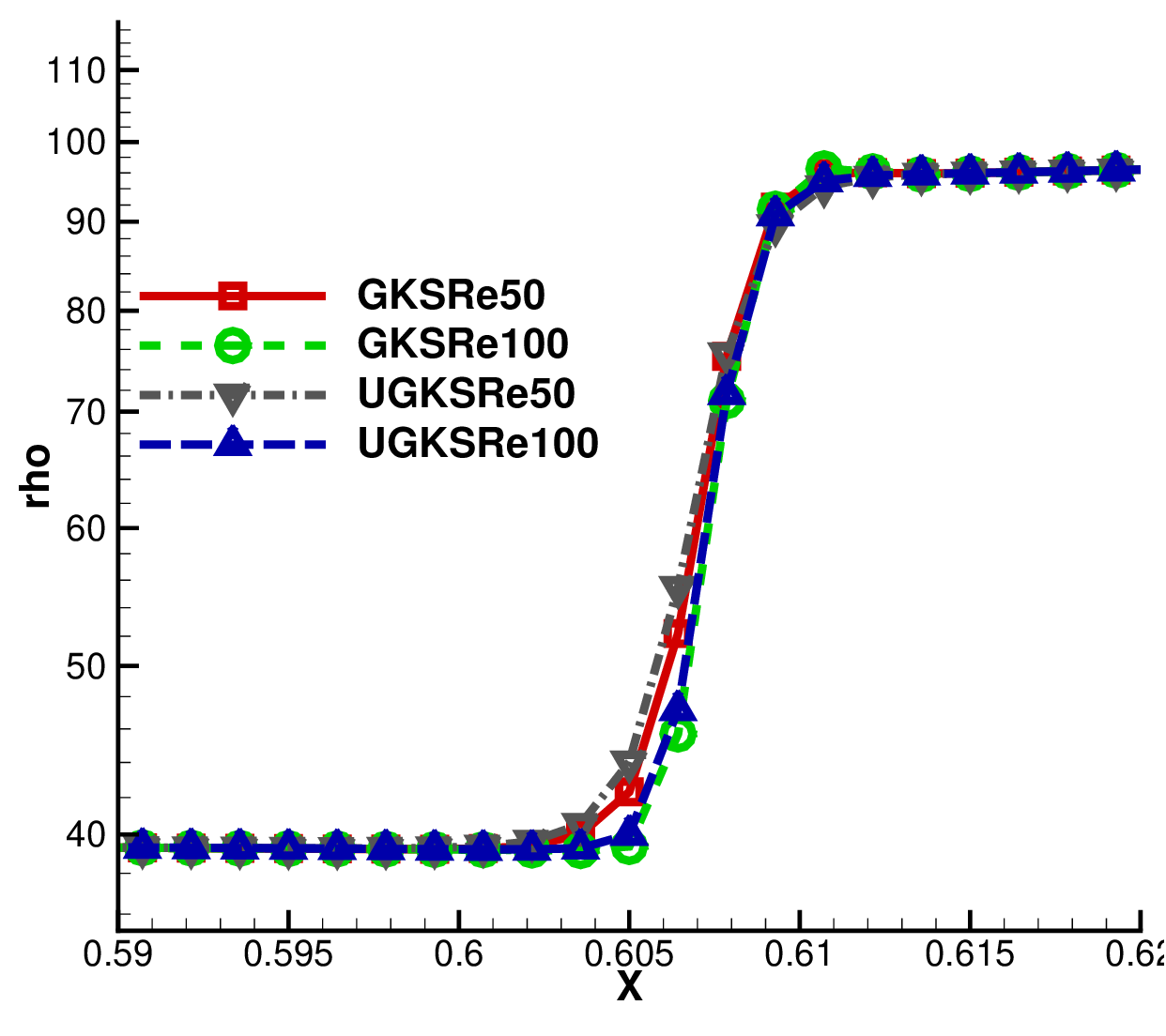}}
  \caption{The density profile of Reynolds numbers 50 and 100 by GKS and UGKS at different times.}
  \label{fig:DensityDifferentTimes}
\end{figure}

The density profile at different times is shown in Figure \ref{fig:DensityDifferentTimes}.
At $t=0.175$, the shock wave has not reached the right wall, so the density profile is not affected by it. The GKS and UGKS predict the same position of the shock wave and the contact discontinuity, which, for $Re=50$, the UGKS gives a thickness of the shock wave that is slightly thicker than the GKS results. At the time of $t=0.4$, the shock wave has reflected off the right wall and interacts with the contact discontinuity. At Reynolds number 100, the GKS and UGKS results are close. At Reynolds number 50, the UGKS density is higher than the GKS results at the front of the shock wave. At the times of $t=0.6$ and $t=1.0$, the shock wave interacts with the rarefaction wave. As shown in the figure enlargement, the density is higher in front of the shock wave and remains so at subsequent times. And little oscillation is observed in the shock structure at $Re=100$ by GKS.

\section{Viscous shock tube}
\label{sec:2Dresults}

Whereas the previous section analyzed the evolution of the primary flow structures in the one-dimensional case, this section focuses on two-dimensional configurations to investigate the interaction between the main flow and the boundary layer. Transitioning to a two-dimensional framework enables the resolution of more complex flow phenomena, such as vortices. When the incident shock wave impinges on the right wall and reflects, the total pressure in the boundary layer ahead of the shock is lower than that in the region behind it. Consequently, the boundary layer cannot smoothly penetrate the reflected-shock region; instead, it separates from the surface, forming a separated flow region near the wall. The fluid passing through the oblique foot and tail shocks retains greater forward momentum than the fluid decelerated by the main reflected shock, resulting in the formation of a wall jet. The bifurcated foot and tail shocks merge with the main reflected shock at the triple point, forming a $\lambda$-shock wave. Finally, the entropy difference between the gas traversing the bifurcation zone and the gas processed by the single-shock system generates a distinct slip line or shear layer.

\begin{figure}[!htbp]
  \centering
  \subfigure[]{\includegraphics[width=0.45\textwidth]{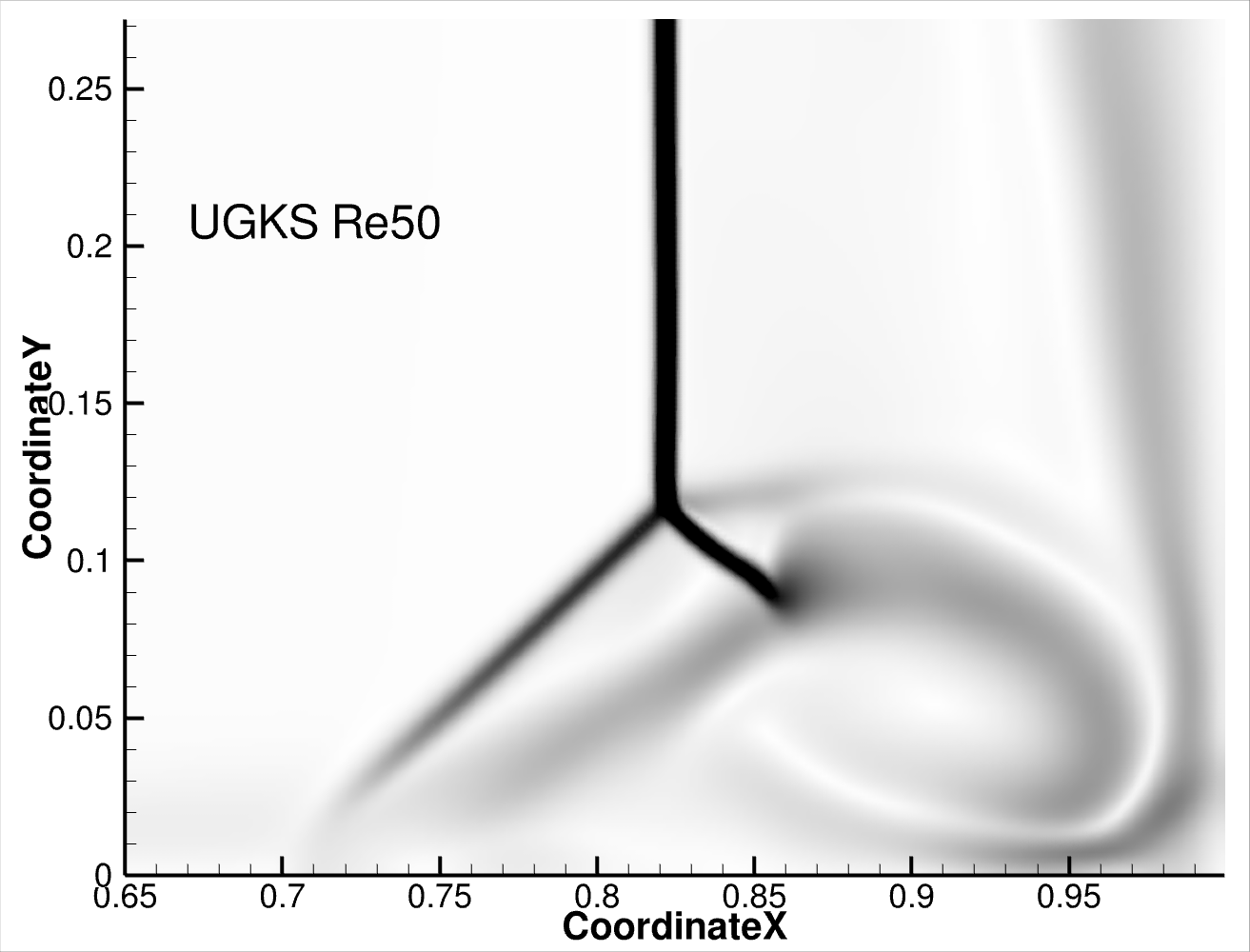}}
  \subfigure[]{\includegraphics[width=0.45\textwidth]{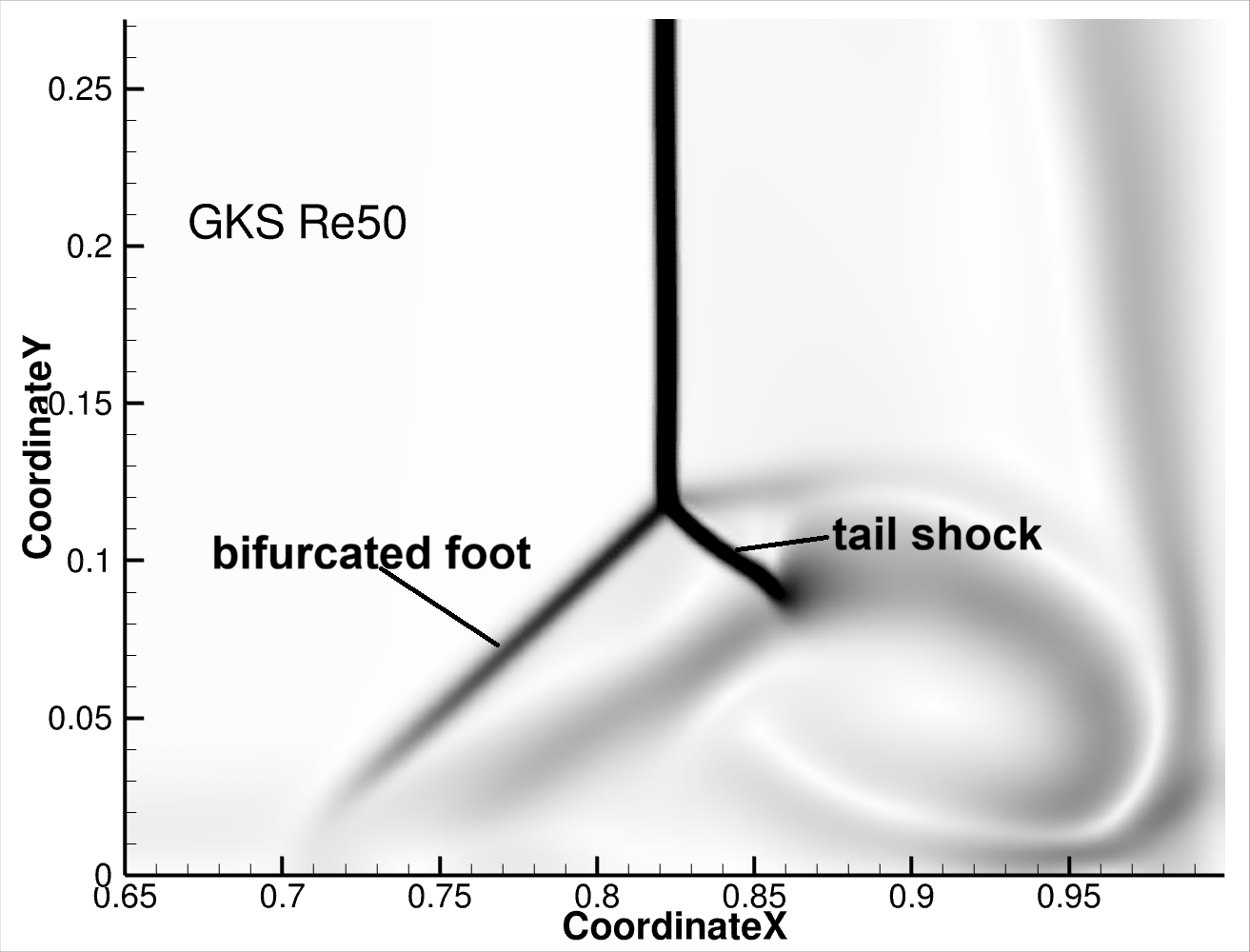}}
  \subfigure[]{\includegraphics[width=0.45\textwidth]{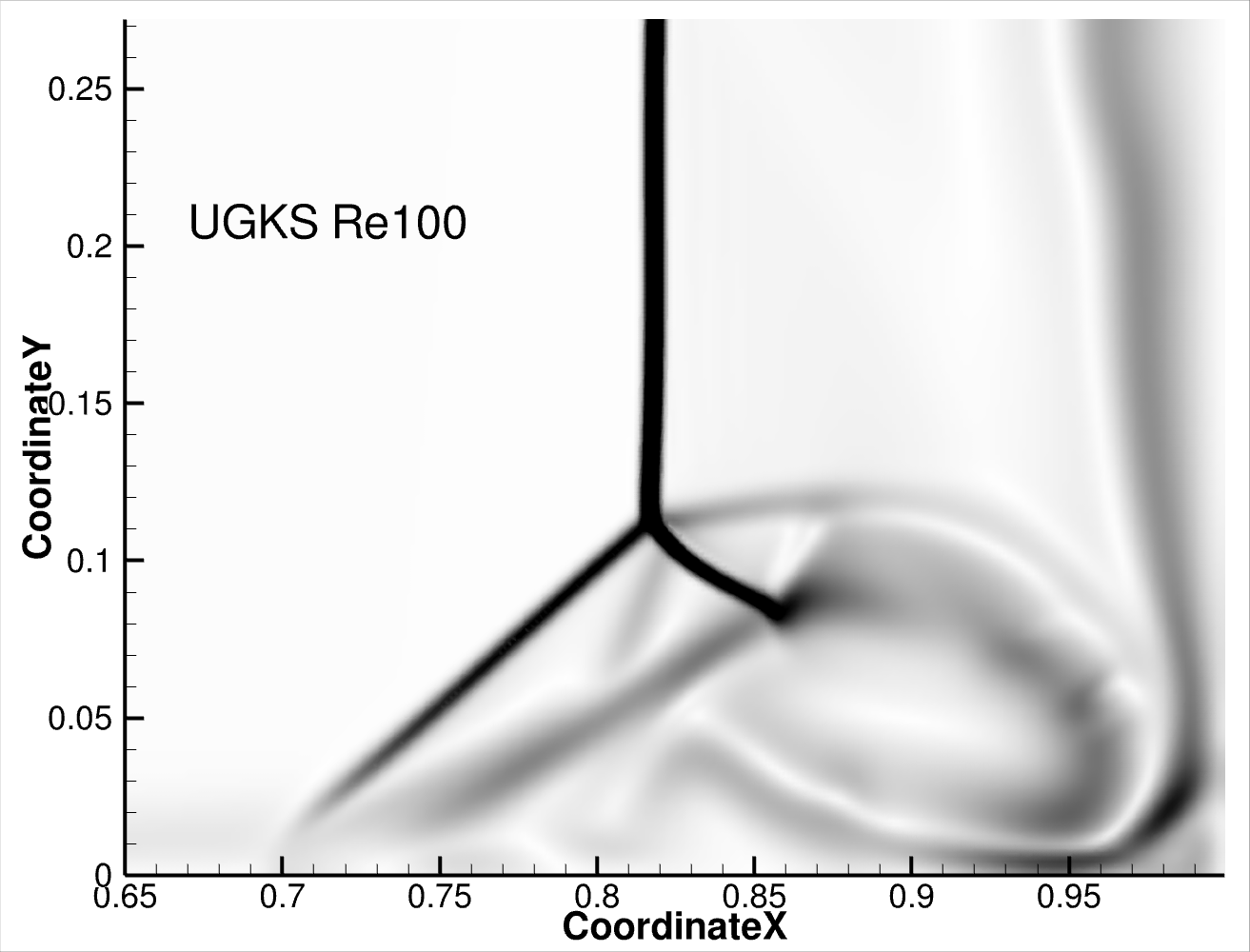}}
  \subfigure[]{\includegraphics[width=0.45\textwidth]{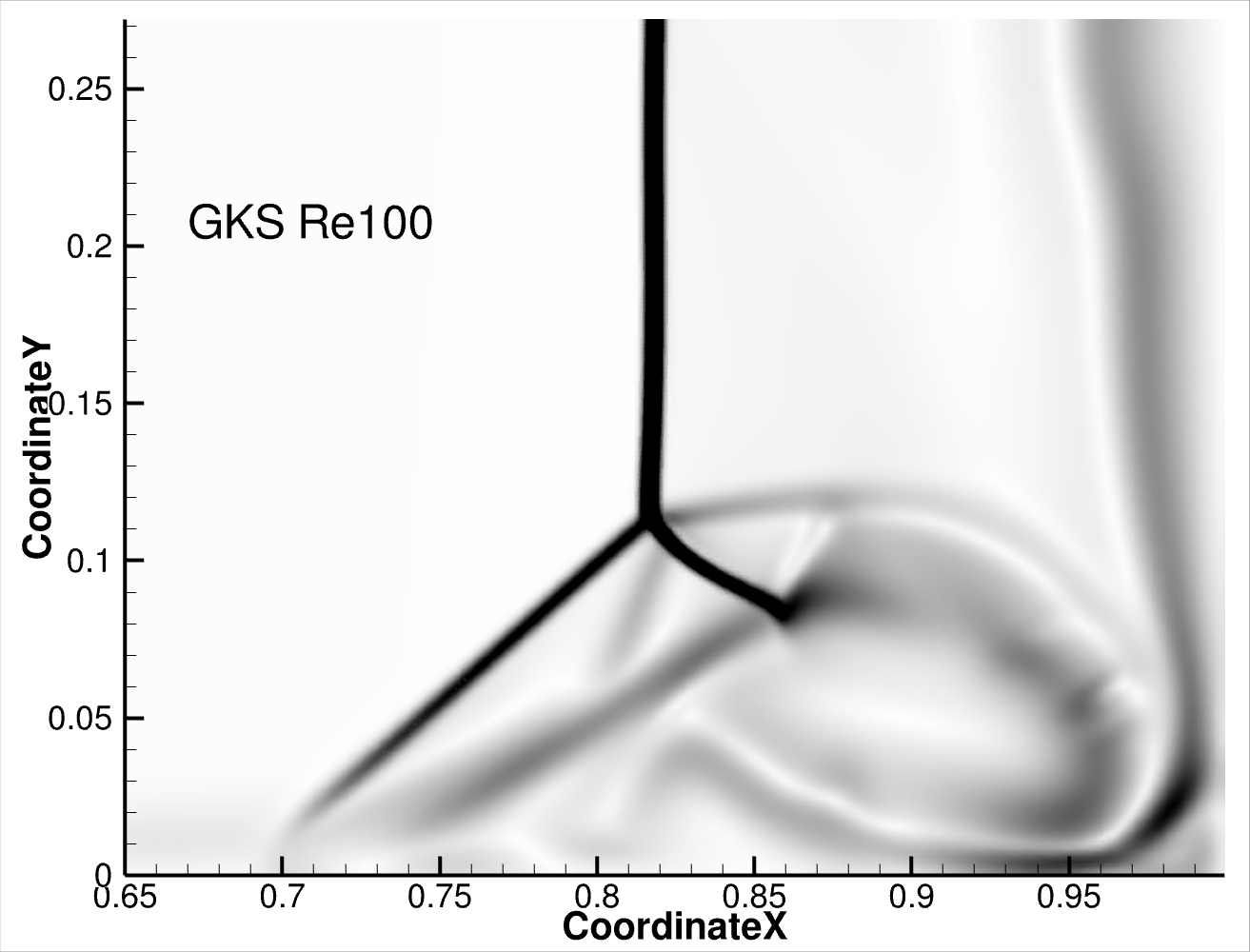}}
  \caption{The contour of density gradient by GKS and UGKS at the time of $t=0.6$. The first two figures are the results of Reynolds number 50, and the last two figures are the results of Reynolds number 100. For each Reynolds number, the first figure is the result of UGKS, and the second figure is the result of GKS.}
  \label{fig:densityGradientt0.6}
\end{figure}

\begin{figure}[!htbp]
  \centering
  \subfigure[]{\includegraphics[width=0.45\textwidth]{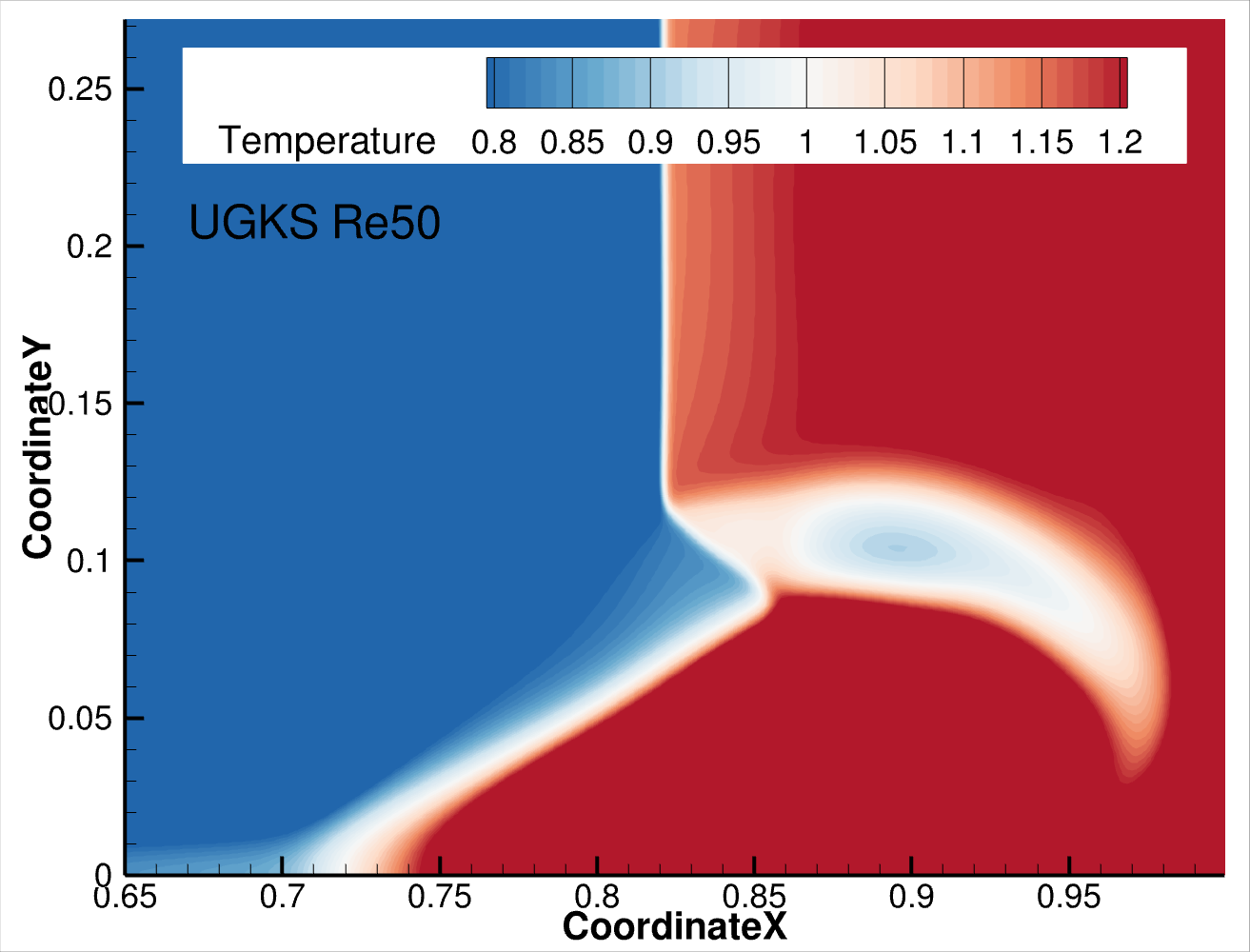}}
  \subfigure[]{\includegraphics[width=0.45\textwidth]{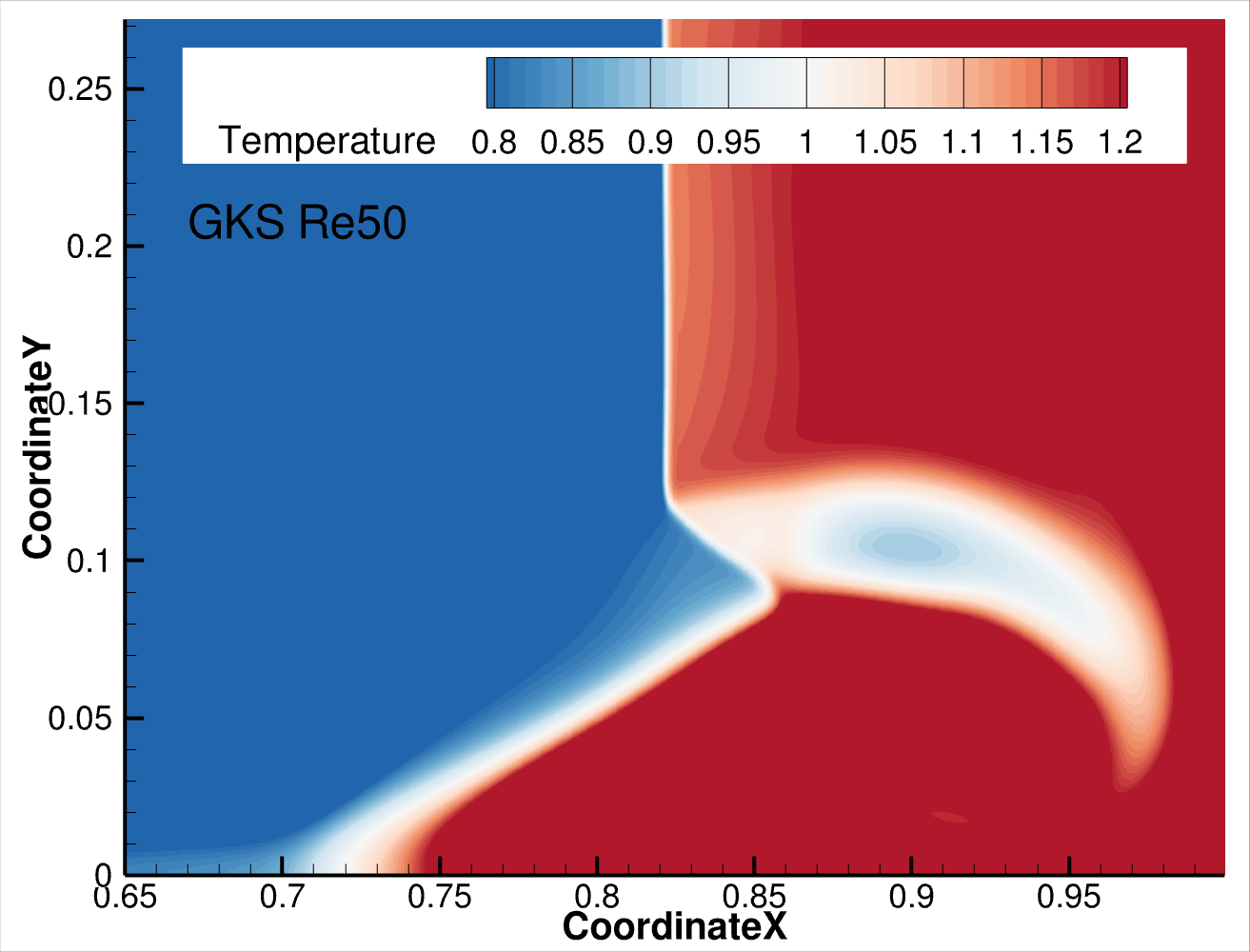}}
  \subfigure[]{\includegraphics[width=0.45\textwidth]{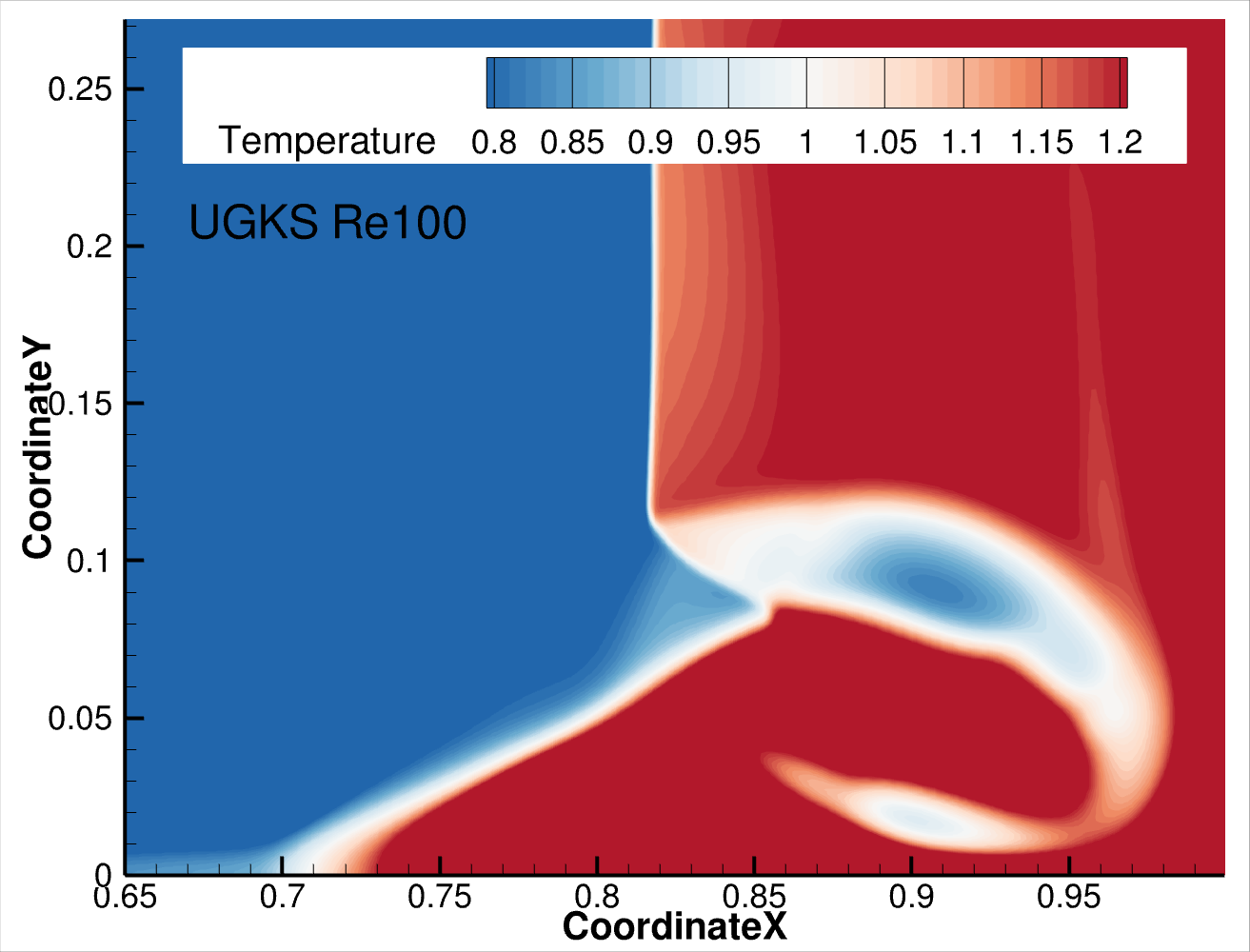}}
  \subfigure[]{\includegraphics[width=0.45\textwidth]{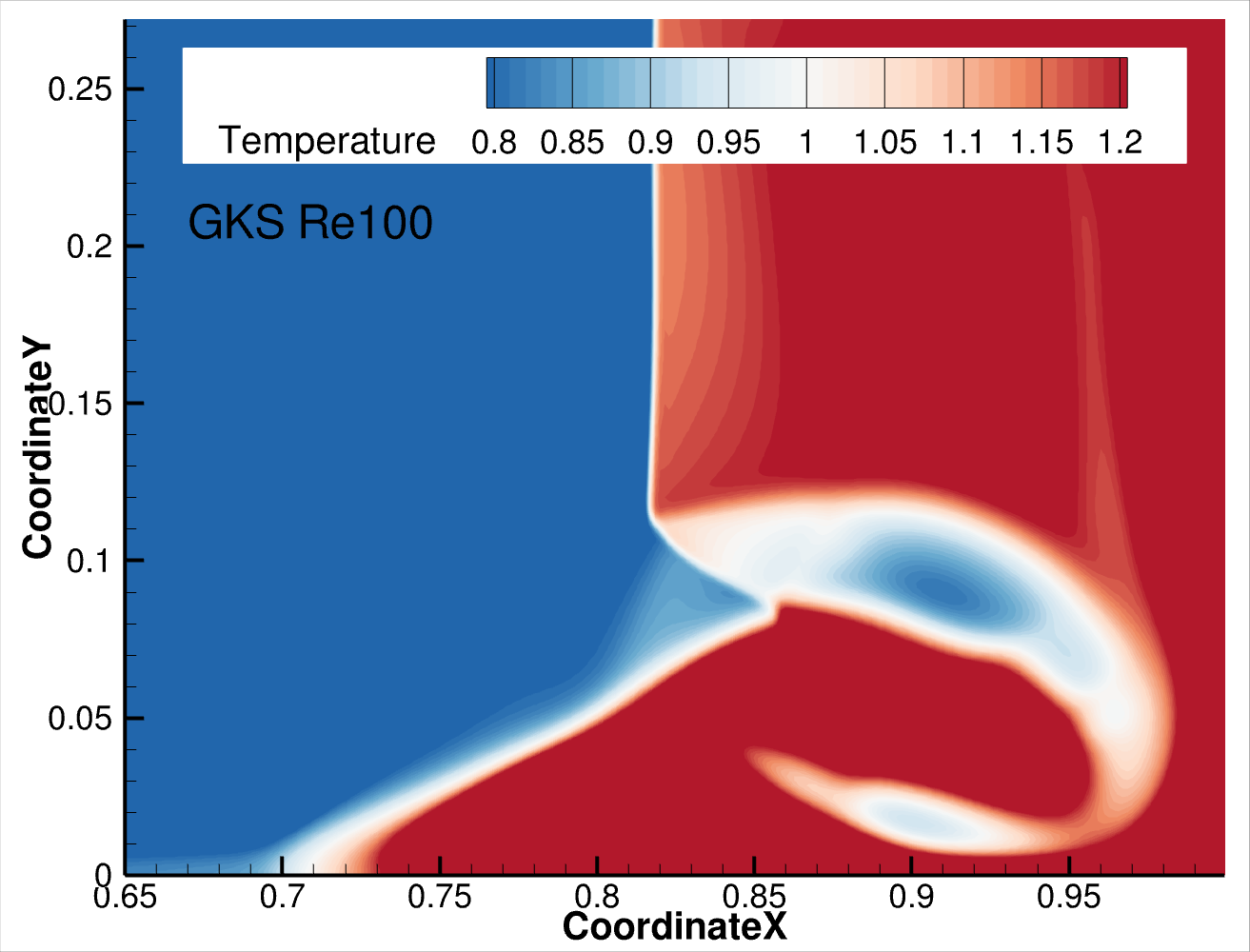}}
  \caption{The temperature contour of GKS and UGKS at the time of $t=0.6$. The first two figures are the results of Reynolds number 50, and the last two figures are the results of Reynolds number 100. For each Reynolds number, the first figure is the result of UGKS, and the second figure is the result of GKS.}
  \label{fig:temperaturet0.6}
\end{figure}

The density gradient of the viscous shock tube problem with GKS and UGKS at the time of $t=0.6$ is shown in Figure \ref{fig:densityGradientt0.6}.
At this time, the $\lambda$-shock wave was formed, and other $\lambda$-shock waves were formed by the interaction of the tail shock wave and the separated flow region. The second $\lambda$-shock wave was more obvious at the Reynolds number of 100. In addition, the triple point of the $\lambda$-shock predicted by UGKS is lower than the GKS results.
\begin{figure}[!htbp]
  \centering
  \subfigure[]{\label{fig:temperatureProfileLiney0.095t0.6}\includegraphics[width=0.35\textwidth]{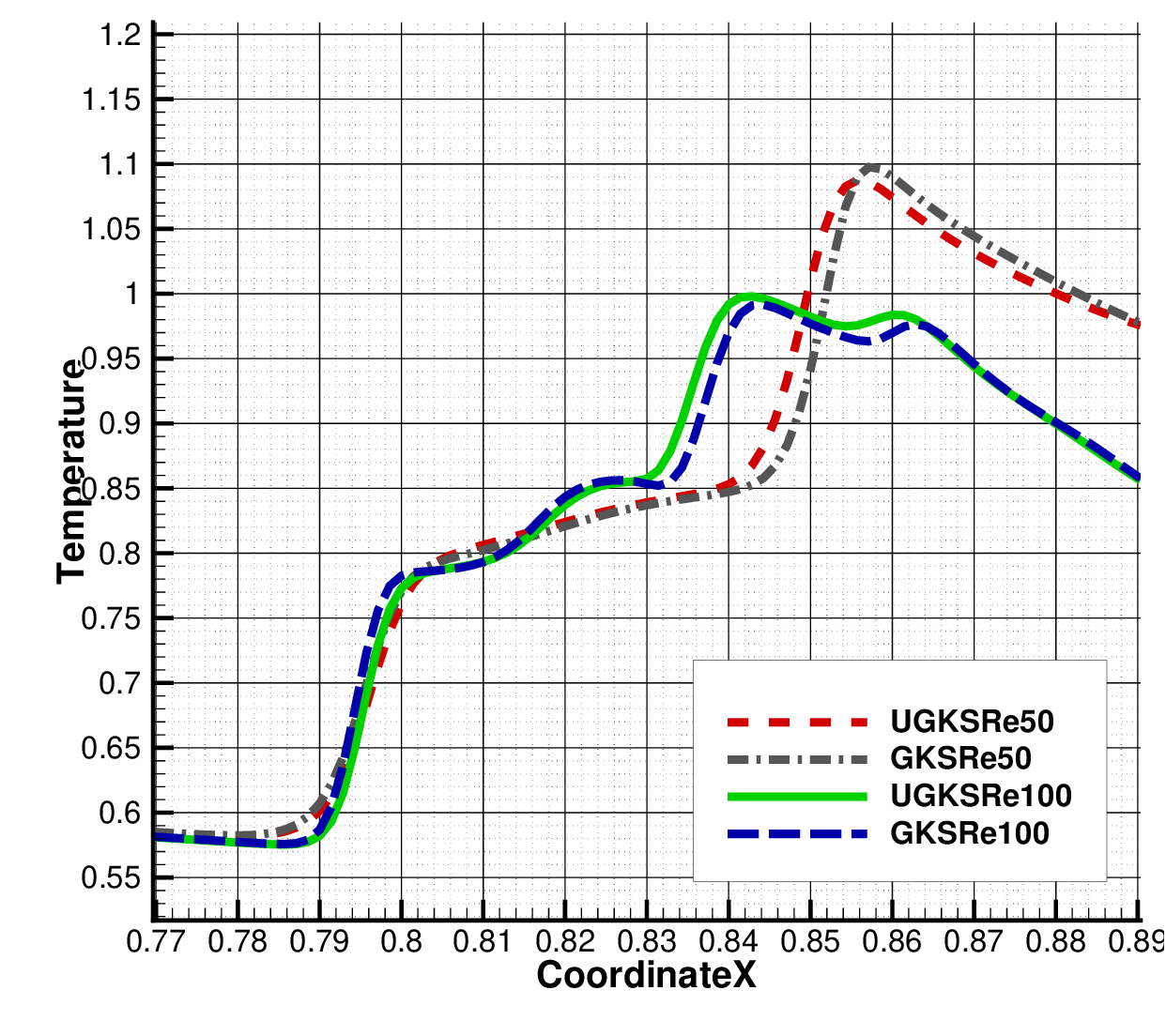}}
  \subfigure[]{\label{fig:xVelocityProfileLinex0.9t0.6}\includegraphics[width=0.35\textwidth]{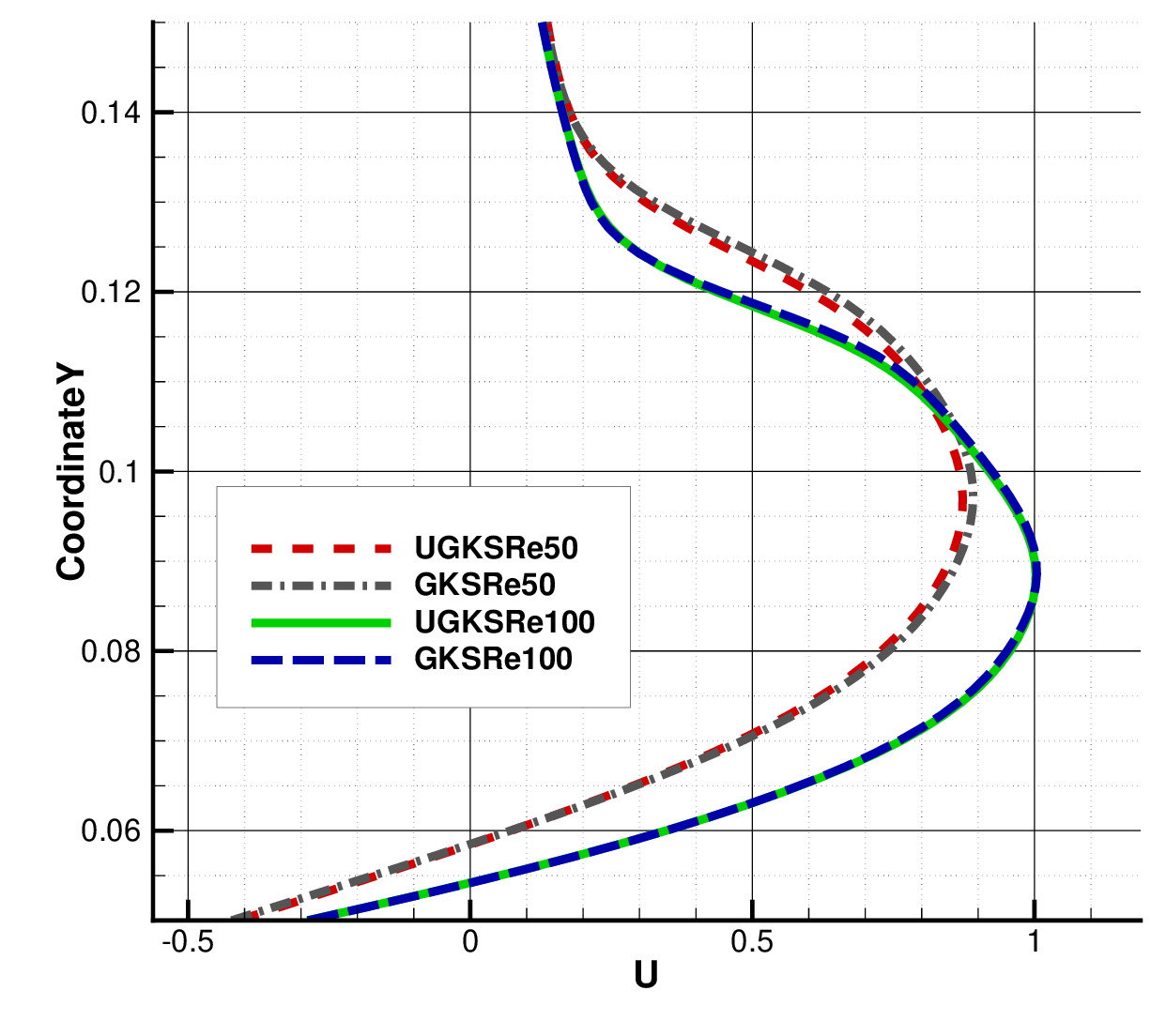}}
  \caption{The line profile at the time of $t=0.6$. Left: The temperature profile along the line of $y=0.095$, right: The $x$-direction velocity profile along the line of $x=0.9$.}
\end{figure}
The temperature profile along the line of $y=0.095$ at the time of $t=0.6$ is shown in Figure \ref{fig:temperatureProfileLiney0.095t0.6}.
From the figure, the different positions of the tail shock wave by GKS and UGKS are observed. The position of the tail shock wave predicted by UGKS is behind the GKS results, which is caused by the different positions of the triple point of the $\lambda$-shock wave predicted by GKS and UGKS, and the difference is more obvious at the Reynolds number of 50, which means that at lower Reynolds numbers, the rarefied effect is apparent and UGKS can resolve it. For the Reynolds number of 100, three shock waves caused by the two $\lambda$-shock waves are observed, while for the Reynolds number of 50, only two shock waves are observed. However, in the GKS results at Reynolds number 50, a flat temperature region is located at $x=0.84$, indicating that the second $\lambda$-shock wave is weakly resolved; in the UGKS results, this flat region is resolved.
The temperature contour of GKS and UGKS at the time of $t=0.6$ is shown in Figure \ref{fig:temperaturet0.6}.
The temperature contours show the low-temperature jet from the gas that encounters the oblique foot and tail shocks.
The $x$-direction velocity profile along the line of $x=0.9$ at the time of $t=0.6$ is shown in Figure \ref{fig:xVelocityProfileLinex0.9t0.6}.
This figure shows the velocity profile of the low-temperature jet from the gas that encounters the oblique foot and tail shocks. The velocity predicted by GKS is higher than that by UGKS, consistent with the temperature profile, and the difference is more pronounced at a Reynolds number of 50.

\begin{figure}[htbp]
  \centering
  \subfigure[]{\includegraphics[width=0.45\textwidth]{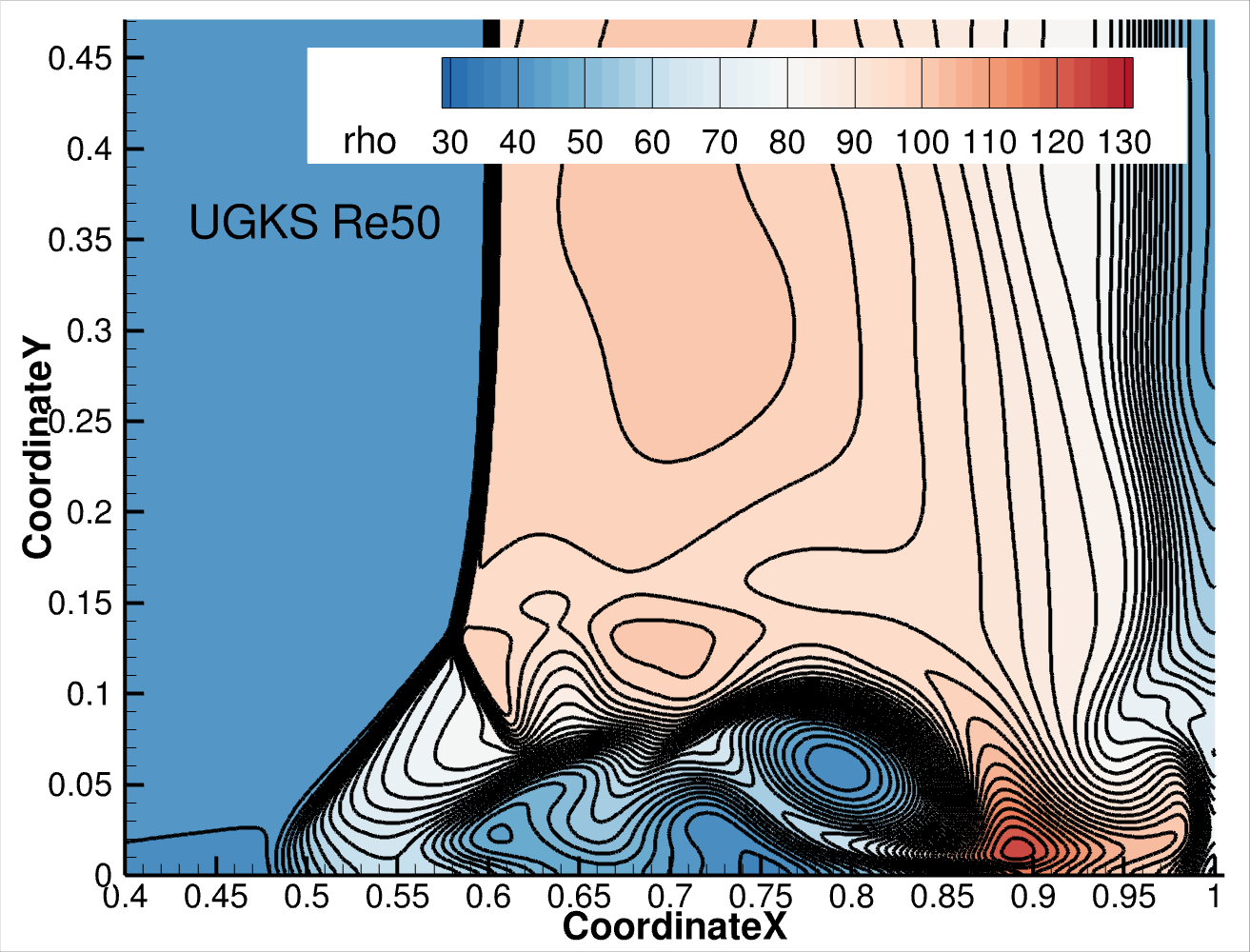}}
  \subfigure[]{\includegraphics[width=0.45\textwidth]{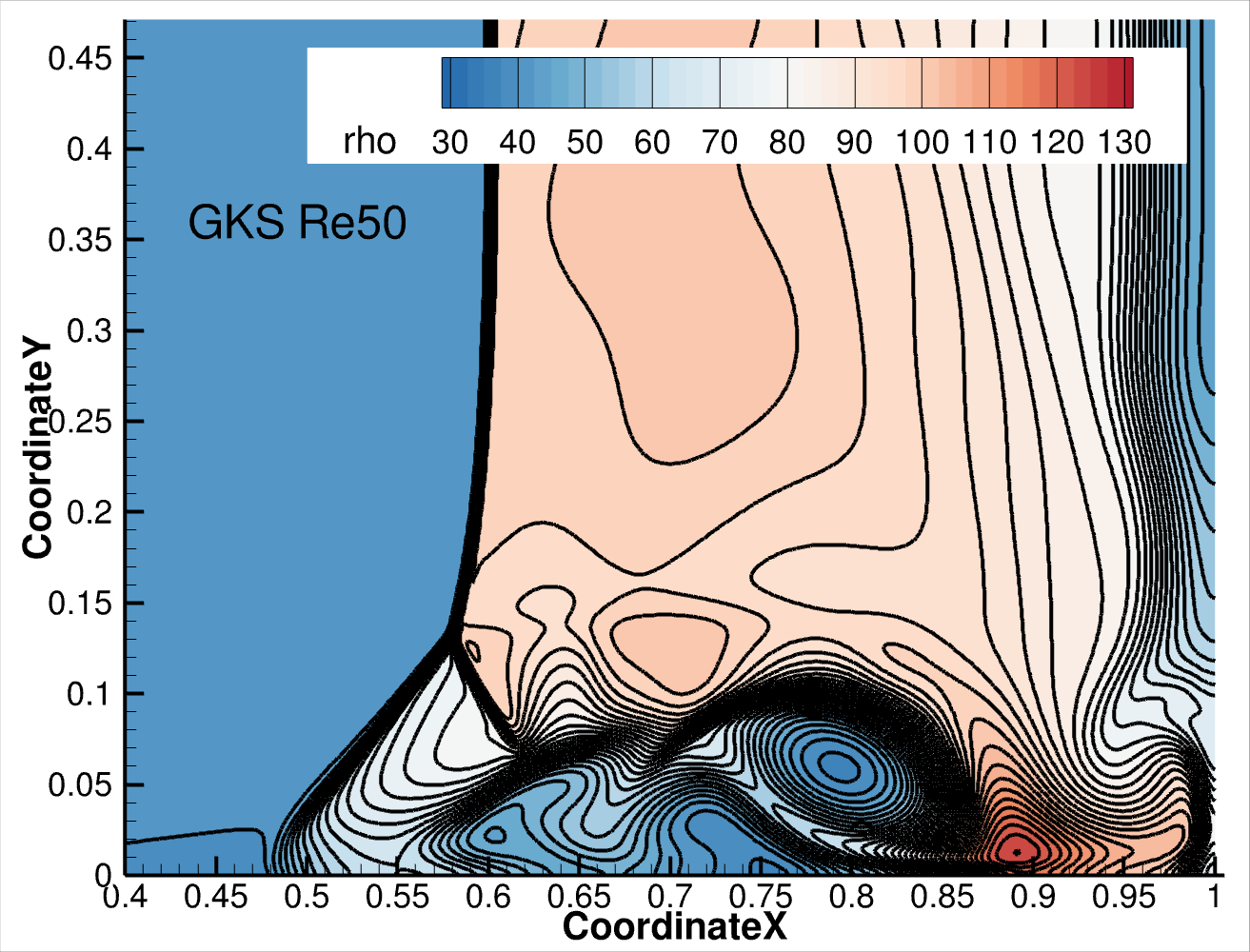}}
  \subfigure[]{\includegraphics[width=0.45\textwidth]{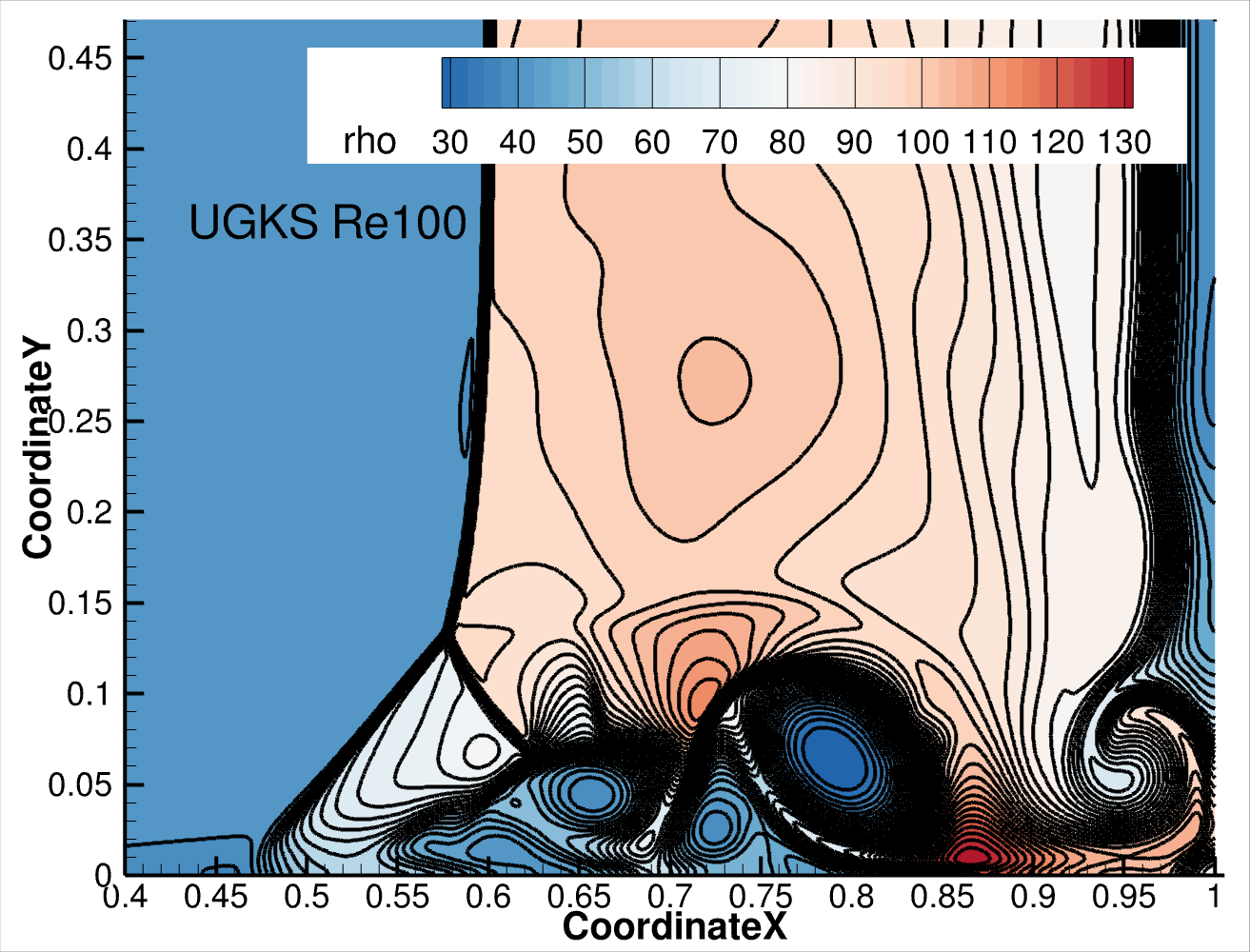}}
  \subfigure[]{\includegraphics[width=0.45\textwidth]{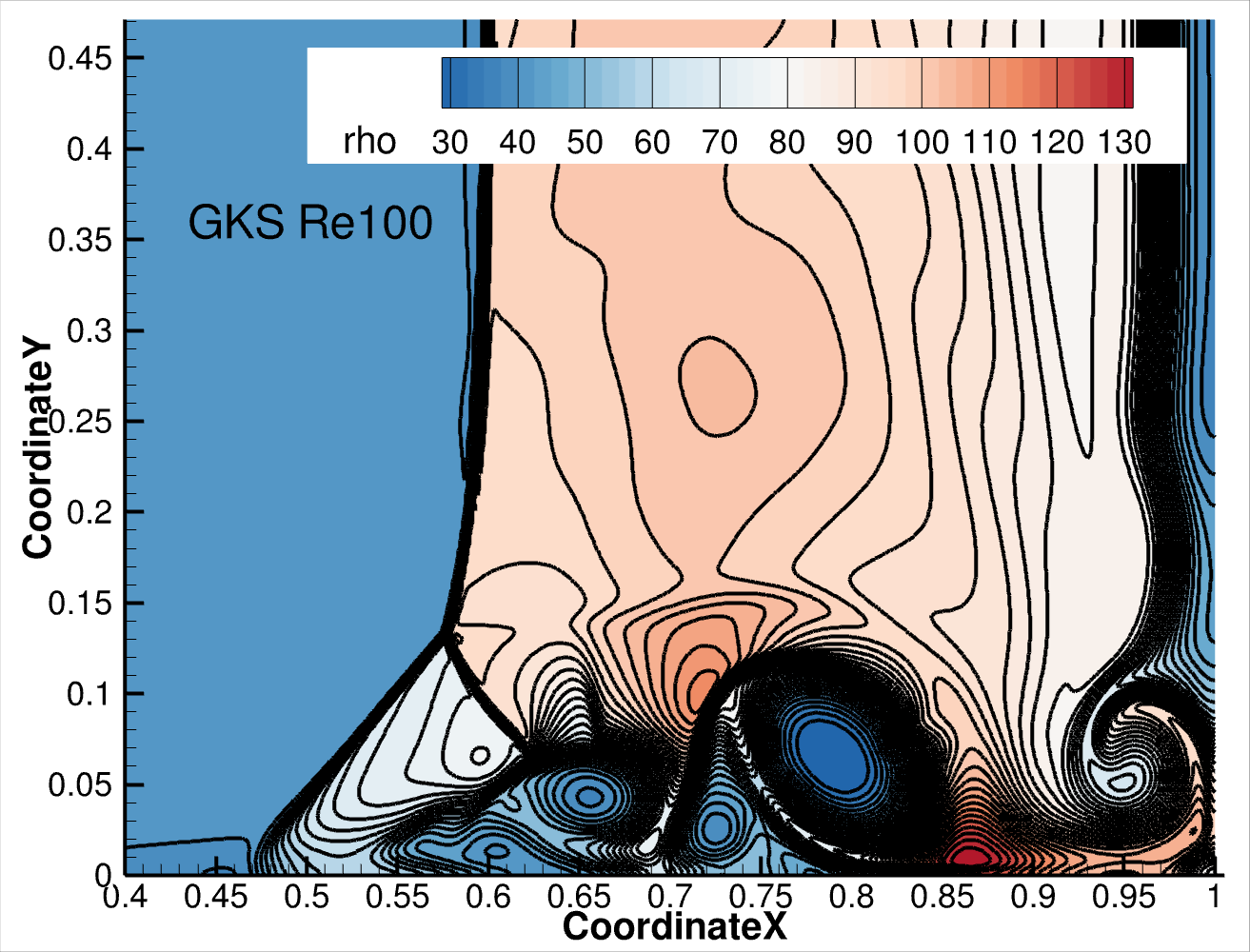}}
  \caption{The density contour of GKS and UGKS at the time of $t=1.0$. The first two figures are the results of Reynolds number 50, and the last two figures are the results of Reynolds number 100. For each Reynolds number, the first figure is the result of UGKS, and the second figure is the result of GKS.}
  \label{fig:densityContourt1.0}
\end{figure}

\begin{figure}[!htbp]
  \centering
  \includegraphics[width=0.45\textwidth]{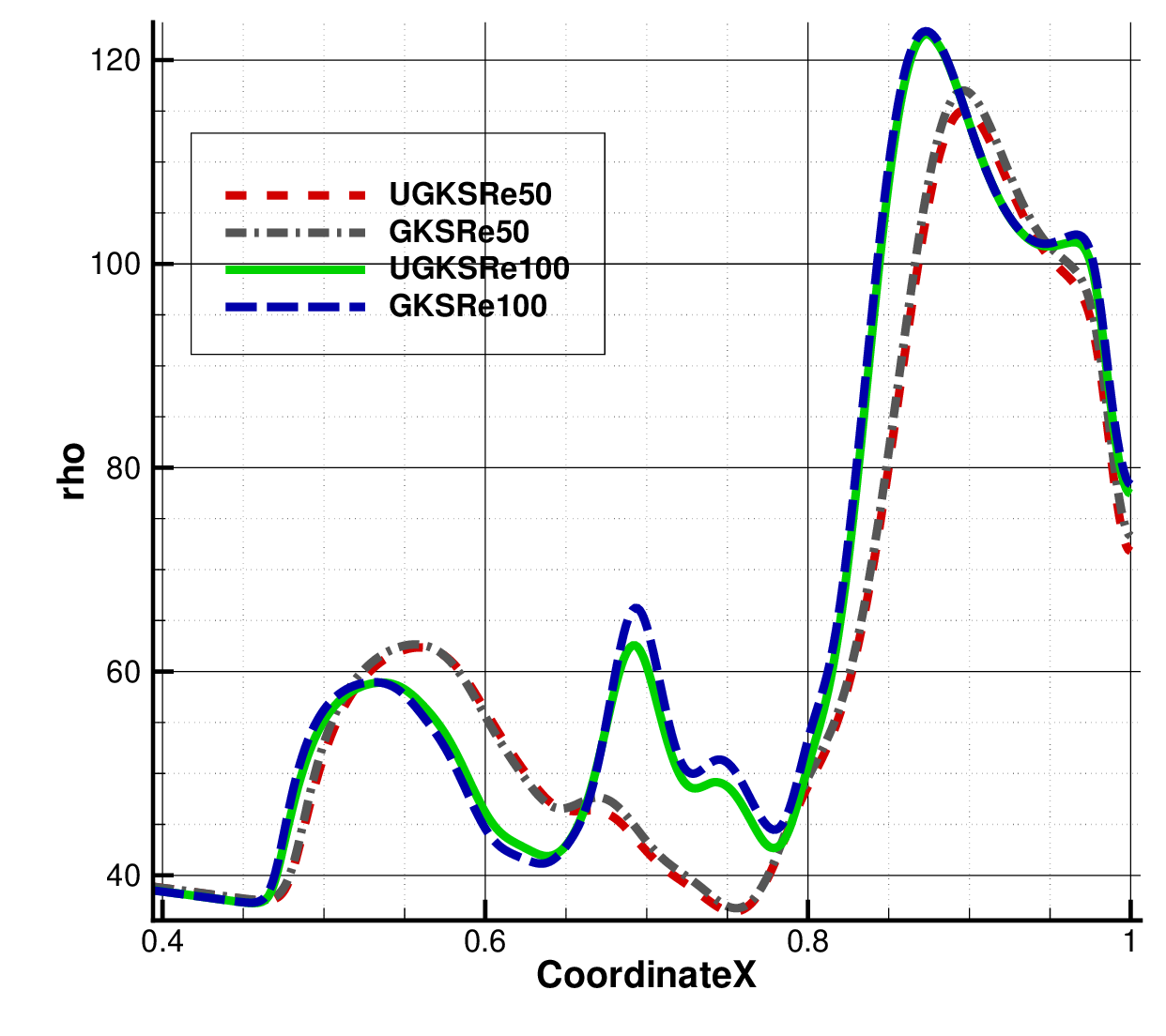}
  \caption{The density profile along the line of $y=0.0$ at the time of $t=1.0$.}
  \label{fig:densityProfileLiney0t1.0}
\end{figure}

\begin{figure}[!htbp]
  \centering
  \subfigure[]{\includegraphics[width=0.45\textwidth]{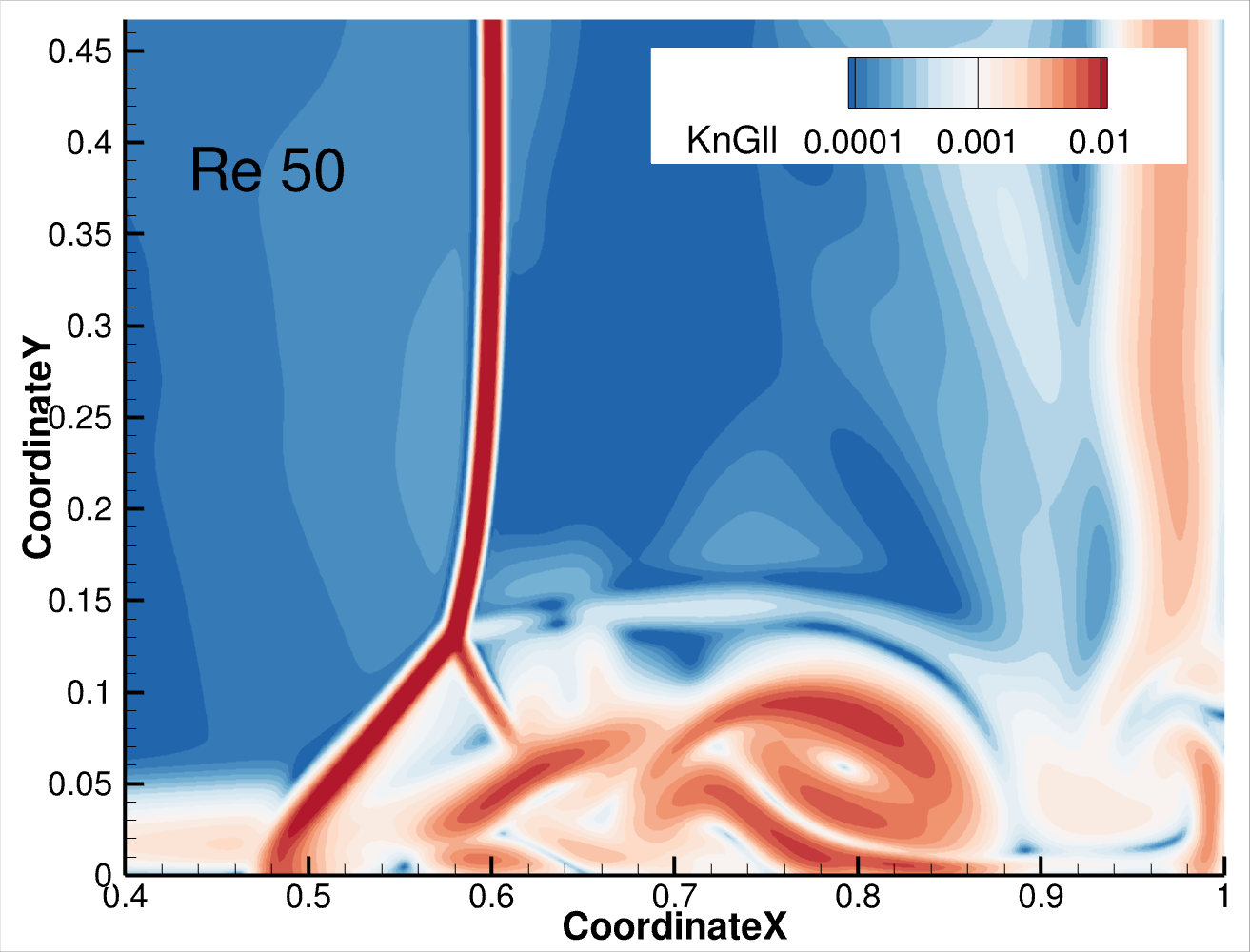}}
  \subfigure[]{\includegraphics[width=0.45\textwidth]{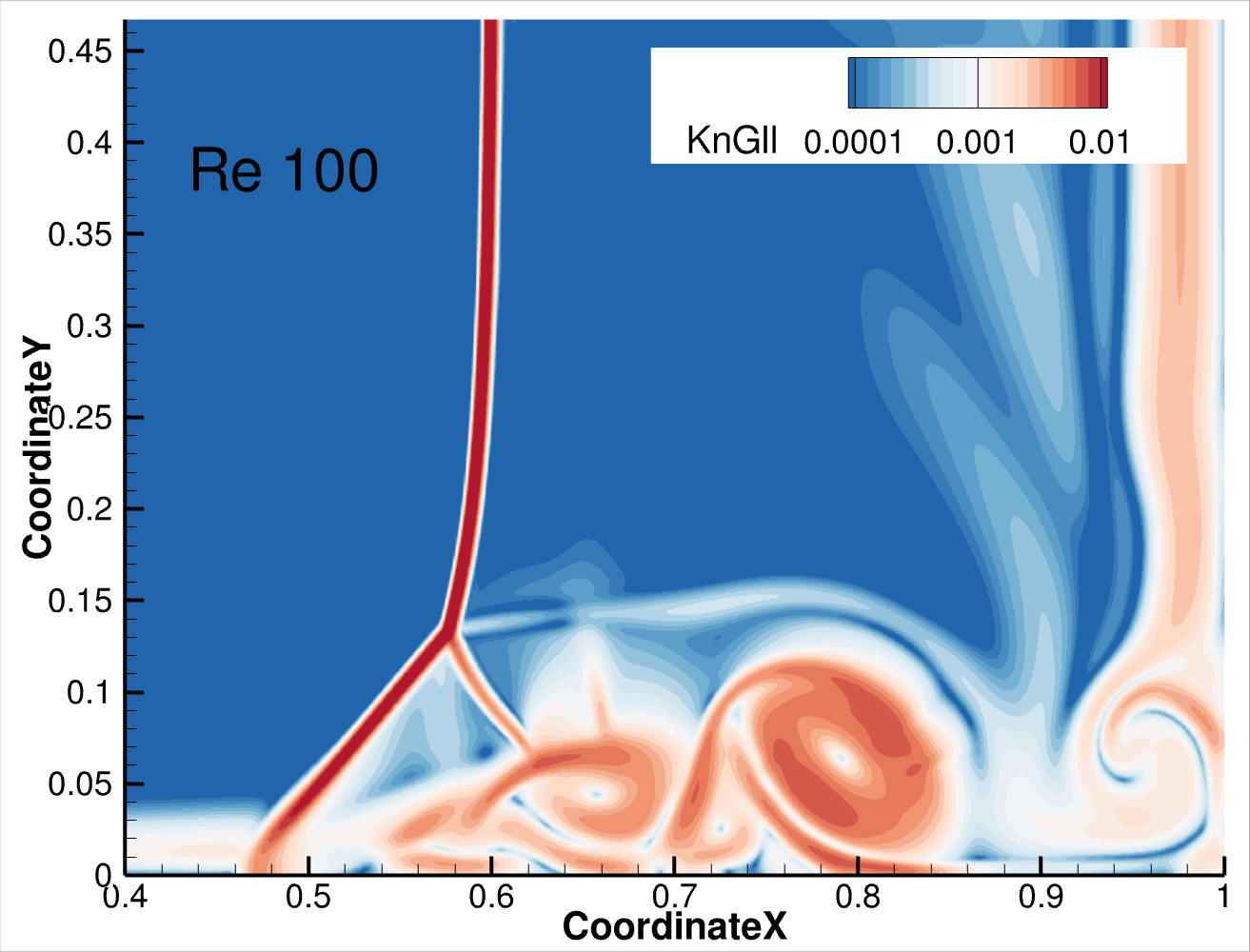}}
  \caption{The Knudsen number $\text{Kn}_{Gll}$ contour at the time of $t=1.0$. (a) The results of Reynolds number 50, (b) The results of Reynolds number 100.}
  \label{fig:KnudsenNumberContourt1.0}
\end{figure}

At the time of $t=1.0$, shock waves move away from the wall, and more complicated flow structures are observed at the boundary layer. The density contours of GKS and UGKS at the time of $t=1.0$ are shown in Figure \ref{fig:densityContourt1.0}.
In addition, the density profile along the line of $y=0.0$ at the time of $t=1.0$ is shown in the Figure \ref{fig:densityProfileLiney0t1.0}.
This profile is usually set as the reference profile to evaluate the accuracy of the numerical method. However, an unusual phenomenon is observed. The difference between GKS and UGKS is more obvious at the Reynolds number of 100, which is consistent with the density contour. The reason is that at a Reynolds number of 100, even though the flow condition is more continuous, the non-equilibrium effect is more apparent than at a Reynolds number of 50. At the Reynolds number of 50, the viscous effect is more apparent, but the flow is smoother; thus, the difference is not as obvious as at the Reynolds number of 100. As shown in the Figure \ref{fig:KnudsenNumberContourt1.0}, the Knudsen number ($\text{Kn}_{Gll}=l|\nabla \rho|/\rho$) contours of UGKS with Reynolds numbers of 50 and 100 are shown, which indicates the non-equilibrium effect. The maximum Knudsen number is 0.077 for Reynolds number 50 and 0.053 for Reynolds number 100.
The non-equilibrium effect appears in the region of the main shock wave, $\lambda$-shock wave, and the boundary layer. Although the non-equilibrium effect of the main shock wave and $\lambda$-shock wave is not as obvious as the Reynolds number of 50, the non-equilibrium effect of the boundary layer is more apparent at the Reynolds number of 100. The main difference between GKS and UG25S at the Reynolds number of 100 is caused by the non-equilibrium effect of the boundary layer. This case shows that, in the continuum flow regime, common computational fluid dynamics methods do not account for non-equilibrium effects, leading to deviations between numerical and physical results.

\begin{figure}[!htbp]
  \centering
  \subfigure[]{
  \includegraphics[width=0.3\textwidth]{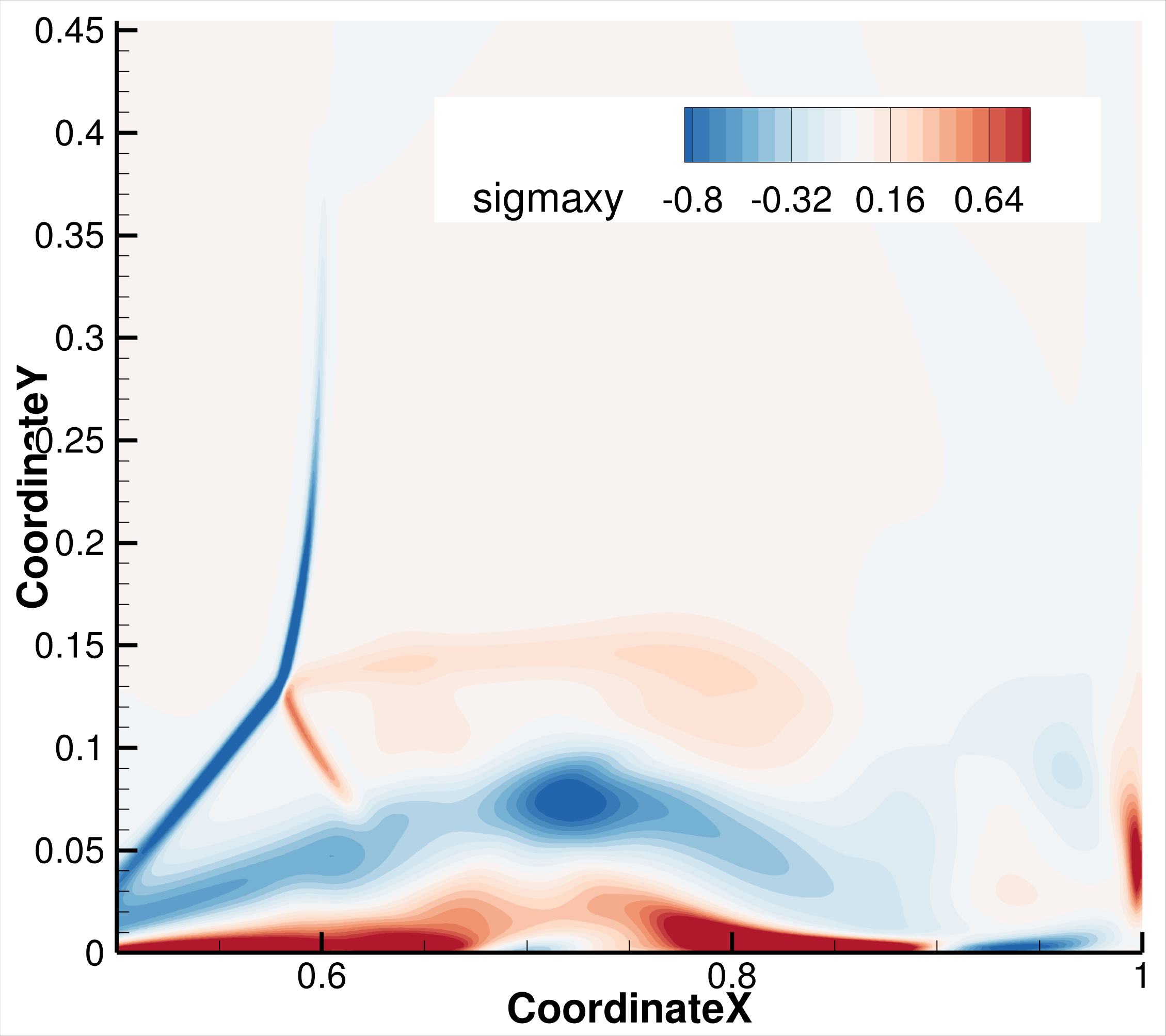}
  \includegraphics[width=0.3\textwidth]{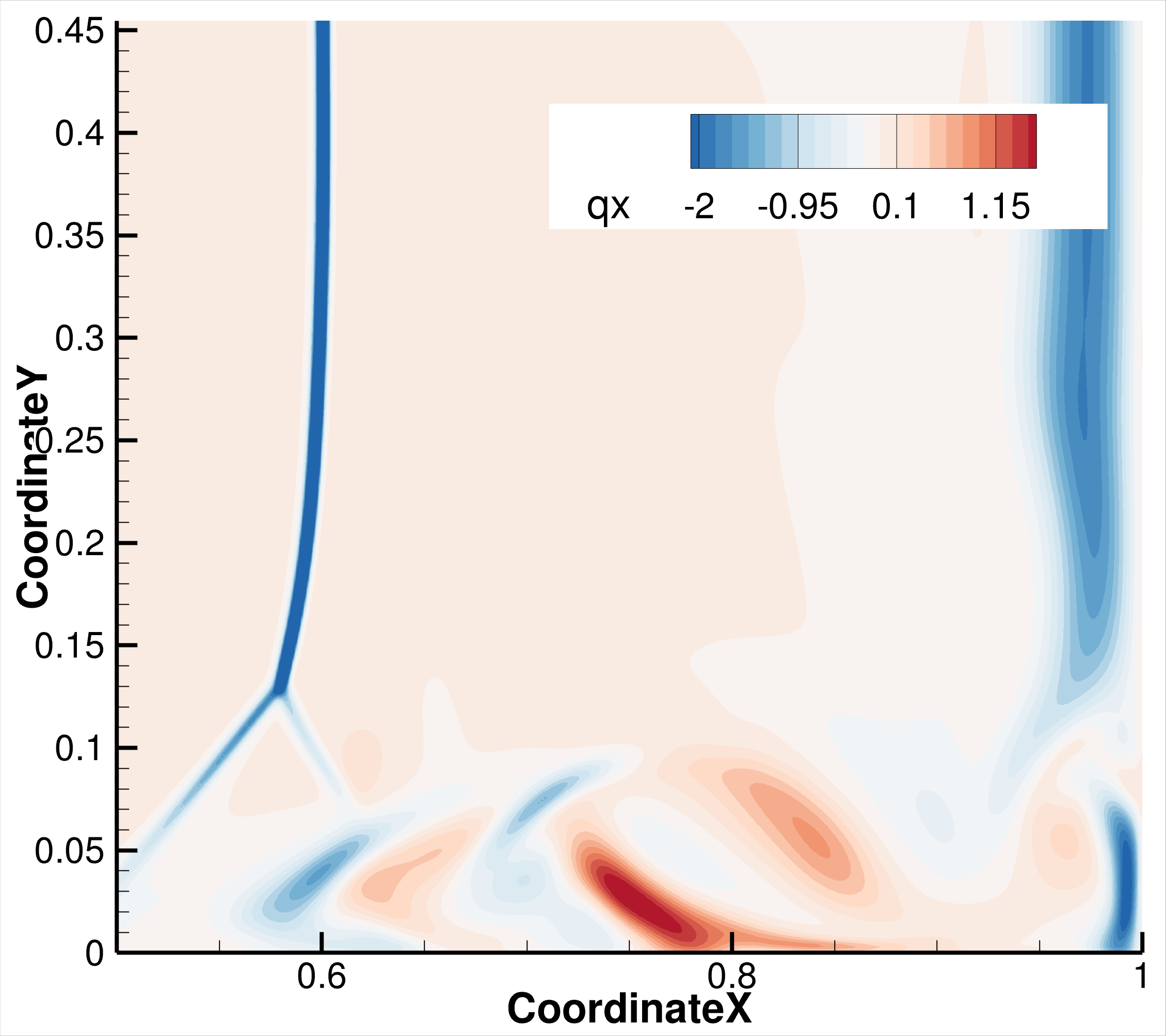}
  \includegraphics[width=0.3\textwidth]{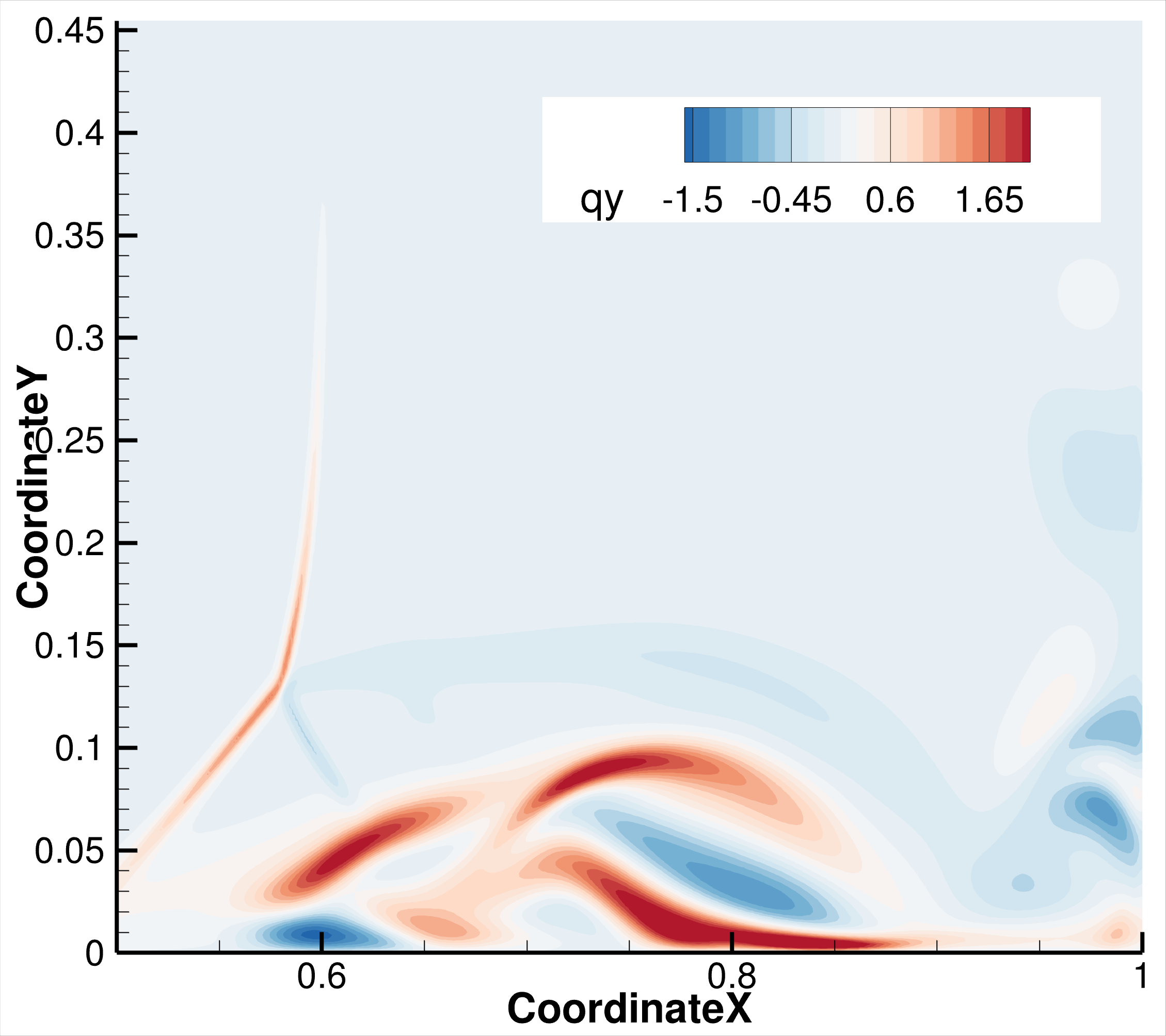}
  }
  \subfigure[]{
  \includegraphics[width=0.3\textwidth]{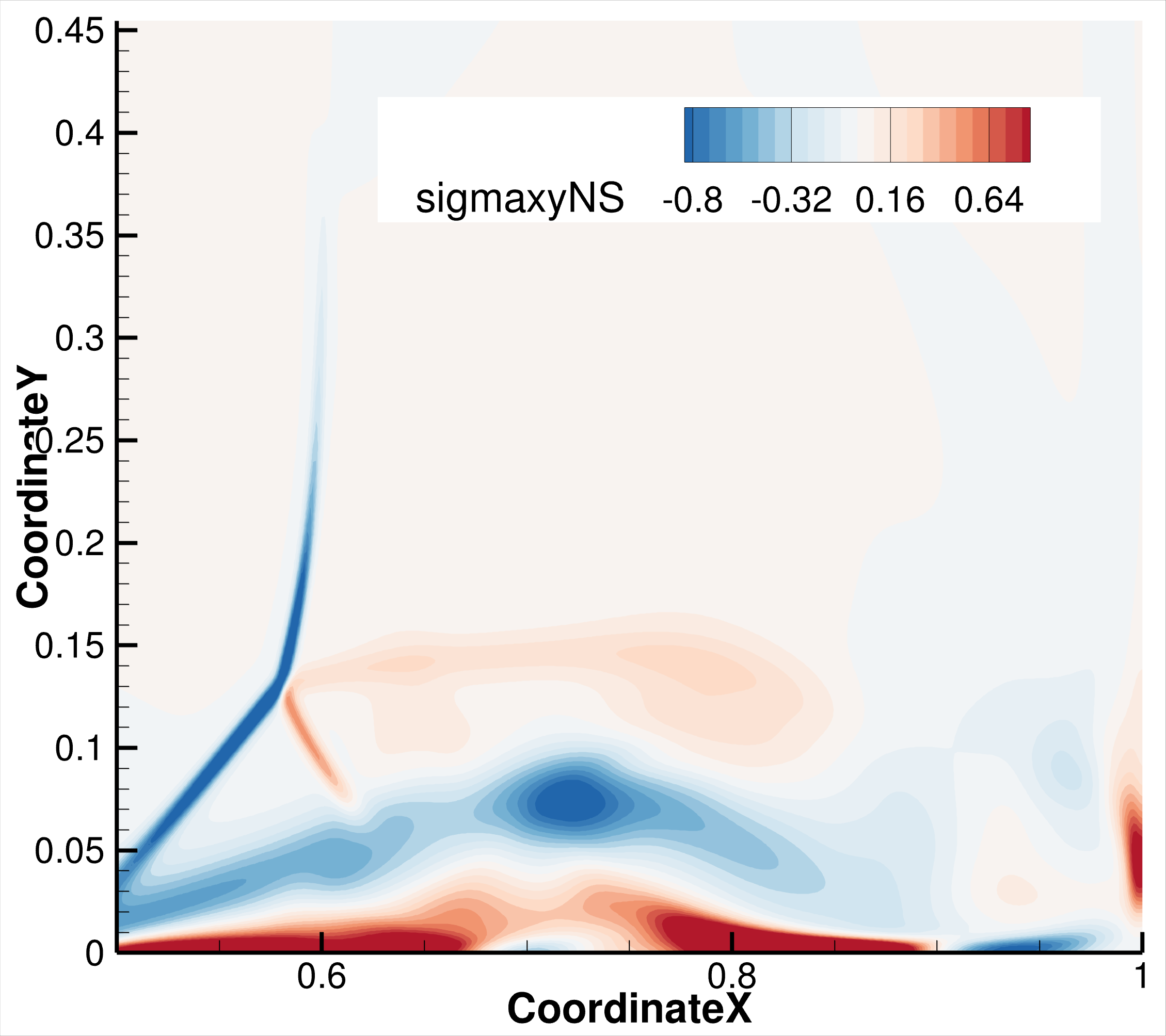}
  \includegraphics[width=0.3\textwidth]{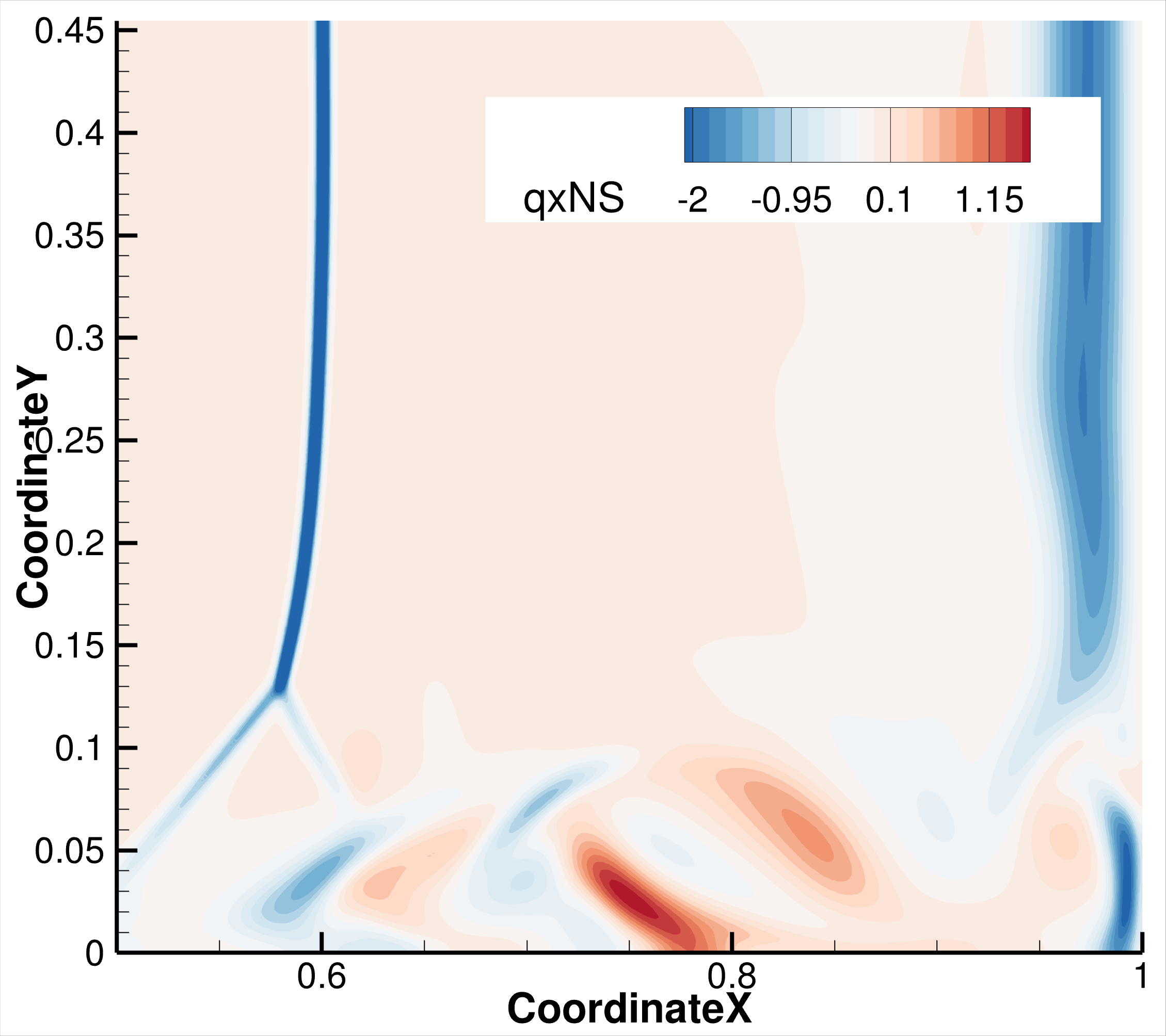}
  \includegraphics[width=0.3\textwidth]{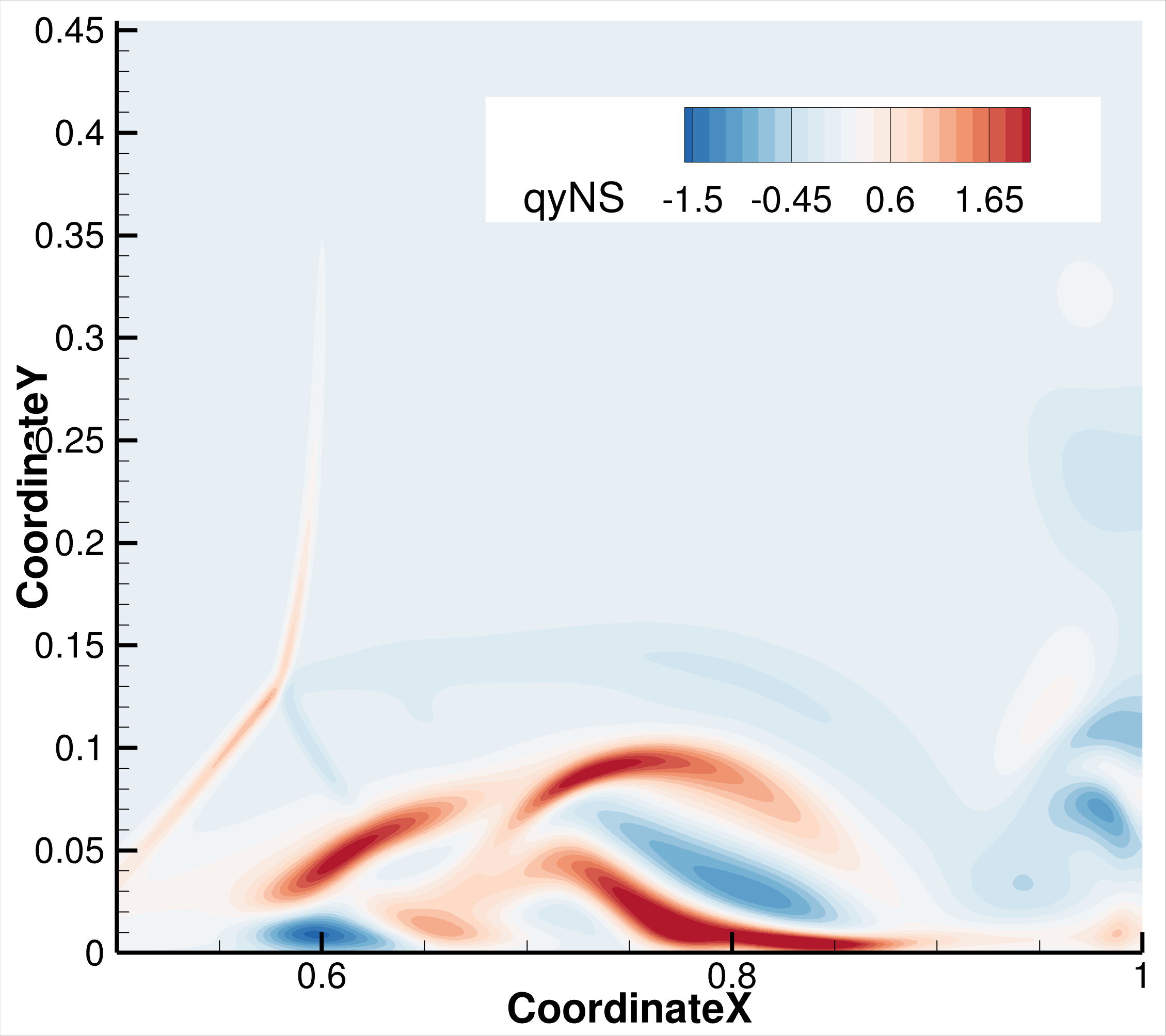}
  }
  \subfigure[]{
  \includegraphics[width=0.3\textwidth]{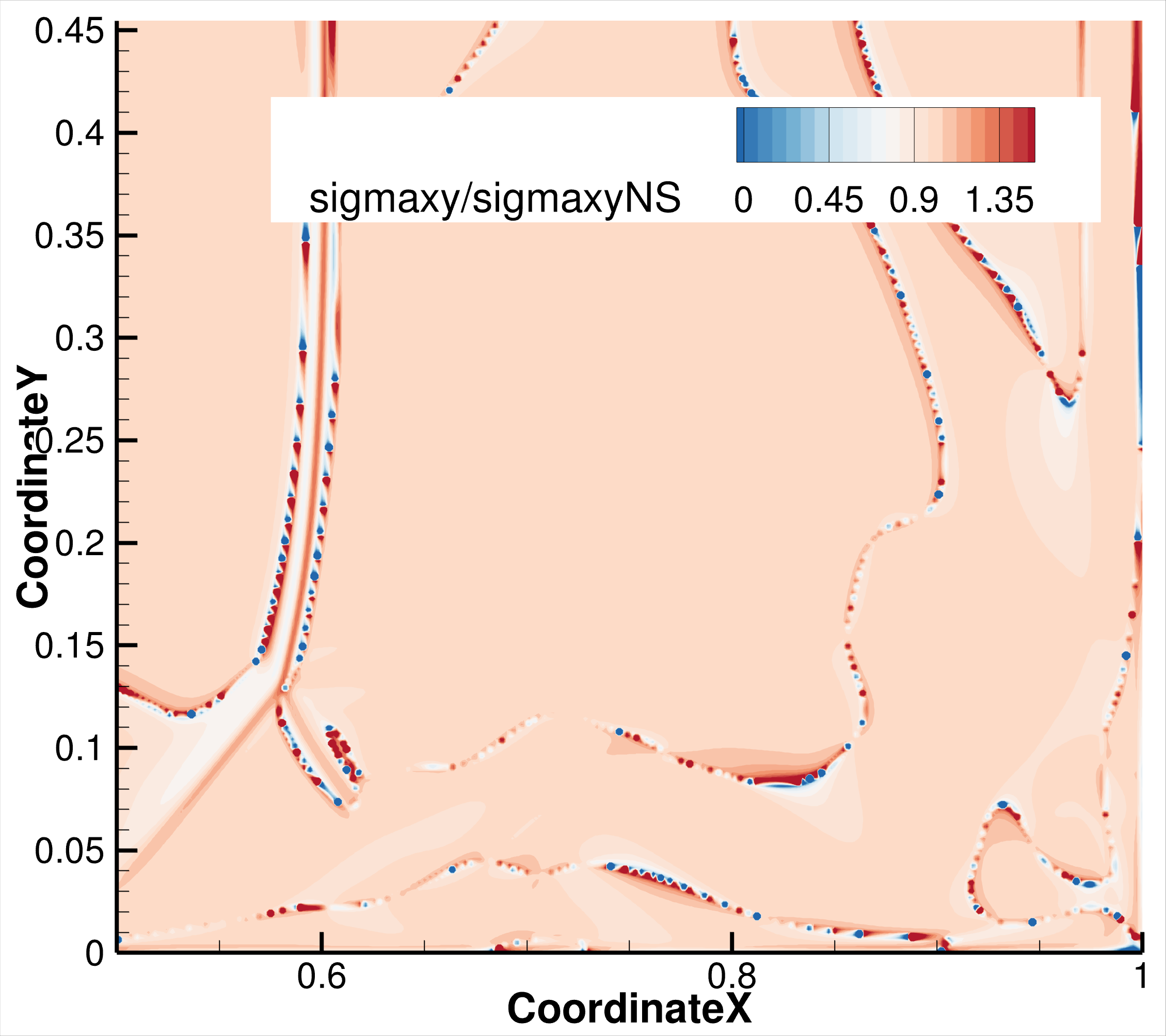}
  \includegraphics[width=0.3\textwidth]{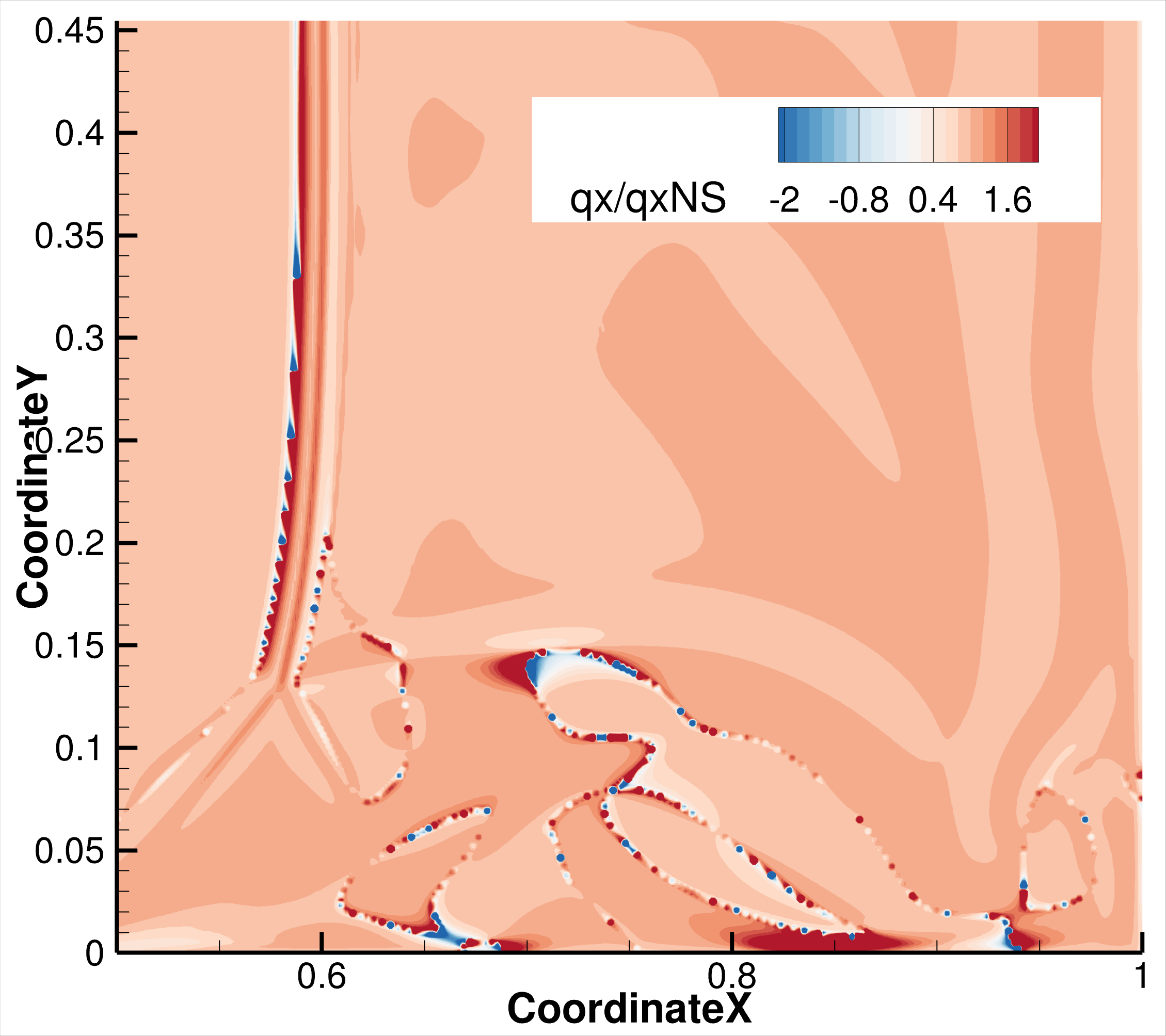}
  \includegraphics[width=0.3\textwidth]{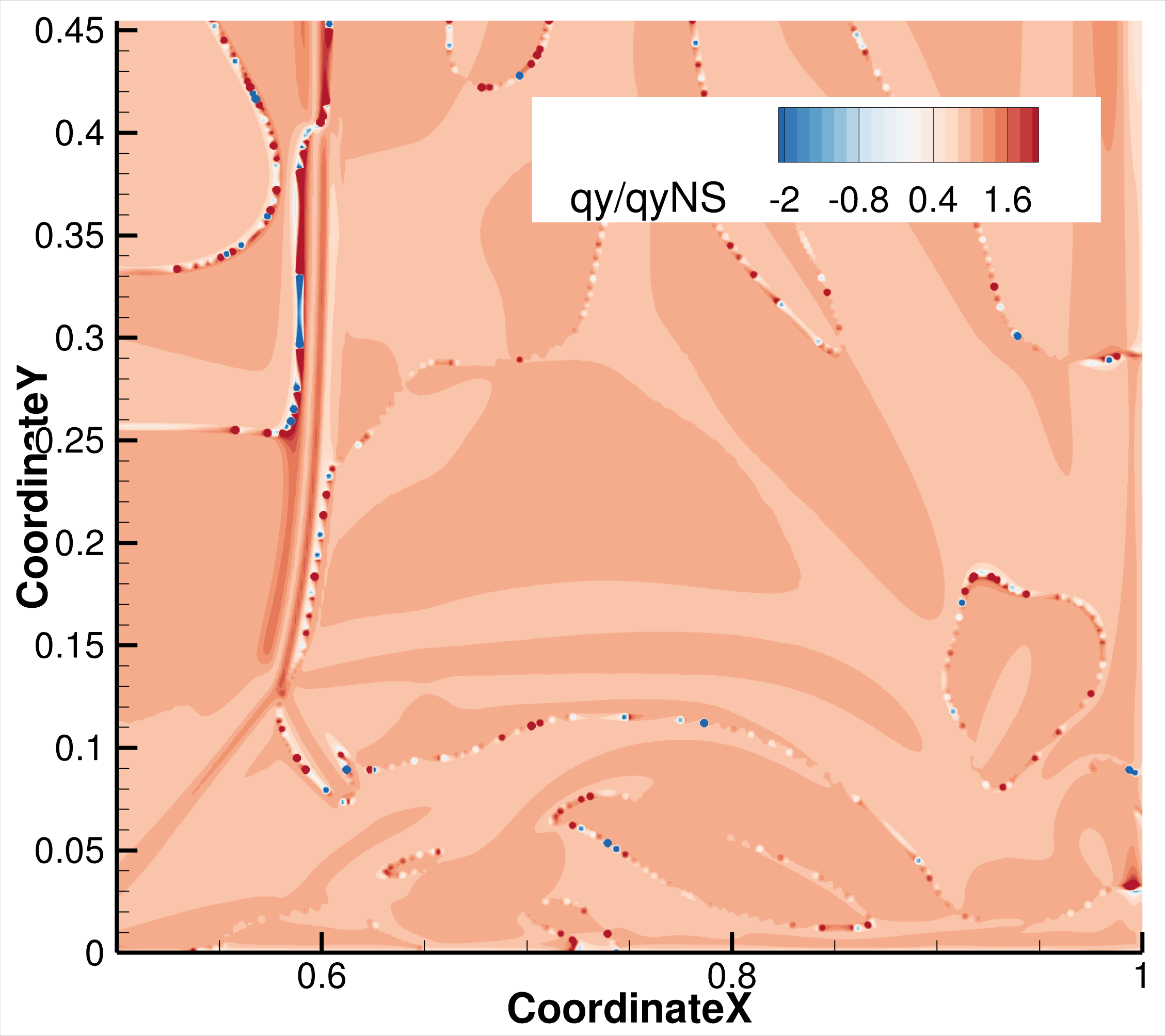}
  }
  \caption{The $\sigma_{xy}$ and $q_x$ and $q_y$ contour of UGKS at the time of $t=1.0$ at the Reynolds number of 50. (a) Calculated by the momentum of distribution function, (b) Calculated by the Newton stress tensor and Fourier's law of heat conduction, (c) The ratio of (a) and (b).}
  \label{fig:sigmaxyContourt1.0}
\end{figure}

\begin{figure}[!htbp]
  \centering
  \subfigure[]{
  \includegraphics[width=0.3\textwidth]{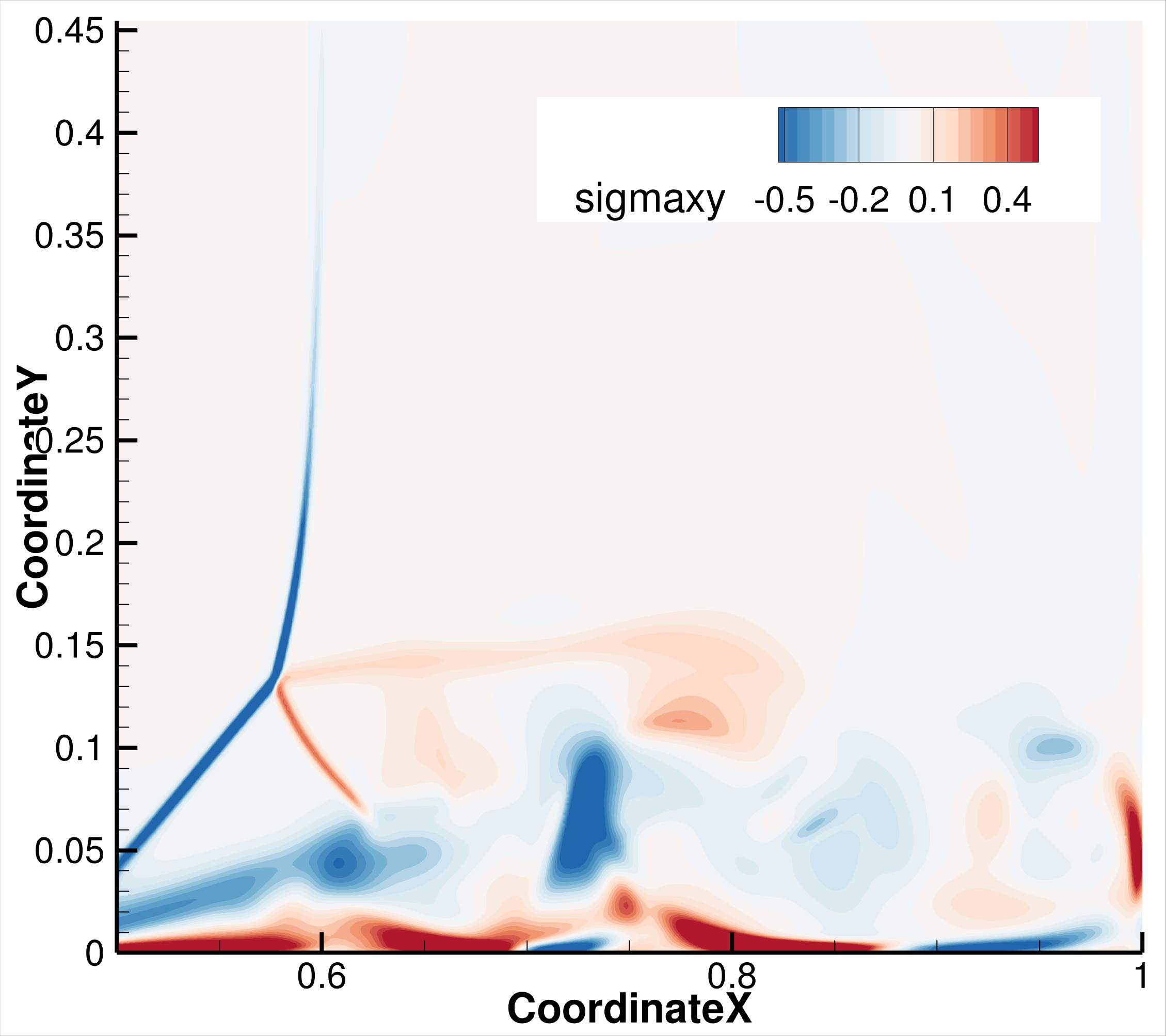}
  \includegraphics[width=0.3\textwidth]{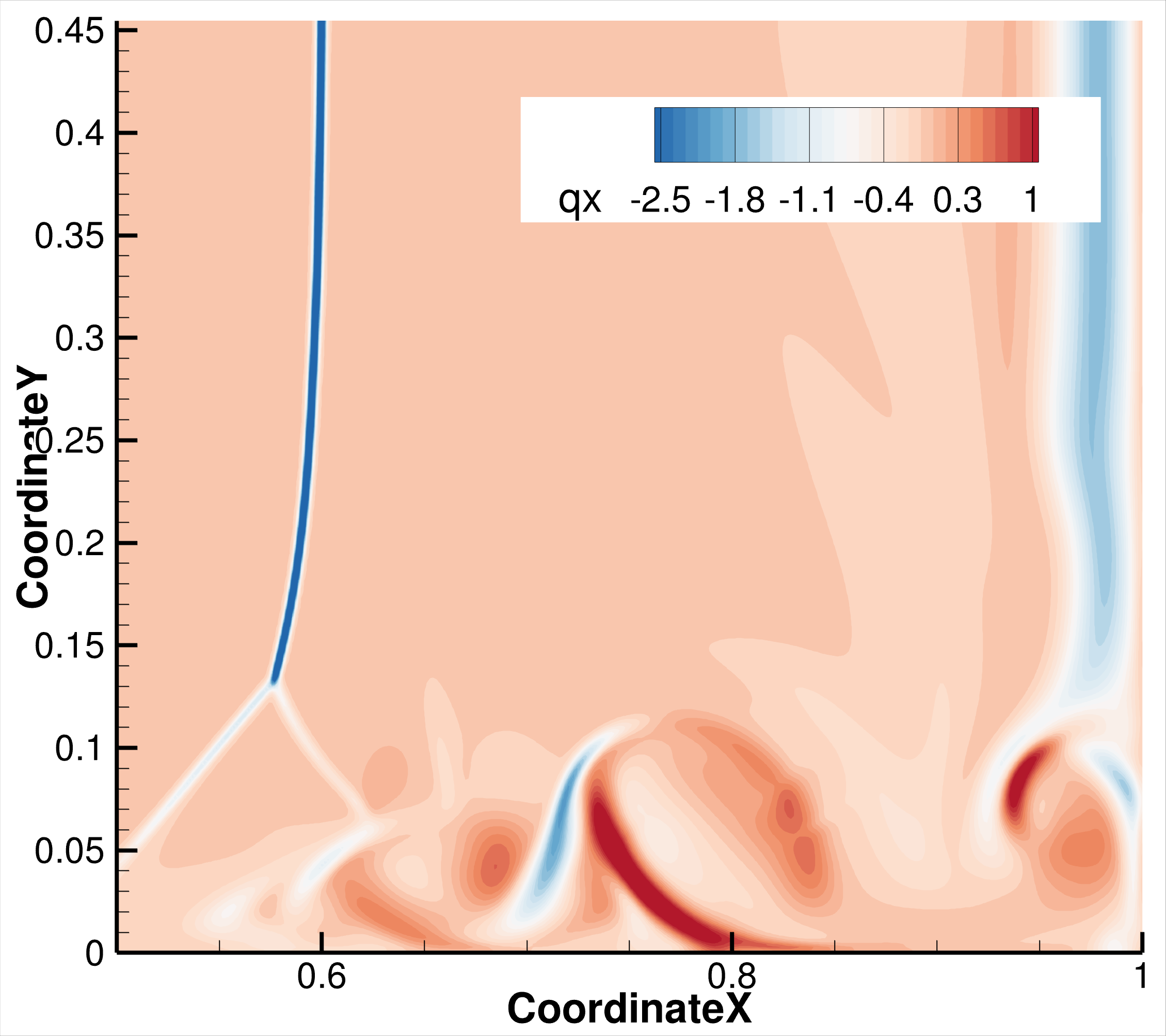}
  \includegraphics[width=0.3\textwidth]{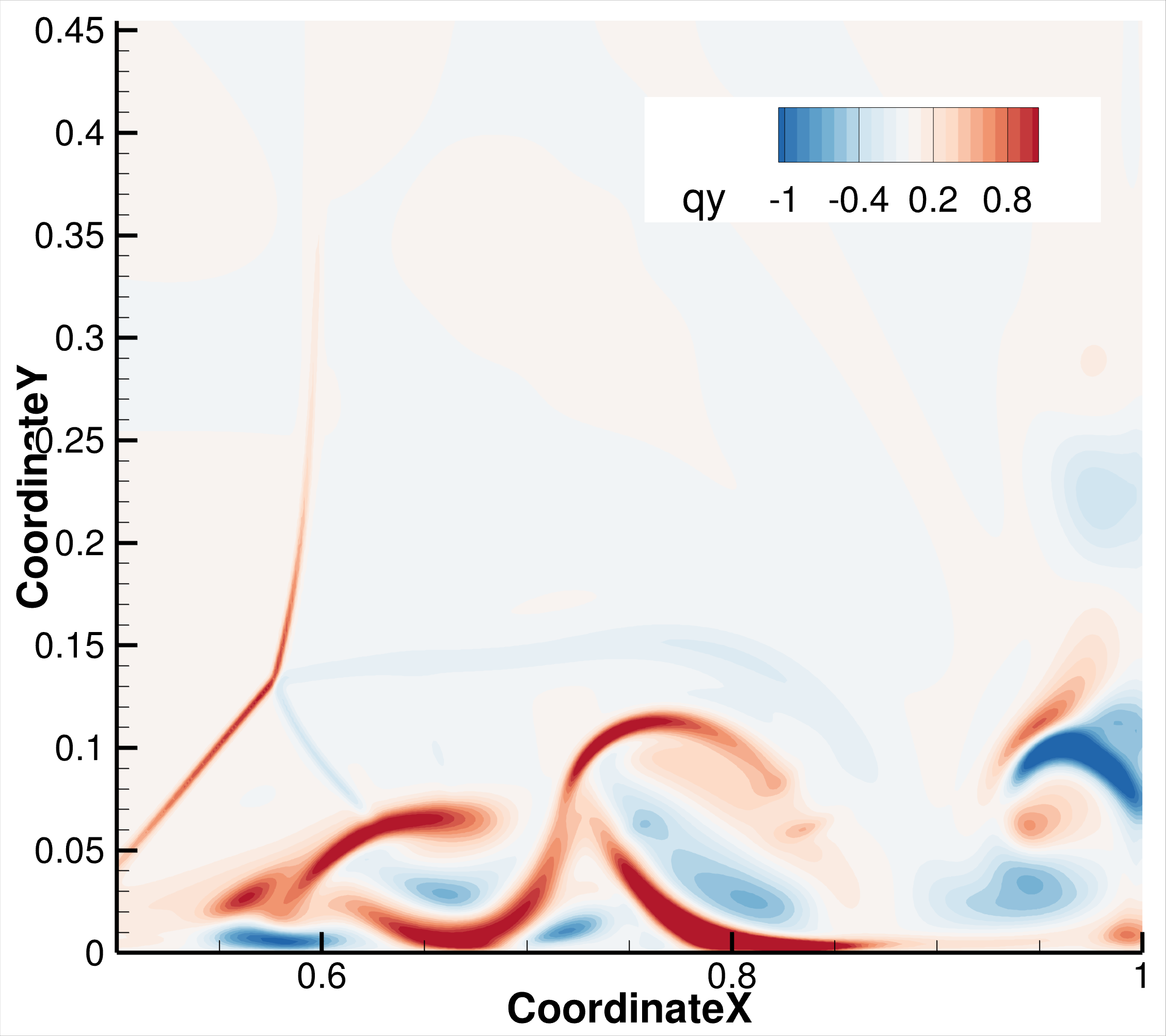}
  }
  \subfigure[]{
  \includegraphics[width=0.3\textwidth]{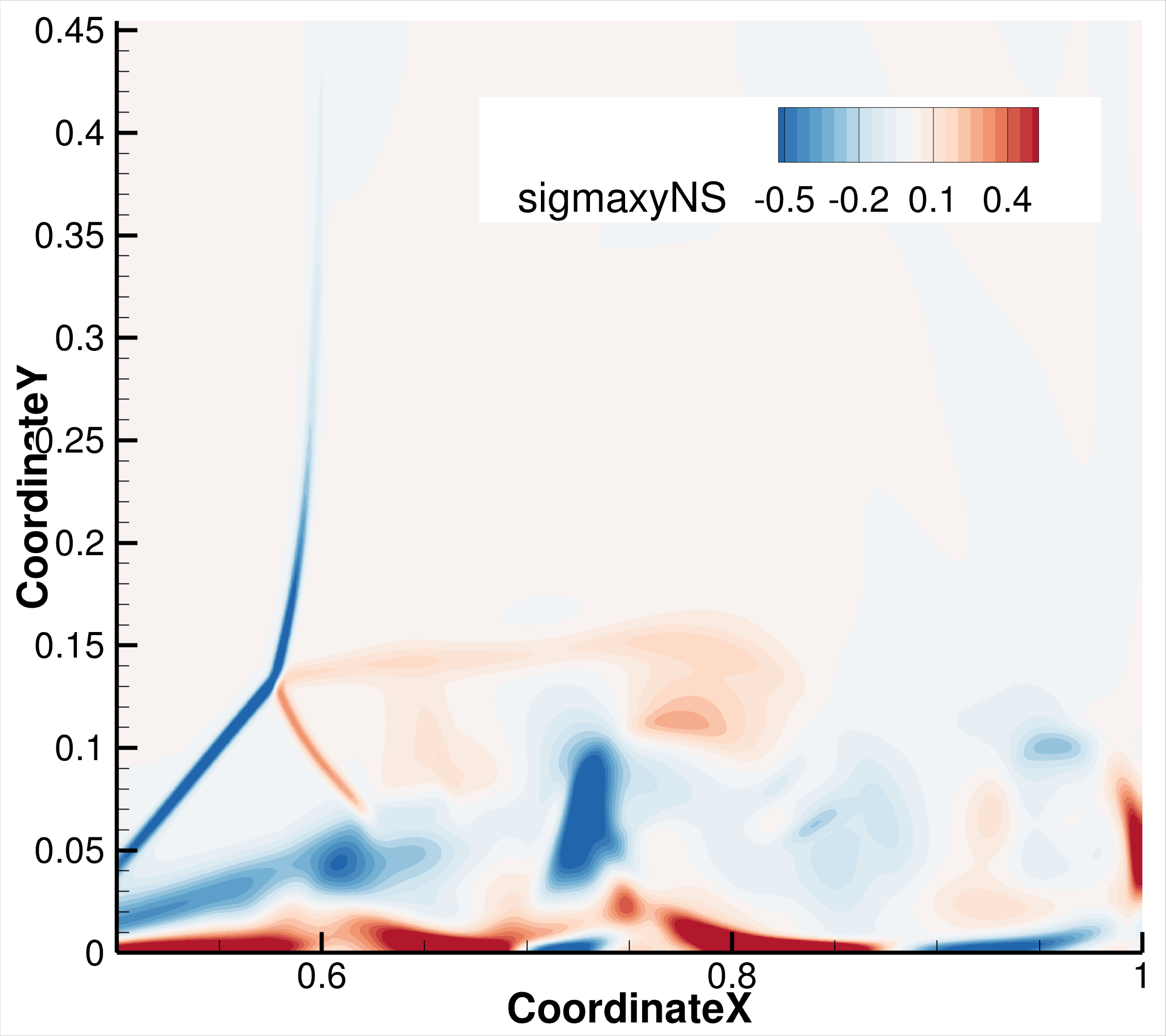}
  \includegraphics[width=0.3\textwidth]{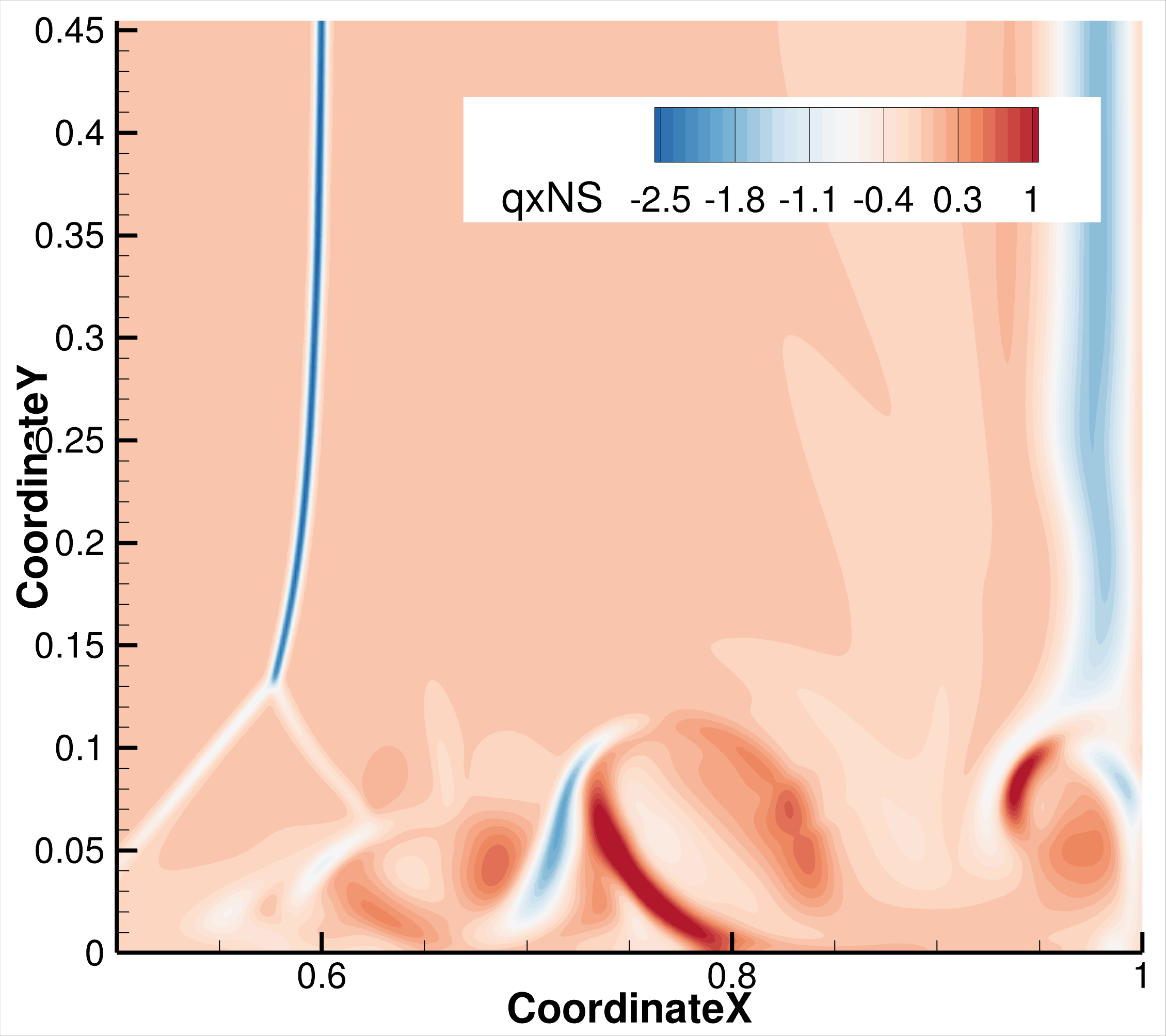}
  \includegraphics[width=0.3\textwidth]{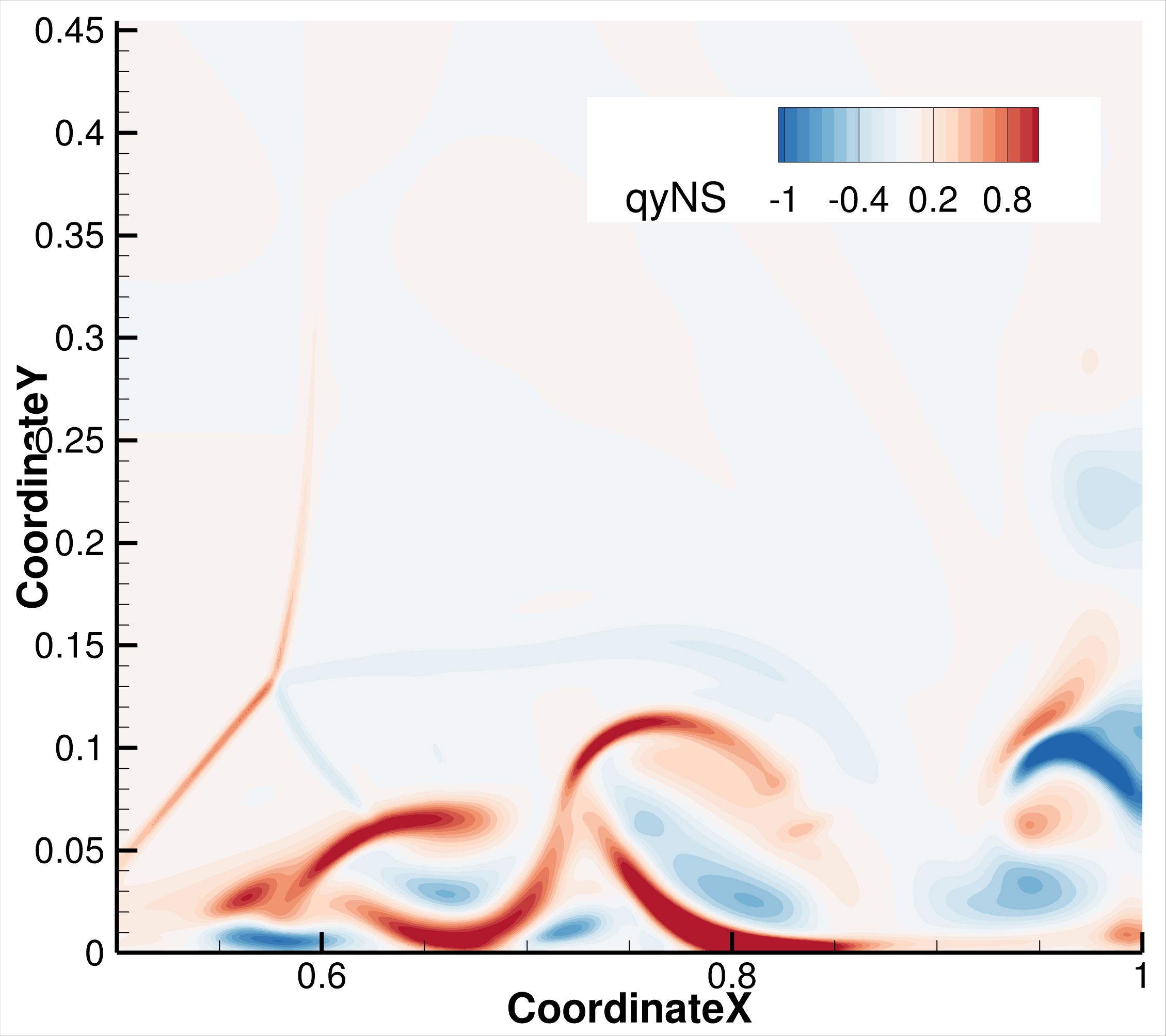}
  }
  \subfigure[]{
  \includegraphics[width=0.3\textwidth]{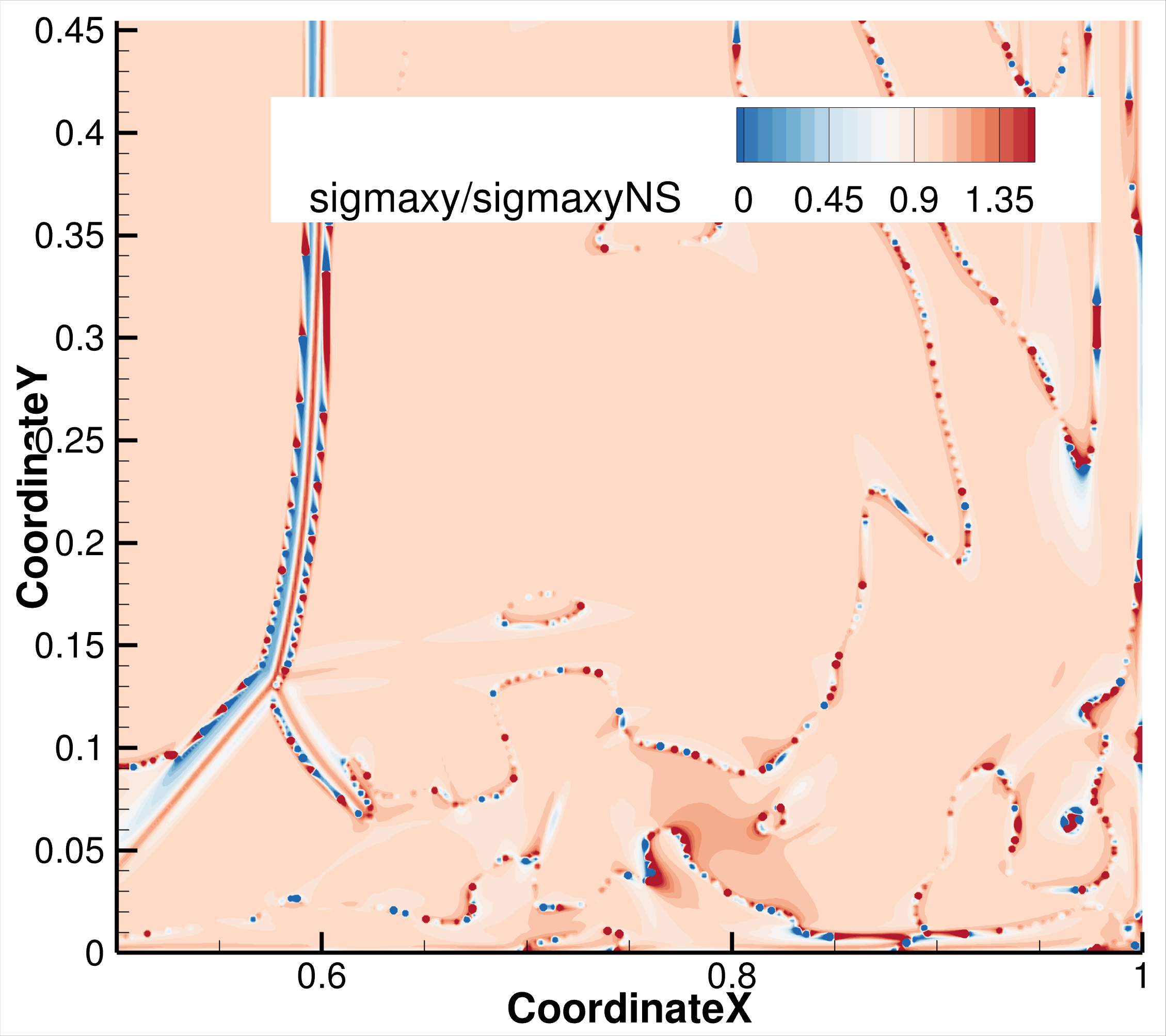}
  \includegraphics[width=0.3\textwidth]{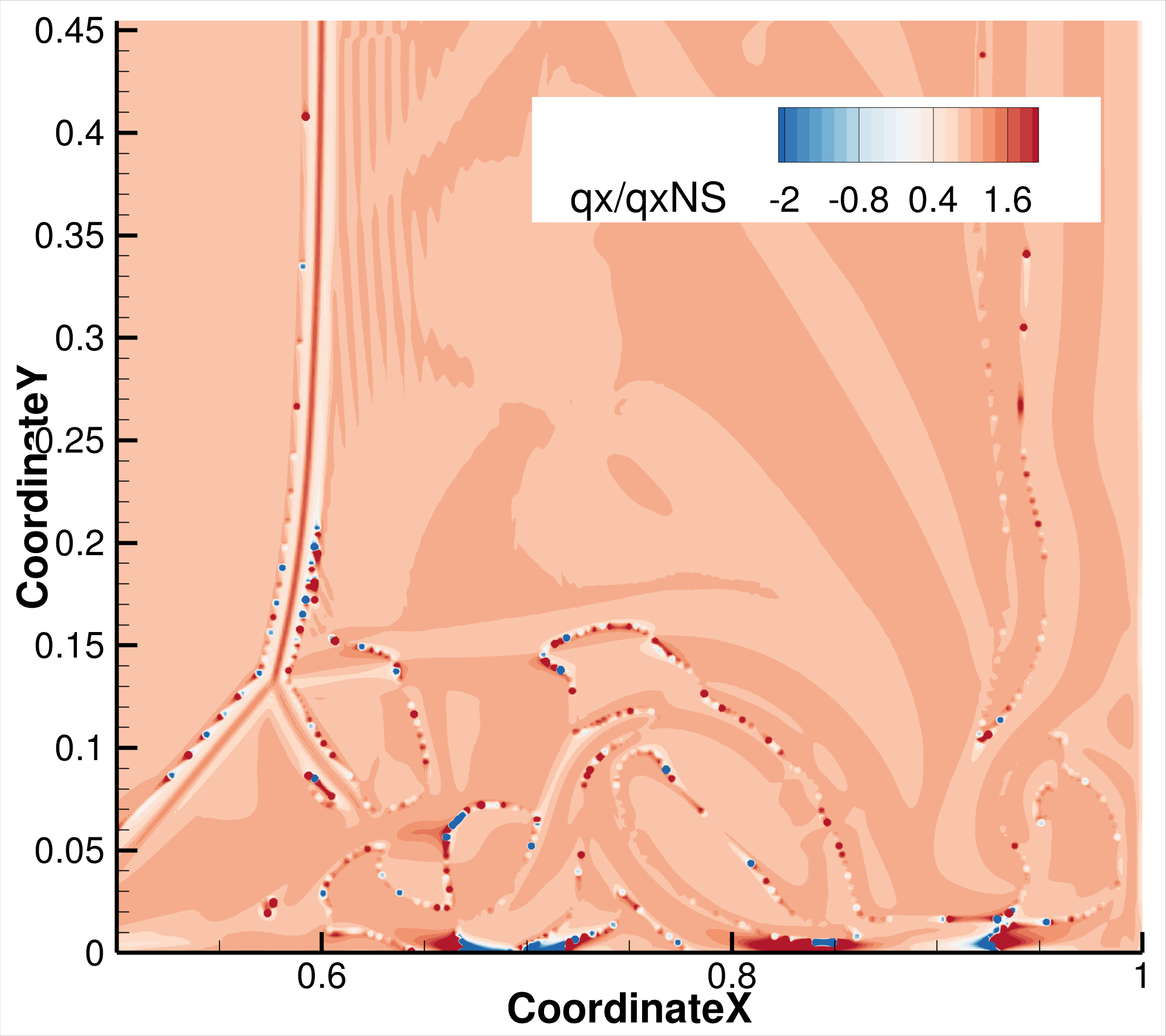}
  \includegraphics[width=0.3\textwidth]{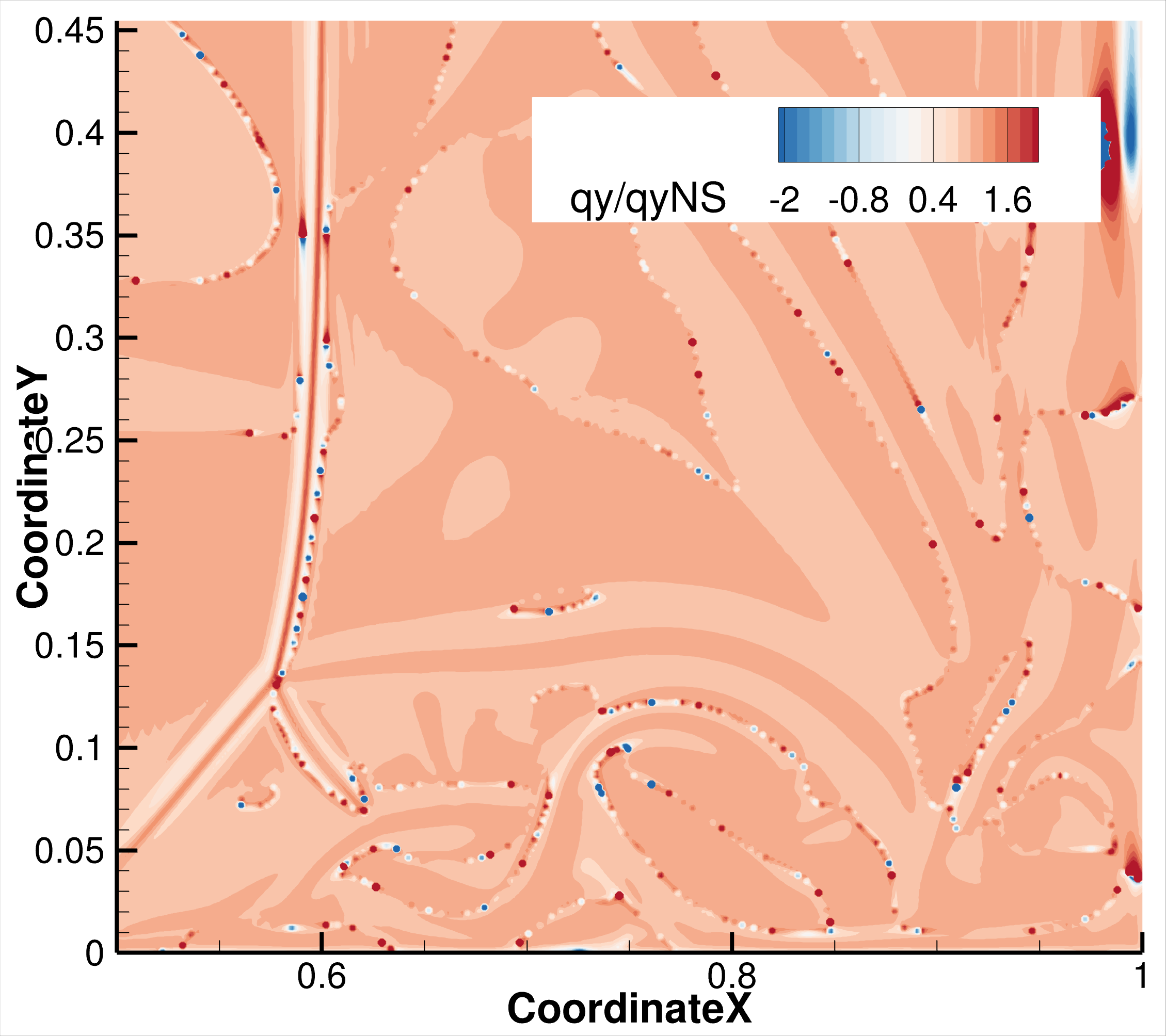}
  }
  \caption{The $\sigma_{xy}$ and $q_x$ and $q_y$ contour of UGKS at the time of $t=1.0$ at the Reynolds number of 100. (a) Calculated by the momentum of distribution function, (b) Calculated by the Newton stress tensor and Fourier's law of heat conduction, (c) The ratio of (a) and (b).}
  \label{fig:sigmaxyContourt1.0Re100}
\end{figure}

Figures \ref{fig:sigmaxyContourt1.0} and \ref{fig:sigmaxyContourt1.0Re100} present the stress and heat flux tensors obtained using the unified gas-kinetic scheme (UGKS) at $t=1.0$ for Reynolds numbers of 50 and 100. Two formulations for these macroscopic quantities are compared: one derived directly from the moments of the velocity distribution function, and the other calculated using the Newtonian viscous stress model and Fourier's law of heat conduction. The most significant deviations between the two methods occur at the shock waves, where strong non-equilibrium effects render the continuum hypothesis inadequate for accurately approximating stress and heat flux. Furthermore, at Re = 50, substantial discrepancies arise both near the wall and within the interaction region between the vortex and the main flow. In certain areas, the two methods even predict opposite signs for the stress and heat flux, a phenomenon particularly pronounced in the
$x$-direction heat flux profiles. At Re = 100, however, the differences are primarily concentrated near the wall, which aligns with the discrepancies observed in the near-wall density distribution shown in Figure \ref{fig:densityProfileLiney0t1.0}.

\section{Conclusion}
\label{sec:conclusion}

In this study, we investigated the viscous shock-tube problem in the low-Reynolds-number regime, demonstrating that non-equilibrium effects can be significant even within the nominal continuum flow regime. To this end, we compared a continuum solver, the gas-kinetic scheme (GKS), with a multiscale solver, the unified gas-kinetic scheme (UGKS), across both one- and two-dimensional configurations. The results indicate that non-equilibrium effects are most pronounced in the one-dimensional case at
$Re=50$ and during the early stages of the two-dimensional evolution, primarily driven by non-equilibrium behavior within the shock structure. In contrast, during the later stages of the two-dimensional evolution, the discrepancies between the solvers become more pronounced at
$Re=100$, where non-equilibrium effects associated with the developing boundary layer dominate. Overall, these findings reveal that the viscous shock-tube problem—commonly employed as a standard benchmark in the continuum regime—exhibits inherently multiscale characteristics. Consequently, non-equilibrium effects must be carefully accounted for in numerical simulations, even when the flow is traditionally classified as a continuum. This observation challenges conventional assumptions and underscores the necessity of multiscale kinetic approaches for accurate flow prediction.

\section{Acknowledgments}
We would like to thank Xing Ji for the helpful discussion and valuable suggestions. This work was supported by the National Key R\&D Program of China (Grant No. 2022YFA1004500), the National Natural Science Foundation of China (92371107), and the Hong Kong Research Grant Council (16208324).

% %% The Appendices part is started with the command \appendix;
% %% appendix sections are then done as normal sections
% %% \appendix

% %% \section{}
% %% \label{}

\bibliographystyle{elsarticle-harv}
\bibliography{jfm}
%\printbibliography
\end{document}